\documentclass[twoside,english,brazilian]{UNISINOSmonografia}
\usepackage[utf8]{inputenc} 
\usepackage[T1]{fontenc} 
\usepackage{graphicx} 
\usepackage{bibentry} 
\usepackage{xcolor} 
\usepackage{booktabs} 
\usepackage{multirow} 
\usepackage{rotating}
\usepackage{mathtools}
\usepackage{amsmath, amssymb}
\usepackage{pdfpages}
\usepackage{scalefnt}

\unisinosbst

\autor{Bischoff}{Vinicius}
\titulo{FMIT: uma técnica de integração de modelos de \textit{features}}
\orientador[Prof.~Dr.]{Silva Farias de Oliveira}{Kleinner}
\coorientador[Prof.~Dr.]{Barbosa}{Jorge Luis Victória }
 
\unidade{Unidade Acadêmica de Pesquisa e Pós-Graduação}
\curso{Programa de Pós-Graduação em Computação Aplicada}
\nivel{Nível Mestrado}
\natureza{%
Dissertação apresentada como requisito parcial para a obtenção
do título de Mestre pelo Programa de Pós-Graduação em Computação
Aplicada da Universidade do Vale do Rio dos Sinos --- UNISINOS
}
\local{São Leopoldo}
\ano{2017}

\palavrachave{brazilian}{Engenharia de software}
\palavrachave{brazilian}{Software – Reutilização}
\palavrachave{brazilian}{Software – Desenvolvimento}

\palavrachave{english}{Software Engineering}
\palavrachave{english}{Software - Reuse}
\palavrachave{english}{Software - Development}

\begin{document}
\includepdf[pages=1]{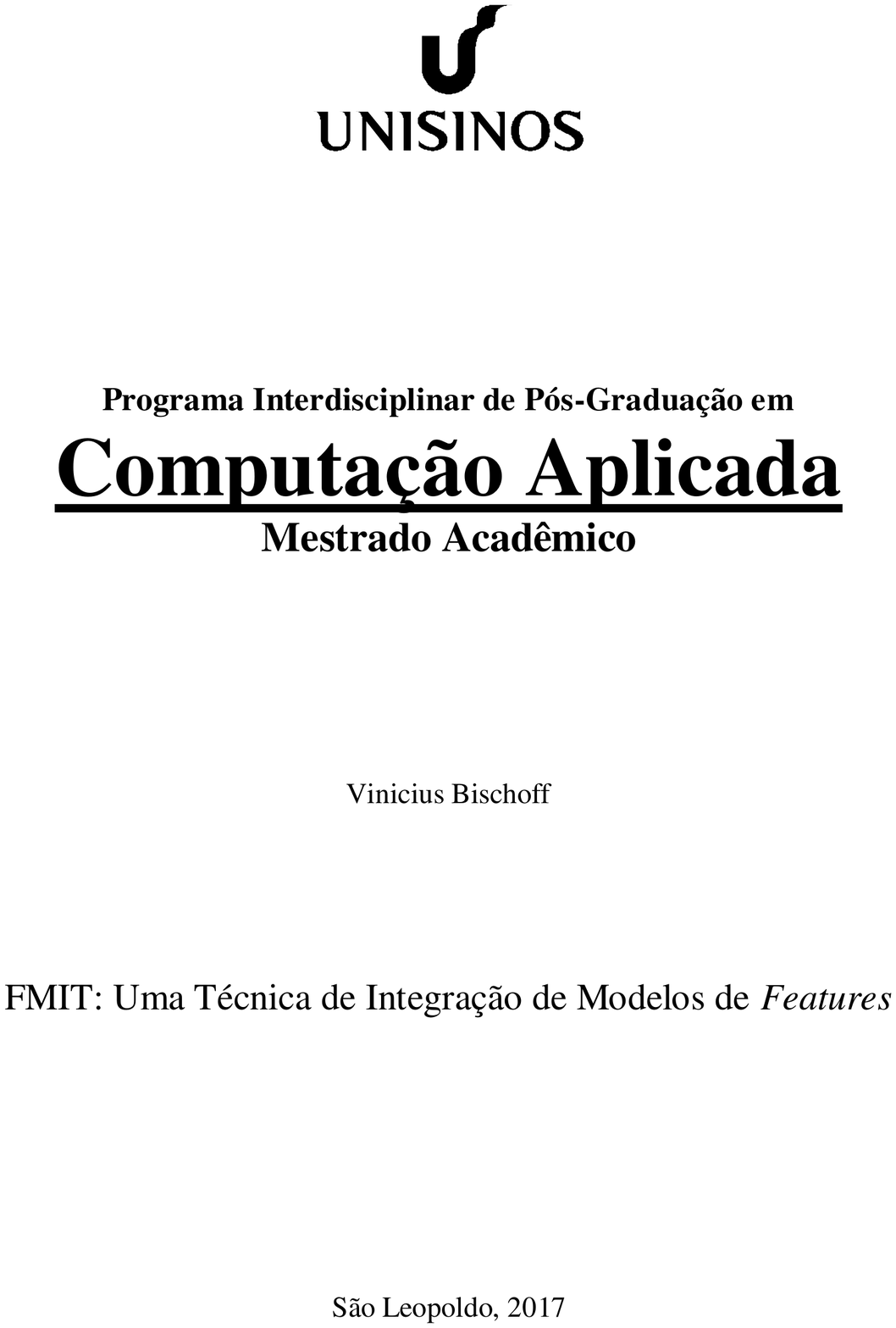}
\folhaderosto
\includepdf[pages=1]{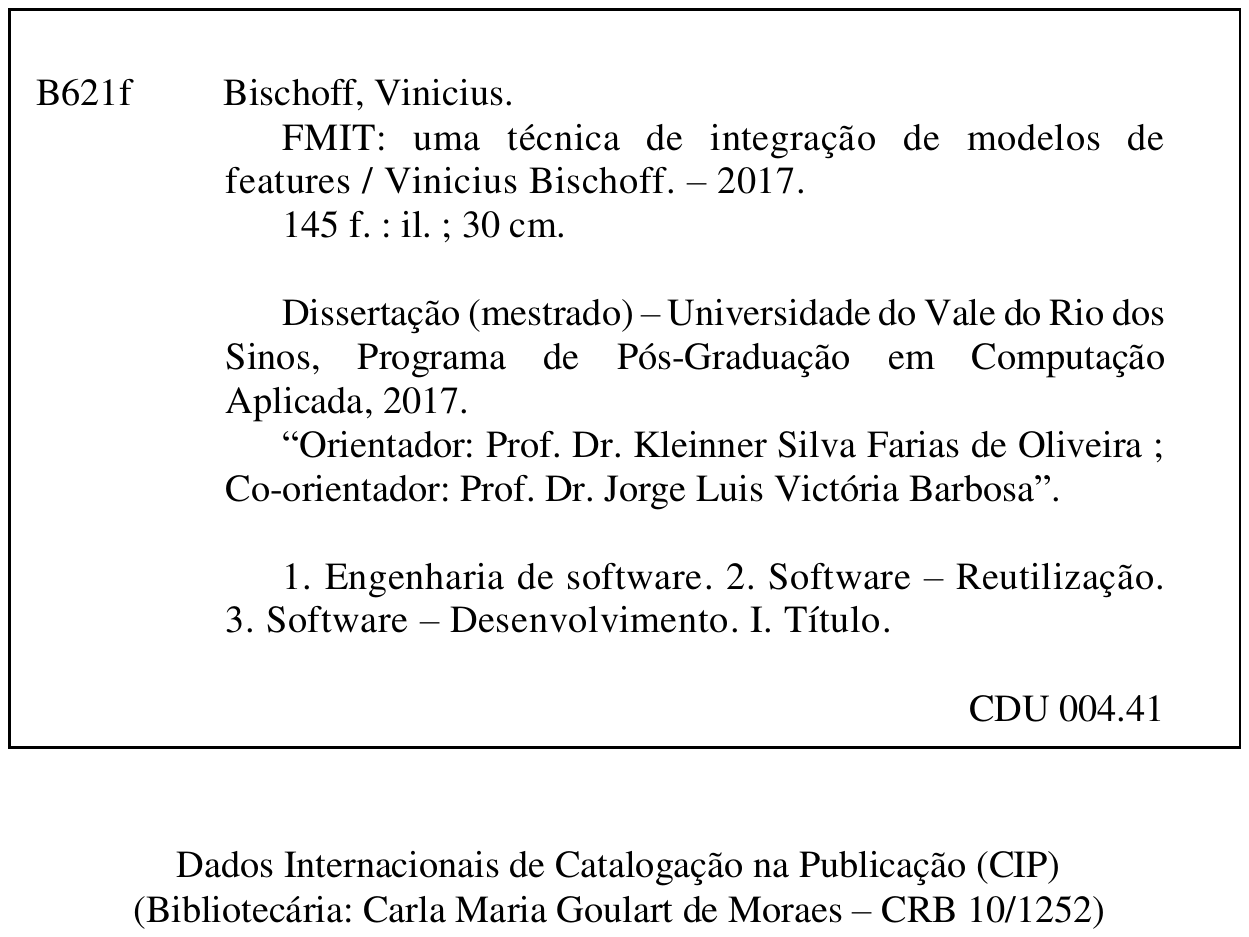}
\includepdf[pages=2]{capa.pdf}

%
\begin{dedicatoria}
\textit{Dedico este trabalho a todos que acreditam que 
\\a construção do conhecimento se dá pela interação, 
\\troca de saberes e compartilhamento de experiências.} 
\begin{itshape} 

\end{itshape}

\end{dedicatoria}

\begin{agradecimentos}
A Deus, por me oferecer oportunidades maravilhosas na minha vida, uma fonte inesgotável de força, paz e sabedoria.
 
Agradeço em especial à minha família, aos meus pais e irmãos, minha querida esposa, Arlete e meu filho, João Vicente, pela atenção, carinho e incentivo durante todo o momento. Obrigado pela dedicação e paciência em compreender a dura e árdua jornada enfrentada.

Agradeço a todos que de uma forma ou outra contribuíram para o desenvolvimento deste trabalho, entre eles, professores, colegas de mestrado.

Ao meu orientador  Prof. Dr. Kleinner Silva Farias de Oliveira e co-orientador Prof. Dr. Jorge Luis Victória Barbosa que depositaram confiança no meu trabalho e me incentivaram todo o momento. Obrigado pela atenção, dedicação e conhecimento passado. Seus ensinamentos me ajudaram a ver a vida de um modo diferente.
Meus sinceros agradecimentos.

Ao amigo e colega de mestrado, Lucian José Gonçales,  pela motivação, troca de saberes e compartilhamento de experiências.
 
À CNPq pelo apoio financeiro.
\end{agradecimentos}

%

%
\begin{abstract}
	Embora os modelos de \textit{features} sejam amplamente utilizados na prática, por exemplo, representando a variabilidade nas linhas de produto de software, a sua integração ainda é um desafio. Muitas técnicas de integração têm sido propostas, entretanto nenhuma destas se mostrou totalmente eficaz.  A integração de modelos de \textit{features} se torna uma tarefa difícil,  custosa e propensa a erros. Uma vez que a sua transição ocorre de forma generalizada e automatizada, as técnicas aplicadas para compor os modelos acabam originando um modelo final, em muitos casos indesejado, sem levar em consideração as necessidades especificas oriundas dos requisitos determinados pelos analistas e desenvolvedores. Portanto, este trabalho propõe a FMIT, uma técnica de integração de modelos de features. A FMIT é baseada em estratégias de integração contemporâneas de modelos, visando aumentar a precisão e a qualidade do modelo de \textit{feature} integrado. Dessa forma, será possível identificar o grau de similaridade entre diagramas de \textit{features} compostos, verificar sua precisão, bem como identificar conflitos. Além disso, este trabalho propõe o desenvolvimento de um protótipo fundamentado no conjunto de estratégias, empregados para a tomada de decisões conforme os requisitos estabelecidos durante a integração de modelos de \textit{features} seja este de modo semiautomático ou automático. Para avaliar a FMIT, estudos experimentais foram realizados com 10 participantes incluindo  estudantes e profissionais. Os participantes  executaram 12 cenários de integração, sendo 6 utilizando a FMIT e 6 de forma manual. Os resultados obtidos sugerem que a FMIT melhorou a precisão 43\% dos casos, bem como reduziu o esforço em 70\% para realizar as integrações.
\end{abstract}

%
\begin{otherlanguage}{english}
\begin{abstract}
Although feature models are widely used in practice, for example, representing variability in software product lines, their integration is still a challenge. Many integration techniques have been proposed, although none of these have proven to be fully effective. Integrating feature models becomes a difficult, costly, error-prone task. Since their transition occurs in a generalized and automated way, the techniques applied to compose the models end up giving rise to a final model, in many cases undesired, without taking into account the specific needs arising from the requirements determined by the analysts and developers. Therefore, this work proposes FMIT, a technique for integrating feature models. The FMIT is based on contemporary model integration strategies to increase the accuracy and quality of the integrated feature model. In this way, it will be possible to identify the degree of similarity between composite feature diagrams, to verify their accuracy, as well as to identify conflicts. In addition, this work proposes the development of a prototype based on the set of strategies, used to take decisions according to the requirements established during the integration of feature models, whether this is semi-automatic or automatic. To evaluate FMIT, experimental studies were conducted with 10 participants, including students and professionals. Participants performed 12 integration scenarios, 6 using the FMIT and 6 manually. The results suggest that FMIT improved accuracy by 43\% of the cases, as well as reduced the effort by 70\% to perform the integrations.   
\end{abstract}
\end{otherlanguage}

\listoffigures

\listoftables

%


%
\begin{listadesiglas}{PIPCA} 
\item[ES]  Engenharia de Software
\item[LPS] Linha de Produto de Software
\item[MF]  Modelo de Features
\item[FODA] Feature-Oriented Domain Analysis
\item[FORM] Feature-Oriented Reuse Method 
\item[CBFM] Cardinality-based feature model
\item[FRSEB]Feature Reuse-Driven Software Engineering Business 
\item[PLUSS] Product Line UML-Based Software Engineering 
\item[CVL] Common Variability Language
\item[TVL] Text-based Variability Language 
\item[FOPLE] Feature Oriented Product Line Software Engineering  
\item[FPA] Famílias de Produtos da Álgebra 
\item[BDD] Diagramas de Decisão Binária
\item[SAT] Problema de Satisfação Booleana 
\item[CSP] Problemas de Satisfação de Restrições 
\item[FNC] Forma Normal Conjuntiva 
\item[FND] Forma Normal Disjuntiva
\item[EMF] Eclipse Metamodel Framework
\item[MOF] Meta Object Facility
\item[GQM] Goal/Question/Metrics
\item[ID] Identificação
\item[CEE]Calculo de Efetividade Estratégica
\item[IDE] Integrated Development Environmen
\item[FMIT] Feature Model Integration Tool
\item[PIPCA]Programa Interdisciplinar de Pós-Graduação em Computação Aplicada
\end{listadesiglas}

%


\tableofcontents

\chapter{Introdução} \label{I_introdução}

Nos últimos anos a Engenharia de \textit{Software} (ES) vem procurando aumentar a capacidade produtiva no desenvolvimento de \textit{software} e tem buscado alternativas que auxiliem a remover as dificuldades. A reutilização de \textit{software} é uma das estratégias empregadas pela engenharia de \textit{software} no desenvolvimento de sistemas com a finalidade reduzir tempo e custo de produção \cite{pohl2005}.

As linhas de produto surgiram com esta motivação, isto é, para criar famílias de produtos com características comuns entre si e sistematizar a reutilização. A combinação dos conceitos de famílias de produtos e customização originou a Linha de Produto de \textit{Software} (LPS), um conjunto de sistemas de \textit{software} definidos sobre uma arquitetura comum que compartilham um mesmo conjunto de recursos \cite{czarnecki2000}. Assim, no processo de engenharia de uma LPS, o passo inicial consiste em identificar necessidades similares e variabilidades dentro de um domínio de aplicação. A partir da similaridades e variabilidades encontradas no domínio, é possível derivar diferentes produtos. Esta organização poderá possibilitar melhorarias no tempo de entrega e qualidade final dos produtos bem como nos seus custos \cite{pohl2005, braga1999}. O tripé, tempo, qualidade e custos procura direcionar e coordenar como as técnicas de variabilidade e gerenciamento tendem a auxiliar nas estratégias condizentes a uma melhoria continua no processo de produção de \textit{software}.

Um dos principais artefatos nas LPS é o modelo de \textit{features} \cite{segura2008}. O modelo de \textit{features} descreve as características identificadas nos possíveis produtos a serem produzidos e as relações existentes entre as mesmas, representando todas as similaridades e variabilidades em uma LPS \cite{benavides2010}. Deste modo, o modelo consiste em um diagrama, onde cada nó representa possíveis \textit{features} de sistemas que serão desenvolvidos na linha de produto. Uma \textit{feature}, por sua vez, é uma característica comum e variável entre sistemas de um determinado domínio de aplicação, possibilitando a geração de diversos produtos com características diferentes \cite{batory2005, apel2009, becan2015}. As \textit{features} expressam portanto, similaridades e variabilidades de produtos de um domínio, podendo significar algo diferente para diferentes linhas de produto e domínio, tais como um requisito, uma funcionalidade ou um aspecto de qualidade. 

A Linha de Produto de \textit{Software} é representada por diversas técnicas como, por exemplo, os métodos FODA \cite{kang1990},  FORM \cite{kang1998}, CBMF \cite{czarnecki2005}, FeatuRSEB \cite{favaro2009}, Pluss \cite{eriksson2005}, Odyssey \cite{braga1999}.  As técnicas propõem ou se utilizam de uma notação ou modelo para representar a variabilidade do domínio ou arquitetura.  O modelo de variabilidade consiste em demonstrar as funcionalidades de um domínio através de suas características assim como seus respectivos relacionamentos e interdependências por meio de uma estrutura hierárquica \cite{czarnecki2000, batory2005, benavides2010}. A forma de representar a variabilidade de uma Linha de Produto de \textit{Software} é através do Modelo de \textit{Features} (MF). A adoção do modelo de \textit{features} emerge em decorrência da internacionalização dos processos de produção, tornando-se comum em projetos de desenvolvimento de \textit{software} na indústria \cite{beuche2007, segura2008, acher2009}. 
 
Pesquisadores e profissionais têm amplamente utilizado modelos de \textit{features} para diferentes propósitos, tais como: (1) gerenciar a variabilidade no contexto de LPS, ajudando a descrever conceitos de domínio em termos de suas semelhanças e diferenças dentro de uma família de sistemas de \textit{software} \cite{acher2010}; (2) especificar as \textit{features} e suas dependências, derivando automaticamente os produtos da LPS \cite{beuche2007}; (3) descrever variabilidade em LPS, auxiliando a derivação de diferentes produtos da linha \cite{segura2008}; (4) documentar as \textit{features} e suas combinações válidas para possibilitar o reuso estratégico dos seus artefatos \cite{benavides2010}; ou mesmo auxiliar os desenvolvedores a integrar as \textit{features} de uma família de sistemas de \textit{software} \cite{acher2009, thum2009, acher2010}. 

A integração de modelos, tanto na academia como na indústria de \textit{software}, tem sido uma preocupação constante dos pesquisadores \cite{benavides2010, berger2013,apel2014}, os quais têm procurado elaborar técnicas para apoiar a integração de modelos de \textit{features} heterogêneas. Sem esse suporte técnico, a produção de modelos de \textit{feature} desejada torna-se uma tarefa propensa a erros e que consome grande esforço \cite{farias2015, kleinner2012,acher2009}. Além disso, a integração de modelos de \textit{features} tem sido amplamente investigada na prática, dado seu papel fundamental para apoiar a evolução das LPS \cite{batory2005, benavides2010, berger2015}.

No entanto, as técnicas propostas na literatura, tais como, \cite{czarnecki2000, batory2005, kolovos2006, segura2008, acher2009} demonstraram ser insuficientes para apoiar a integração de modelos de \textit{features}. Por exemplo, as técnicas de: Programação Generativa (PG) baseia-se nas LPS e centra-se na automatização do desenvolvimento com base no reuso de componentes. Este trabalho propõe melhorias para as técnicas de formalização do modelo \textit{features}, bem como discute o suporte de ferramentas de modelagem os quais não auxiliam nas configurações e composições de modelos de \textit{features} \cite{czarnecki2000}; apesar dos anos de evolução, as ferramentas muitas vezes fornecem suporte limitado para restrições de \textit{features} e oferecem pouco ou nenhum suporte para depurar modelos de \textit{features} \cite{batory2005}; discutem os requisitos sintáticos, bem como a comparação e composição de modelos. Aplicam uma abordagem baseada em regras para realizar a comparação automatizada em modelos MOF e EMF, podendo ser expandida para outros modelos. Outra possível extensão citada é usar o dicionário de sinônimos \cite{kolovos2006}; os autores apresentam um catálogo de regras visuais para automatização dos modelos independente da tecnologia utilizada, esta abordagem aplica a técnica de análise de par crítico para detectar conflitos entre as regras de composição, entretanto as regras são estabelecidas somente para os relacionamentos, pois entre as \textit{features} sua aplicação é automática, o que a torna inflexível, nem sempre produzido modelos corretos \cite{segura2008}; os autores propõem a integração de dois modelos \textit{feature},  aplicando mecanismos de comparação sintáticos e semânticos, bem como abordam a utilização de um conjunto de operadores, isto é, união e interseção para os compor modelos, no entanto falta verificar sua aplicabilidade e usabilidade dos operadores proposto \cite{acher2009}. 

Assim, o processo de integração de modelos de \textit{features}, está sujeito às limitações resultantes da inflexibilidade na evolução de um novo modelo de \textit{features}. As técnicas propostas para automatizar a integração citadas anteriormente apresentam algumas lacunas, como: (1) suporte ferramental que auxiliem as integrações \cite{czarnecki2000}, (2) suporte limitado para comparar modelos de \textit{features} \cite{batory2005}, (3) falta do dicionário de sinônimos \cite{kolovos2006}, (4) catálogo de regras visuais flexível \cite{segura2008}, (5) ampliar operações de composição (diferença e complemento)\cite{acher2009}, os quais acabam em muitos casos produzindo uma derivação incorreta que pode comprometer o modelo, gerando retrabalho para as equipes de desenvolvimento causando um impacto direto no esforço, nos custos de produção e, principalmente, na qualidade do produto gerado.

Deste modo há um desencontro com o alinhamento das estratégias de integração recomendadas na literatura, bem como os propósitos estabelecidos através das Linhas de Produto de Software, ou seja, as indústrias passam a reduzir uma parcela de sua capacidade produtiva e competitiva. Neste contexto, o presente trabalho, procura superar essas deficiências visando aperfeiçoar o desenvolvimento das técnicas de integração de modelos de \textit{features} e auxiliar os desenvolvedores durante o processo de integração destes modelos.  

\section{Problemática}

Considerando que os Modelos de \textit{Features} (MF) podem ser criados em colaboração por diferentes equipes de desenvolvimento de \textit{software} \cite{vcavrak2012,werner2003}, em algum momento os modelos criados em paralelo devem ser integrados para formar uma visão geral das variablilidades da LPS. Por esta razão, várias técnicas de integração de modelos de \textit{features}, têm sido propostas nos últimos anos, tais como, \cite{benavides2010, acher2010, thum2009, apel2014}. 

A integração de modelos de \textit{features} pode ser brevemente definida neste trabalho como uma operação em que um conjunto de tarefas deve ser executado em dois modelos de \textit{features} de entrada, (MF$_{A}$) e (MF$_{B}$), a fim de produzir um modelo de saída, (MF$_{AB}$). Enquanto MF$_{A}$ representa o modelo base (origem), MF$_{B}$, consiste no modelo delta (destino) com todos os incrementos que devem ser inseridos, excluídos ou alterados em MF$_{A}$ para transformá-lo em MF$_{AB}$. As técnicas de integração normalmente produzem um modelo integrado, (MF$_{IA}$) que muitas vezes não corresponde ao modelo de saída pretendido, MF$_{AB}$, isto é, MF$_{IA}$ $ \neq $ MF$_{AB}$. 

Este fato ocorre porque os elementos dos modelos de entrada, (MF$_{A}$) e (MF$_{B}$) geralmente apresentam informações conflitantes, e essas técnicas acabam sendo incapazes de solucionar todos os conflitos corretamente.

\begin{figure}
	\caption{Exemplo de uma integração de modelo de \textit{feature}.}
	\label{fig:conflito}
	\centering%
	\begin{minipage}{.8\textwidth}
		\includegraphics[width=\textwidth]{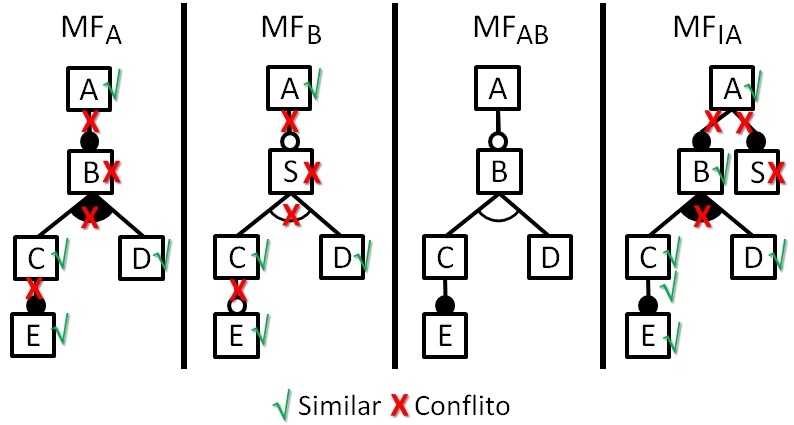}
		\fonte{Elaborado pelo Autor.}
	\end{minipage}
\end{figure}

Essa situação está representada na Figura \ref{fig:conflito}, a qual exibe a dificuldade da integração dos modelos, em um cenário simples formado por dois modelos de \textit{features} de entrada, MF$_{A}$ e MF$_{B}$, produzindo o modelo de saída  MF$_{AB}$, ou seja, o modelo pretendido e o modelo integrado, MF$_{IA}$, o qual exibe sua respectiva saída. Pode-se observar os conflitos semânticos e sintáticos entre os modelos, MF$_{A}$ e MF$_{B}$, ao realizar a comparação sintática há um único conflito  a \textit{features} [B] e a \textit{features} [S] ambas assinaladas com um "X" e ao comparar os relacionamentos entre as \textit{features} são identificados três conflitos, (obrigatório, um circulo preto; opcional, um circulo branco; ou-inclusivo, arco preto e, por fim,  ou-exclusivo, arco branco) ambos identificados com um "X", já as \textit{features} assinaladas com um "V" são similares. Neste exemplo exibi-se duas saídas: (1) o modelo pretendido MF$_{AB}$, isto é, o modelo que deveria ser produzido e o (2) modelo integrado, MF$_{IA}$, percebe-se as inconsistências decorridas de uma má integração assinaladas com um "X", já as assinaladas com um "V" são similares. Assim, surge a necessidade de promover melhorias nas técnicas aplicadas e nas ferramentas que auxiliam  as equipes de desenvolvimentos na condução de se obter o modelo pretendido.

Na verdade, é muito difícil solucionar conflitos automaticamente \cite{farias2015, kleinner2012}, porque a resolução destes depende de um entendimento que o modelo de \textit{features} realmente significa, e tal informação raramente é representada de uma maneira formal, afetando o comportamento de outras \textit{features} de forma inadequada. Consequentemente, analistas e desenvolvedores acabam por ter de investir algum esforço extra para rever e alterar modelos inconsistentes e torná-los compatível como o modelo desejado. 

Conforme mencionado anteriormente, a literatura reporta que as técnicas de integração de modelos de \textit{features} disponíveis na academia não são precisas. Isto é, as técnicas disponíveis são inflexíveis, não permitindo a interação durante o processo de integração, tornando esta atividade complexa, dispendiosa e condizente a propensão de erros, bem como eleva o esforço das equipes de desenvolvimento em sua correção.

Deste modo, surgem demandas que caracterizam os problemas abordados, a seguir.
\begin{itemize}
\item \textbf{P-1:} A integração de modelos de \textit{features} conduzida de forma manual, bem como as ferramentas propostas na literatura que dão suporte a composição, acabam  em muitos casos produzindo como saída um modelo indesejado, devido a sua inflexibilidade durante o processo de integração, gerando assim, modelos imprecisos e possivelmente propagando erros para outras fases do desenvolvimento, impactando diretamente na qualidade final dos produtos.

\item \textbf{P-2:} Há uma serie de ações necessárias na correção dos modelos produzidos inadequadamente, reduzindo desempenho da força de trabalho e elevando esforços na detecção das inconsistências anteriormente produzidas, ocorrendo a execução deste processo manualmente, o que o torna ineficiente, portanto reduzindo a capacidade produtiva das equipes de desenvolvimento, elevando os custos de produção.
\end{itemize}

\section{Questões de Pesquisa} \label{IntQuestõesdePesquisa}
A partir dos problemas investigados fica evidente a necessidade de propor melhorias nas técnicas de integração e no aprimoramento das técnicas de comparação de modelos de \textit{features}, bem como a obtenção de soluções funcionais que auxiliem as equipes de desenvolvimento. Os trabalhos investigados na literatura representam um avanço não só para a academia, mas também para a indústria de \textit{software}, contudo há algumas limitações nas abordagens utilizadas. Sendo assim, a questão de pesquisa chave desta proposta é apresentada a seguir.\\ 

\colorbox{lightgray}{
\begin{minipage}[t]{0.9\linewidth}
\textbf{Questão de Pesquisa Geral:} Como aprimorar as técnicas de integração de modelos de \textit{features} para obter como saída o modelo mais próximo do desejado, sendo esta técnica uma facilitadora para as equipes de desenvolvimento?
\end{minipage}
}\\

Depois de definir a questão chave, surge a primeira questão de pesquisa, onde realizou-se uma análise entre as abordagens  manual e semiautomática. Visto que a técnica automática acaba em alguns casos produzindo modelos imprecisos os quais necessitam de ajustes para se obter o modelo pretendido. Sendo este processo executado de forma manual, surge a seguinte questão de pesquisa:
\begin{itemize}
\item \textbf{QP-1:} Aplicação da técnica semiautomática em relação a técnica manual melhora a corretude dos modelos de \textit{features} integrados?
\end{itemize}

A segunda questão de pesquisa procura identificar e reduzir os esforços na resolução de conflitos, assim como verificar a compreensão das equipes de desenvolvimento na tomada de decisão. 

\begin{itemize}
\item \textbf{QP-2:} Qual o impacto da aplicação técnica semiautomática em relação a técnica manual no que diz respeito ao esforço na resolução de conflitos?
\end{itemize}

\section{Objetivos}\label{I_Objetivos}
Após contextualizar este estudo, apurar os principais problemas pesquisados, e abordar algumas das limitações dos trabalhos investigados, se estabeleceu o objetivo principal deste trabalho, podendo a partir deste elaborar as ações necessárias para o seu cumprimento, isto é, os objetivos específicos.\\

\fbox{
  \parbox {0.9\textwidth}{
    \textbf{Objetivo principal:} propor uma técnica que auxilie as equipes de desenvolvimento na integração de modelos de \textit{features}, reduzindo os esforços de produção, a propensão a erros, e por fim, elevar a precisão dos modelos integrados.
  }
}\\

Além do objetivo principal, definiu-se os seguintes objetivos específicos: (1) obter um panorama atual sobre o estado da arte, procurando entender como esta área tem se desenvolvido e quais as principais abordagens são utilizadas em sua implementação; (2) realizar estudos experimentais para avaliar a abordagem da técnica manual, ou seja, o modelo integrado em comparação ao modelo pretendido, para analisar como os analistas e desenvolvedores investem os esforços na composição dos modelos e na resolução de conflitos, bem como mensurar inconsistências surgidas durante a integração; (3) propor uma técnica para melhorar a precisão dos modelos integrados, com ênfase nas técnicas já documentadas na literatura que compreende as atividades de identificação de similaridade entre os elementos do modelo composto; (4) projetar e desenvolver um protótipo, com a finalidade de carregar os modelos, persistir, comparar sua equivalência e integrar os modelos de \textit{features}. Ao implementar o protótipo, espera-se que o mesmo seja um facilitador na tomada de decisões, visto sua flexibilidade podendo adaptar-se de acordo com as necessidades, quando há algum elemento conflitante; (5) realizar estudos empíricos para avaliar e mensurar a efetividade da técnica proposta, tendo em vista produzir conhecimento sobre sua aplicação.

\section{Metodologia}  
A viabilização deste trabalho ocorreu através de uma metodologia cuja primeira etapa consiste na realização de uma revisão da literatura através de um mapeamento sistemático sobre as técnicas de integração de modelos de \textit{features}, respeitando um conjunto de passos, ou seja, uma revisão planejada com o intuito de responder questões específicas e que utilizam métodos explícitos e sistemáticos para selecionar, identificar e avaliar os estudos incluídos na revisão \cite{kitchenham2010, kitchenham2011}. Essa revisão tem por objetivo analisar as técnicas disponíveis na academia e na indústria, bem como identificar as principais dificuldades enfrentadas por pesquisadores e profissionais, com a finalidade de elencar novas oportunidades de pesquisa. De acordo com \citetexto{Marconi2003} a revisão da literatura é usada para reunir informações sobre determinados tópicos, uma vez que procura propiciar maior familiaridade com o tema, além de aprofundar seus conceitos preliminares. 

A segunda etapa consiste em estudos empíricos de integração de modelos de \textit{features} manualmente. Os estudos procuram os fatores que afetam o processo de composição de modelos quando conduzido de forma manual a partir da realização experimentos com a finalidade de mensurar o esforço que as equipes de desenvolvimento investem para integrá-los, assim como verificar a ocorrência de conflitos nos modelos produzidos.

Na terceira etapa ocorre a especificação da arquitetura proposta, detalhando seus componentes principais tendo como finalidade determinar quais os elementos são necessários para o modelo e a interação entre estes. Em seguida, na quarta etapa foi desenvolvido um protótipo com todos os componentes necessários á sua execução. Na quinta etapa foram definidos possíveis cenários para aplicação da técnica de integração, bem como a definição de sua execução, isto é, um experimento controlado.

Finalmente, a última etapa compreende a execução de novos experimentos de integração de modelos de \textit{features} com o suporte do protótipo desenvolvido, com a finalidade de analisar dados comparativos do esforço aplicado e a corretude entre a integração manual e a integração semiautomática de modelos \textit{features}. Conforme \citetexto{gil2002}, os experimentos caracterizam a pesquisa como quantitativa quanto ao método, visto que expõem a comparação de medidas e uso de técnicas estatísticas para avaliação dos resultados.

\section{Organização do Trabalho} 
Este trabalho está organizado em sete capítulos, conforme descrito a seguir: o Capítulo \ref{FundamentaçãoTeórica} que relaciona os conceitos fundamentais para o entendimento deste trabalho. O Capítulo \ref{TrabalhosRelacionados} apresenta as principais abordagens que possuem relação com a proposta desta pesquisa, relacionando suas principais características e descreve as oportunidades de pesquisa. O Capitulo \ref{TécnicaDeIntegração} apresenta as técnicas de integração propostas neste trabalho. O Capítulo \ref{AspectosDeImplementação} introduz o protótipo de integração de modelo de \textit{feature} que implementa a técnica descrita no Capitulo \ref{TécnicaDeIntegração}. O capitulo \ref{AvaliaçãoDaSolução} descreve a avaliação da técnica e do protótipo proposto. Por fim, a conclusão deste trabalho é apresentada no Capítulo \ref{Conclusões}, bem como as contribuições as limitações e sugestões de trabalhos futuros desta pesquisa. 

\chapter{Fundamentação Teórica} \label{FundamentaçãoTeórica}
Este capítulo tem como objetivo apresentar os principais conceitos relacionados ao entendimento da proposta deste trabalho. A Seção \ref{FT_Reutilização} apresenta os principais conceitos relacionados à reutilização de software. Em seguida, a Seção \ref{FT_LPS} descreve os conceitos de linha de produto de \textit{software} de forma geral, bem como os aspectos relacionados ao seu desenvolvimento. A Seção \ref{FT_Variabilidade} aborda a variabilidade assim como é apresentado os conceitos envolvidos sobre modelos de \textit{features} e por fim a Seção \ref{FT_Configuradores} apresenta a configuração automática e os conceitos sobre integração entre modelos de \textit{features}.

\section{Reutilização de \textit{Software}}  \label{FT_Reutilização}

A reutilização de \textit{software} é amplamente utilizada na indústria de \textit{software} tendo com princípio  reduzir os esforços aplicados no processo produtivo de desenvolvimento de \textit{software}, assim como minimizar os custos de operação. De acordo com \citetexto{krueger1992}, reutilização de \textit{software} é o processo de criação de \textit{software} a partir de um \textit{software} já existente, ao invés de sua construção inicial. Algumas motivações em se reutilizar \textit{software} são a redução de tempo e esforço no desenvolvimento, impactando diretamente na qualidade do produto gerado \cite{krueger2006, pohl2005,krueger2010}. 

Porém, a realidade nos ambientes empresariais, demostrou que os desafios a serem enfrentados são mais abrangentes do que a simples execução das técnicas de reutilização \cite{desouza2006}.  Tendo em vista esta afirmação, a reutilização de \textit{software} não apresenta sua efetividade na prática, a grande questão é que a reutilização é realizada de forma \textit{ad hoc}, ou seja, não sistemática, dependente de iniciativa e do conhecimento intelectual individual dos colaboradores. As empresas não implantam de forma consistente as técnicas propostas, e estão sujeitas a pouco ou sequer a nenhum tipo de controle e planejamento gerencial \cite{ezran2002}. 

A reutilização sistemática consiste no entendimento sobre como é possível contribuir para os objetivos do negócio, na definição das estratégias técnicas e gerencias para se extrair o máximo da reutilização, na integração com os processos de \textit{software} e de melhoria que fazem com que a reutilização ocorra de forma controlada \cite{ezran2002,krueger2010}.  

\section{Linhas de Produto de Software } \label{FT_LPS}

As Linhas de Produto de \textit{Software} (LPS) permitem a reutilização sistemática de \textit{software}, através do desenvolvimento de famílias de produtos que compartilham características comuns \cite{berg2005}. Segundo \citetexto{pohl2005}, as LPS referem-se à técnica de engenharia para a criação de sistemas de software similares a partir de um conjunto compartilhado de partes do \textit{software}, usando uma forma sistemática para a construção de aplicações.  A definição de \citetexto{clements2002}, descreve a LPS como um conjunto de sistemas intensivos de \textit{software} e que compartilham um conjunto comum de \textit{features} gerenciáveis, as quais satisfazem necessidades específicas de um segmento, que são desenvolvidas a partir de um conjunto de recursos comuns segundo um processo definido.  Uma \textit{feature} é definida como uma propriedade do sistema relevante para os analistas e que é usada para capturar semelhanças e variabilidade entre produtos de uma LPS \cite{czarnecki2002,czarnecki2005}.  

As diferenças entre os produtos são denominados de variabilidade. Para \citetexto{berg2005} o  nível de variabilidade determina a capacidade de um sistema ser personalizado de acordo com um contexto específico, possibilitando a produção em larga escala.  Produtos que incorporam variabilidade podem apresentar vantagens como abordar vários segmentos e fornecer conjuntos de diversas características para diferentes necessidades.

O desenvolvimento de uma LPS envolve três atividades básicas \cite{clements2002}: (1) o \textbf{desenvolvimento de recursos do núcleo} (\textit{core assent development}), os recursos de núcleo são os artefatos e recursos reusáveis que formam a base da LPS e podem incluir a arquitetura, documentação, especificações, componentes reusáveis de software e casos de teste. O objetivo desta atividade é o estabelecimento de uma capacidade produtiva, conhecida por engenharia de domínio; (2) o \textbf{desenvolvimento do produto} (\textit{product development}), aborda a criação de um produto que atenda a uma determinada demanda de mercado ou consumidores. É composto por três fatores, o escopo de linha de produção a base de recursos do núcleo e do plano de produção, conhecida por engenharia de aplicação; e por fim, (3) o \textbf{gerenciamento} (\textit{management}), responsável pela supervisão, coordenação e distribuição de recursos.  Abrange o gerenciamento organizacional o qual identifica as restrições de produção, assim como suas estratégias e o gerenciamento técnico, que por sua vez supervisiona as demais atividades apara manter o controle de todo o processo. 

\begin{figure}[h]
	\caption{Atividades para o desenvolvimento de uma LPS.}
	\label{fig:atividadeLPS}
	\centering%
	\begin{minipage}{.4\textwidth}
		\includegraphics[width=\textwidth]{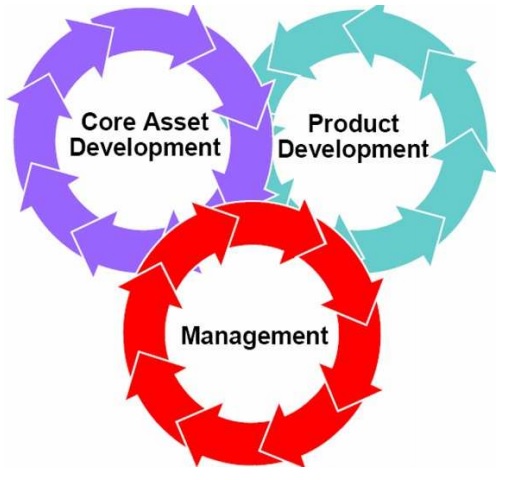}
		\fonte{Clements; Nortthrop, 2002.}
	\end{minipage}
\end{figure}

Na Figura \ref{fig:atividadeLPS} cada círculo representa uma das atividades necessárias para o desenvolvimento da LPS, as três atividades estão inter-relacionadas, ou seja, não há como determinar qual das atividades vem primeiro. Estes fatos decorrem de que os recursos do núcleo são formados a partir de produtos já existentes (abordagem reativa) e em outras ocasiões os recursos do núcleo que são desenvolvidos antes de qualquer produto (abordagem proativa), ou também se pode executar ambas as combinações intensificando as interações entre as atividades (abordagem extrativa) extraindo características comuns e variáveis de sistemas existentes para formar uma nova LPS \cite{krueger2006,clements2002}.

A adoção dos procedimentos no desenvolvimento das LPS procura identificar melhorarias atreladas a qualidade e manutenibilidade, bem como reduzir o custo e o tempo. Permite aos analistas e desenvolvedores utilizar componentes reutilizáveis e aumentar a capacidade de produção das organizações tendo em vista atender as mudanças de mercado, introduzindo novos produtos de forma mais rápida e eficiente, adequando aos componentes reutilizáveis \cite{reinhartz2011, durscki2004}.

As dificuldades na escolha da abordagem para trabalhar em LPS ocorrem em sua maioria devido à mudança cultural, sendo que os desenvolvedores estão acostumados a trabalhar com um sistema por vez \cite{bergey2000}, bem como os problemas para a extensão do mapeamento do domínio, ou seja, a construção de uma arquitetura básica, devido às inconstâncias entre os componentes desenvolvidos o gerenciamento incorreto do conhecimento, assim como, a evolução dos componentes sem armazenar suas modificações \cite{jansen2004, jansen2005}. 

De acordo com \citetexto{durscki2004}, os problemas de implantação de uma LPS são relevantes. Se destaca: (1) abordagem inadequada frente ao foco de elevar a produtividade; (2) caso os produtos gerados não possuam semelhança não garantem a viabilidade do projeto; (3) interação insuficiente entre as equipes de desenvolvimento, ou seja, a LPS necessita de colaboração entre os envolvidos no projeto.

\section{Variabilidade e Modelos de \textit{Features}} \label{FT_Variabilidade}

A variabilidade é fundamental para o gerenciamento e desenvolvimento de linhas de produto de \textit{software}. Podem conter conceitos relacionados a decisões, \textit{features} ou pontos de variação. Modelo de \textit{features} é uma das notações mais expressivas aplicadas para a modelagem de variabilidade \cite{andersen2012}. Portanto, no desenvolvimento de uma LPS, uma das primeiras atividades a serem executadas é a análise de \textit{feature}, que identifica as características visíveis dos produtos da LPS e as organizam em modelos de \textit{features}  \cite{acher2010,lee2006}.  Através da modelagem de \textit{feature} onde são especificadas as funcionalidades comuns e variáveis da família de produtos a serem produzidas.

O modelo de \textit{features} é considerado um modelo de alto nível empregado para expressar os produtos de uma linha de produto de \textit{software} representando as características de um domínio específico, suas variabilidades e semelhanças, assim como seus relacionamentos \cite{kang1990}. O principal objetivo do modelo de \textit{features} é a modelagem das propriedades comuns e as variáveis dos produtos possíveis de uma linha de produção, incluindo suas interdependências. As \textit{features} representam os atributos da aplicação de um dado domínio, estando diretamente relacionadas e visíveis ao cliente final \cite{kang1990}. Nesta proposta de dissertação, a definição de \textit{feature} é dada como sendo uma funcionalidade de comportamento específico para os analistas e desenvolvedores. 
Diversas definições são apresentadas na literatura para o termo \textit{features} \cite{classen2008}.

\begin{itemize}
	\item "Uma estrutura que amplia e modifica a estrutura de um determinado programa, a fim de satisfazer um requisito das partes interessadas, para implementar e encapsular um decisão de design e oferecer uma opção de configuração" \cite{apel2009};
	\item "Uma característica do produto a partir de visões do usuário ou cliente, determinado a partir de um conjunto de requisitos individuais" \cite{chen2005}; e
	\item "Um aspecto importante para um cliente" \cite{riebisch2002}.
\end{itemize}

O primeiro modelo de \textit{features} foi proposto por \cite{kang1990}, como parte do método \textit{Feature Oriented Domain Analysis – FODA}.  Desde então vários outros métodos foram propostos na literatura,\textit{ Feature-Oriented Reuse Method – FORM}\cite{kang1998},\textit{ Feature Reuse-Driven Software Engineering Business - FeatuRSEB} \cite{griss1998},1998), \textit{Product Line UML-Based Software Engineering – PLUS} \cite{eriksson2005} Cardinality-based Feature Models - CBFM  \cite{czarnecki2005}.

O resultado deste processo é uma representação compacta das possíveis derivações de um determinado domínio ou produto, através de um conjunto de características dispostos hierarquicamente, demostrando os relacionamentos existentes entres os elementos e suas possíveis restrições. 

\begin{figure}[h]
	\caption{Exemplo de um modelo de \textit{features} e seus relacionamentos.}
	\label{fig:exemploMF}
	\centering%
	\begin{minipage}{.3\textwidth}
		\includegraphics[width=\textwidth]{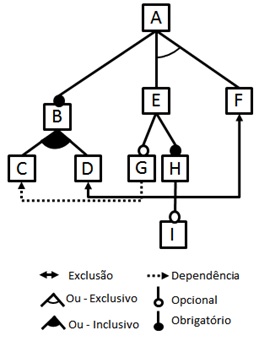}
		\fonte{Elaborado pelo Autor.}
	\end{minipage}
\end{figure}

O modelo de \textit{features} é uma árvore, em que a raiz representa conceito e suas folhas são as \textit{features} conectadas por arestas que representam o seu estado \cite{czarnecki2004}. O seu estado é exibido através de notações intuitivas para representar os pontos de variações. A Figura \ref{fig:exemploMF} apresenta um modelo de \textit{features} e suas notações. O exemplo ilustra as notações tipicamente utilizadas para representar os relacionamentos entre as \textit{features}, obrigatório, opcional, alternativa exclusiva e inclusiva, e os relacionamentos transversais, exclusão e dependência. O relacionamento hierárquico é definido entre uma \textit{feature} ancestral e suas \textit{features} descendentes. Uma \textit{feature} descendente só poderá fazer parte de um produto em que sua \textit{feature} ancestral aparece. 

\begin{itemize}
	\item \textbf{Obrigatória.} Uma \textit{feature} filha que tem um relacionamento obrigatório, ela é incluída em todos os produtos em que \textit{feature} pai aparece.  No exemplo a \textit{features} raiz "A" obrigatoriamente será criada a \textit{feature} "B".
	\item \textbf{Opcional.} A \textit{feature} filha tem uma relação definida como opcional, poderá ser incluída facultativamente em todos os produtos em que sua funcionalidade principal é exibida. No exemplo acima a \textit{feature} “G” poderá ser incluída ou não no modelo de \textit{features }selecionado.
	\item \textbf{Alternativa-Exclusiva.} Um conjunto de \textit{features} filhas é definido como alternativa, quando somente uma \textit{feature} poderá ser selecionada, as demais são excluídas, a \textit{feature} pai faz parte do produto. No exemplo acima poderá ser selecionada somente uma das \textit{features},“E” ou “F”.
	\item \textbf{Alternativa-Inclusiva.} Um conjunto de \textit{features} filhas poderá ser incluído adicionalmente nos produtos em que a \textit{feature} pai aparece. No exemplo acima poderá ser selecionada as \textit{features}, “C” ou  \textit{feature} “D” ou ambas.
	\item \textbf{Dependência.} Quando uma \textit{feature} sempre que selecionada requer a presença de outra.
	\item \textbf{Exclusão.} A seleção de uma \textit{feature }impede a seleção de outra \textit{feature}.
\end{itemize}

De acordo com a Figura \ref{fig:exemploMF}, a \textit{feature} raiz A, representa um conceito, ou seja, um produto A. As \textit{subfeatures} definidas abaixo dela representam as possibilidades de variação existentes neste domínio. Conforme pode ser visualizado, a \textit{feature} B é obrigatória. Implicando que o domínio A, deve definir \textit{feature} B a ser utilizada. Como \textit{subfeatures} de B, têm-se as \textit{features} C e D. Por serem \textit{features} alternativas inclusivas, poderão ocorrer à seleção de ambas as \textit{features}. Porém, as \textit{subfeatures} alternativas exclusivas E e F, quando selecionada uma delas, implica na exclusão de outra.  Como exemplo de \textit{feature} opcional, tem-se a \textit{feature} G, além disso, a notação FODA \cite{kang1990},  permite a utilização dependência e exclusão entre as \textit{features}. 

\section{Configuradores e Integração de Modelos de \textit{Features}} \label{FT_Configuradores}

O modelo de \textit{feature} apresentado na seção anterior protocola a geração de produtos, permitindo o relacionamento entre as \textit{features} e os recursos do núcleo. Essa integração ocorre dos processos de desenvolvimento de domínio e aplicação. As dificuldades inerentes dos processos de criação das LPS ocorridos durante integração decorrentes das fases de desenvolvimento de recursos do núcleo e desenvolvimento de produtos são oriundas da necessidade de uma configuração especifica para cada produto tendo um contexto isolado para os recursos do núcleo causando interesses conflitantes entre as equipes de desenvolvimentos, isto é, o domínio versus aplicação \cite{krueger2006}.

\begin{figure}[h]
	\caption{Configurador de LPS.}
	\label{fig:confLPS}
	\centering%
	\begin{minipage}{.5\textwidth}
		\includegraphics[width=\textwidth]{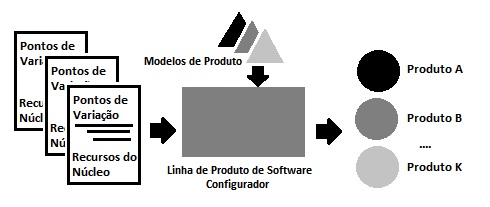}
		\fonte{Adaptado de Krueger (2006).}
	\end{minipage}
\end{figure}

A sistematização da construção destas atividades surgiu para superar estes conflitos (domínio e aplicação) são os configuradores automáticos \cite{krueger2006}. A figura \ref{fig:confLPS} exibe o configurador que recebe duas entradas: as partes centrais e modelos de produtos, criando automaticamente instâncias de produtos.

O configurador automático para a LPS deverá gerar produtos para a LPS de forma segura, verificando as propriedades dos produtos \cite{teixeira2013} para verificar a consistência ou presença de erros. Uma das formas existentes para facilitar o processo de derivação de produtos de uma LPS é através da utilização de ferramentas de instanciação, que facilitam a seleção, a composição e a configuração dos artefatos do núcleo e de suas respectivas variabilidades.  Algumas ferramentas para derivação automática de produtos são propostas destas. Cita-se: Gen-Arc \cite{cirilo2007}, pure::variants \cite{beuche2012} e Gears \cite{krueger2010}.

Os defeitos de um \textit{software} podem ser introduzidos em qualquer instante durante o processo de desenvolvimento. Portanto, as atividades de desenvolvimento do modelo de \textit{features} ocorrem nos estágios iniciais do processo de desenvolvimento de \textit{software} \cite{czarnecki2005}.  A reutilização de modelos \textit{features}, quando executada de forma inadequada poderá gerar inconsistências no modelo pretendido, bem como causar retrabalho.

As dificuldades existentes na verificação de erros e inconsistência do modelo é uma tarefa dispendiosa e inviável para modelos de grande escala manualmente. \cite{teixeira2013, farias2015}.  Contudo, as técnicas e métodos de integração não são suficientemente eficazes e resultam tipicamente em modelos inadequados, apresentando eventuais incoerências e se tornando distinto do modelo desejado. Ou seja, ao realizar um processo de forma automática de integração sobre dois modelos de entrada (MF$_{r}$ e MF$_{c}$), obtido através das iterações, espera-se produzir o modelo pretendido, MF$_{p}$. Entretanto este modelo não é atingido, resultando em um modelo inadequado, contendo inconsistências \cite{farias2015}. As incoerências apresentadas são derivadas de deliberações errôneas ou equivocadas e exige a iteração manual sobre o modelo de \textit{features}, assim determinado o modelo de \textit{features} pretendido.

\begin{figure}[h]
	\caption{Processo de integração de modelo de \textit{features}.}
	\label{fig:exemploIntegraçãoModeloKleinner}
	\centering%
	\begin{minipage}{.8\textwidth}
		\includegraphics[width=\textwidth]{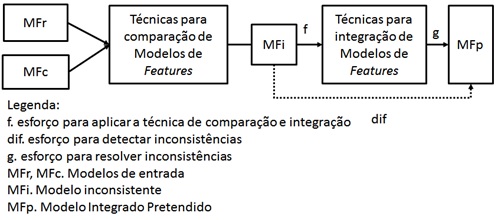}
		\fonte{Adaptado de Oliveira (2012).}
	\end{minipage}
\end{figure}

O processo de integração apresentado na Figura \ref{fig:exemploIntegraçãoModeloKleinner} é executado em duas etapas:A primeira etapa, corresponde em aplicar a (1) técnica de comparação sob o MF$_{r}$, e o MF$_{c}$, obtendo o MF$_{i}$, representado pela equação   $_{f}$(MF$_{r}$, MF$_{c}$).  A realização do esforço adicional para analisar as incoerências do modelo resultante é representada por $_{dif}$(MF$_{i}$, MF$_{p}$). Após, aplicando as (2) técnicas de integração de forma a solucionar os conflitos $_{g}$(MF$_{i}$) \cite{kleinner2012}.

\begin{figure}[ht]
	\caption{Exemplo de integração de modelo de \textit{features}.}
	\label{fig:exemploIntegraçãoUnião}
	\centering%
	\begin{minipage}{.6\textwidth}
		\includegraphics[width=\textwidth]{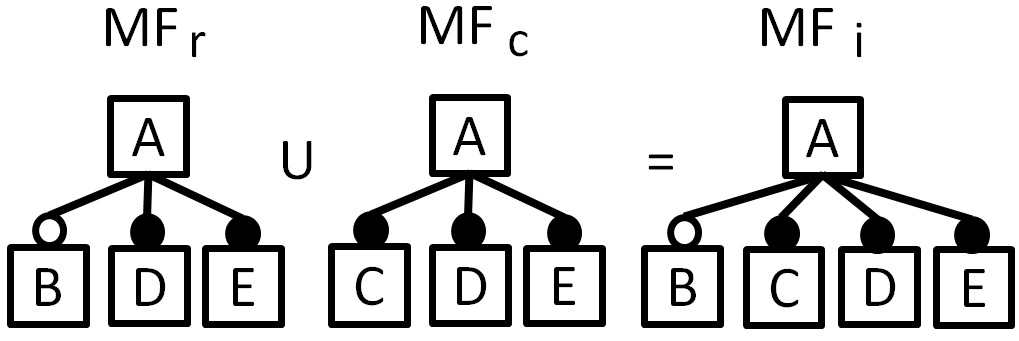}
		\fonte{Elaborado pelo Autor.}
	\end{minipage}
\end{figure}

A integração entre dois modelos de \textit{features} é dada pela comparação entre os modelos de modo a verificar as diferenças dos conjuntos, logo após é aplicado a sua composição, ou seja, a união entre os modelos, gerando um novo modelo, o modelo pretendido. A Figura \ref{fig:exemploIntegraçãoUnião} apresenta um exemplo de integração entre dois modelos, bem como o resultado desta união. Nesse exemplo conforme mencionado anteriormente, o processo de integração ocorre, na entrada de dois modelos (MF$_{r}$ e MF$_{c}$), ou seja, quando pretende-se derivar um novo modelo o MF$_{p}$.  Um fato que deve-se levar em consideração durante o processo de integração é que todas as \textit{subfetures }derivadas da integração destes produtos devem ser obrigatórias, sendo o passo seguinte a comparação entre os dois modelos o qual pode-se observar que o primeiro modelo, MF$_{r}$  a \textit{feature} raiz é representada pelo produto \textbf{A} e \textit{subfeature} \textbf{B}, \textbf{D} e \textbf{E}, que apresenta um relacionamento obrigatório e outro opcional, já o segundo modelo, MF$_{c}$, apresenta  o produto \textbf{A} e \textit{subfeatures} \textbf{C}, \textbf{D} e \textbf{E} ambas obrigatórias, durante esta etapa ocorre a verificação de similaridade entre os modelos.

\begin{figure}[ht]
	\caption{Exemplo de integração de modelo de \textit{features} conflitante.}
	\label{fig:exemploIntegraçãoMFConf}
	\centering%
	\begin{minipage}{.5\textwidth}
		\includegraphics[width=\textwidth]{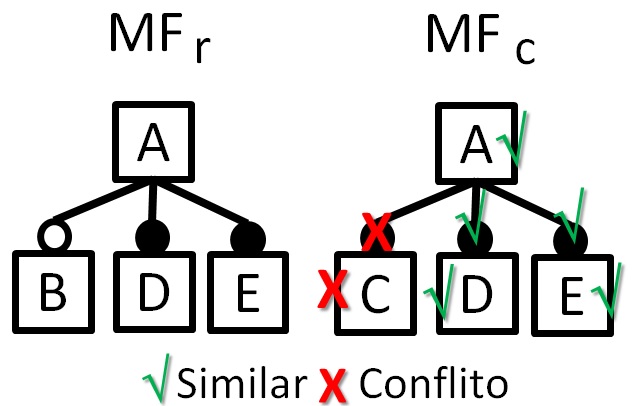}
		\fonte{Elaborado pelo Autor.}
	\end{minipage}
\end{figure}

A Figura \ref{fig:exemploIntegraçãoMFConf} exibe a comparação entre os dois modelos verifica aspectos sintáticos e semânticos; os sistemas de configuração de produto são baseados na variabilidade do modelo e seu desenvolvimento é um processo demorado e sujeito a erros \cite{acher2010, kleinner2012, lesta2015} considerando o desenvolvimento contínuo de produtos deve-se adaptar frequentemente os modelos, elevando a possibilidade de erros em sua integração, gerando produtos indesejáveis.  O primeiro aspecto considerado neste exemplo refere-se à similaridade sintática entre os modelos, MF$_{r}$ e MF$_{c}$. 

As contradições surgidas durante a integração quando executadas automaticamente sem a intervenção humana acabam por propagar erros, impactando diretamente na qualidade do produto, assim como no esforço e custos para refazer a tarefa. Assim, verifica-se a ausência de ferramentas que deem suporte, bem como heurísticas para determinar restrições de integração de modelos \cite{lesta2015}, diagnosticando a similaridade entre os produtos.

O grande diferencial da abordagem da LPS é a apreensão na definição do escopo de produtos derivados de outras linhas conforme as estratégias de mercado e negócios, prospectando vantagens em curto prazo.  Deste modo as familiais de produtos tornam-se atrativas para indústria, conforme é apresentado no \textit{hall of fame}\footnote{http://www.splc.net/fame.html} em \textit{Software Product Line Conferences}, o principal fórum na área de engenharia de \textit{software} de linhas de produto, que fazem parte empresas como Bosch Group, Hewlett Packard, Siemens entre outros. 

\chapter{Trabalhos Relacionados} \label{TrabalhosRelacionados}
Este Capítulo tem como objetivo realizar uma visão panorâmica do estado da arte sobre integração de modelos de \textit{features}, procurando entender como esta área tem se desenvolvido e quais as principais abordagens são utilizadas em sua implementação. A aplicação desta metodologia esta alinhada com os objetivos descritos na Seção \ref{I_Objetivos}. O Capítulo \ref{TrabalhosRelacionados} esta estruturado conforme as seguintes Seções.
 A Seção \ref{TR_MSL} apresenta as principais etapas realizadas na condução do processo de Mapeamento Sistemático da Literatura com a descrição de todos os passos que se seguiram para a realização do mesmo. Na Seção \ref{TR_Resultados} são apresentados os resultados para cada questão de pesquisa, bem como procura-se entender, caracterizar e resumir os estudos primários. A Seção \ref{TR_Discussão} tem duplo objetivo, revelar quando e onde os estudos primários foram publicados e ilustrar uma visão panorâmica sobre o estado da arte. Nesta Seção \ref{TR_Ameaças} demostra as estratégias utilizadas para mitigar algumas ameaças à validade, ou seja, a confiabilidade dos resultados obtidos no processo de investigação. Por fim, a Seção \ref{TR_Conclusões} descreve as considerações finais deste estudo, apresentando as conclusões obtidas com este trabalho e propostas de trabalhos futuros.
 
\section{Mapeamento Sistemático da Literatura } \label{TR_MSL}

No âmbito deste trabalho, aplicou-se um estudo de mapeamento sistemático, metodologia proposta através do paradigma baseado em evidências que orienta na análise e desenvolvimento de um tema de pesquisa \cite{kitchenham2010}, visando apresentar uma visão geral de uma área de pesquisa, identificando o tipo de pesquisa, seus resultados e a quantidade de publicações disponíveis \cite{goncales2015}.

Na execução da busca pelos artigos candidatos, no âmbito de responder as questões de pesquisa localizou-se 775 estudos sendo estes pesquisados em 06 bases cientificas para obter uma classificação com maior relevância, após aplicação dos filtros através de palavras-chave e uma leitura densa dos artigos finais selecionou-se 34 publicações. 

Nas subseções subsequentes apresentamos as diretrizes estabelecidas, detalhado as questões de pesquisa investigadas, definição dos critérios de inclusão e exclusão utilizados para a seleção dos estudos primários, os procedimentos de extração dos dados, assim como, as estratégias de pesquisa descrita no decorrer desta leitura, e por fim, apresentamos os resultados obtidos através desta pesquisa.
 
\subsection{Questões de Pesquisa}\label{TR_SubQuestões}

Para investigar as abordagens existentes sobre integração de modelos de \textit{features}, no contexto de Linhas de Produto de \textit{Software}, especificamente na derivação de produtos, no que tange à definição, do uso de processos e métodos empregados para implantação e execução das atividades de integração entre os modelos, definiram-se seis Questões de Pesquisa (QP) para serem investigadas. Assim, este estudo essencialmente tenta criar uma visão panorâmica do estado da arte sobre integração de modelos de \textit{features}. 
\begin{itemize}
	\item QP1: Quais são as notações usadas para modelar \textit{features}?
	\item QP2: Quais são as estratégias de comparação utilizadas para identificar a equivalência entre os modelos de \textit{features}?
	\item QP3: Quais são as técnicas utilizadas para realizar a integração entre os modelos de \textit{features}?
	\item QP4: Quais são as técnicas utilizadas para configurar e validar inconsistências do modelo de \textit{features}?
	\item QP5: Quais são os tipos de ferramenta que suportam as técnicas de integração de modelos de \textit{features}?
	\item QP6: Quais são os tipos de métodos de pesquisa mais utilizados? 
\end{itemize}

Estas questões de pesquisa são motivadas pela necessidade de: Analisar as notações usadas para representar modelos que expressam variabilidade. (QP1) - Notações de \textit{Feature}; Mapear e descobrir quais técnicas e os aspectos de comparação são usadas para identificar equivalências entre os elementos do modelo de entrada. (QP2)- Técnicas de Comparação; Conhecer e classificar as técnicas para integrar modelos de \textit{features} (QP3) - Técnicas de Integração; Verificar quais as técnicas adotadas para validar e configurar a variabilidade os modelos de \textit{features} (QP4) - Técnicas de Verificação e Validação; Examinar quais técnicas de suporte são empregadas para integrar modelos de \textit{features} (QP5) - Técnicas de Integração; Por fim, identificar os métodos de pesquisa utilizados para investigar os modelos de integração (QP6)- Métodos de Pesquisa.

\subsection{Estratégia de Pesquisa}\label{TR_SubEstratégia}

A próxima etapa realizada foi a busca de estudos primários, sendo que estes estão relacionados diretamente com as questões de pesquisa investigadas, explicadas anteriormente. As estratégias de busca foram imparciais e iterativas sendo estas definidas com base nas orientações estabelecidas na literatura, conforme \cite{kitchenham2010}, que explica os procedimentos relacionados com a construção de \textit{string} de busca e as definições do escopo da pesquisa. A estratégia de pesquisa visa elaborar de forma interativa uma lista de estudos primários, incluindo revisões sistemáticas da literatura, estudos de mapeamento e pesquisas.

Executaram-se as seguintes etapas para definir a \textit{string} de busca: (1) definir as principais palavras-chave; (2) identificar palavras alternativas, sinônimos ou termos relacionados com as principais palavras-chave; (3) verificar se as principais palavras chave estão contidas nas estratégias de busca e categorias de pesquisa; e (4) sinônimos associados, palavras alternativas ou termos com os operadores lógicos "\textit{AND}" e "\textit{OR}". 

As principais palavras chaves investigadas neste estudo são “\textit{Integrate}”, “\textit{Feature}”, “\textit{Model}” e “\textit{Tools}”. Apresentasse a seguir a \textit{string} de busca que retornou os melhores resultados na sua aplicação a seguir: 

\begin{center}
	\textit{((Merging OR Integration OR Composition) AND\\
	(Feature OR Functionality OR Characteristic) AND\\
	(Model OR Design OR Diagram) AND\\
	(Tools OR Applicable OR Procedure OR Techniques))}
\end{center}

Selecionaram-se as seis principais bases eletrônicas (\textit{ACM Digital Library, IEEE, Google Scholar, Scopus, Springer Link e Science Direct}) para realizar a pesquisa para os estudos primários. Ressalta-se que a pesquisa realizada com a string de busca levou em consideração artigos, relatórios técnicos, trabalhos em curso, anais de conferências, revistas, e listas de referência de estudos primários relevantes, portanto, a busca inicial encontrou 775 estudos potencialmente relevantes. As fontes de pesquisa de dados citadas na Tabela \ref{tab:fonteDeDados}, em sua grande maioria destacam-se por serem amplamente utilizadas para pesquisa na área da computação, sendo estas reconhecidas por sua abrangência e relevância significativa na área de estudo.

\begin{table}[ht]
	\caption{Fonte de dados científicos.}
	\label{tab:fonteDeDados}
	\centering%
	\begin{minipage}{.7\textwidth}
		\begin{tabular*}{\textwidth}{ll}
		 \hline
			\textbf{Fontes de Dados}         & \textbf{Endereço Eletrônico }        \\ \hline
			1 - ACM Digital Library & http://portal.acm.org/       \\
			2 - IEEE Xplore         & http://ieeexplore.ieee.org/  \\
			3 - Google Scholar      & https://scholar.google.com   \\
			4 - Scopus              & http://www.scopus.com/       \\
			5 - Springer Link       & http://www.springerlink.com/ \\
			6 - Science Direct      & http://www.sciencedirect.com \\ \hline	
		\end{tabular*}
		\fonte{Elaborada pelo autor.}
	\end{minipage}
\end{table}

\subsection{Critérios de Inclusão e Exclusão}\label{TR_SubCritérios}

Esta seção define os critérios utilizados para incluir e excluir os estudos, considerados relevantes às questões de pesquisa investigados. Como Critérios de Inclusão (CI), foram considerados os estudos que: propõem técnicas de integração de modelos de \textit{features} (CI1); publicações escritas em Inglês (CI2); estudos publicados entre janeiro/2004 e dezembro/2015 (CI3); e disponíveis em bibliotecas digitais e eletrônicas (CI4). 

Em seguida, aplicaram-se os Critérios Exclusão (CE), desconsiderando-se as publicações que: encontrem-se duplicadas em mais de uma fonte de busca, onde será considerada a versão atual; (CE1); publicações que não são publicadas em inglês (CE2);  publicações que não mencionem as palavras-chave da pesquisa no título, resumo ou palavras-chave do artigo (CE3); publicações consideradas como resumo, chamadas de conferência ou patentes (CE4); publicações que estejam fora do contexto de engenharia de linhas de produtos (CE5); e finalmente, publicações que não cumpram a motivação das questões de pesquisa descritas na subseção \ref{TR_SubQuestões}(CE6).

\subsection{Extração de Dados}\label{TR_SubExtração}

Esta etapa está concentrada na extração de dados a partir de estudos preliminares para análise posterior, uma vez que um dos propósitos do estudo de mapeamento sistemático é produzir provas na área sobre o estudo investigado \cite{kitchenham2011}. A extração dos dados visa identificar, avaliar e analisar os estudos primários sobre as questões de pesquisa discutidas anteriormente na subseção \ref{TR_SubQuestões}.

\begin{figure}[ht]
	\caption{Fonte de dados científicos.}
	\label{fig:form}
	\centering%
	\begin{minipage}{.8\textwidth}
		\includegraphics[width=\textwidth]{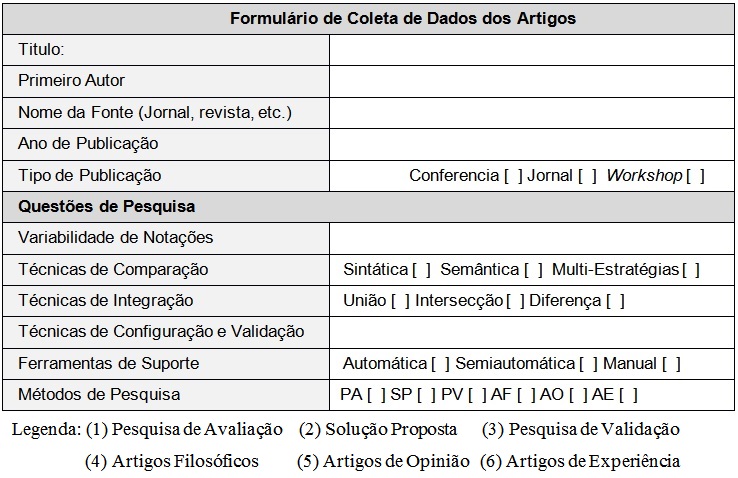}
		\fonte{Adaptado de Fernández (2013).}
	\end{minipage}
\end{figure}

Utilizou-se de planilhas em Excel através de um formulário proposto por \citetexto{fernandez2013} para armazenar os dados coletados com o objetivo de produzir informação para implementar uma análise estatística dos estudos de acordo com a Figura \ref{fig:form}. Conforme mencionado anteriormente, o objetivo deste estudo é entender, caracterizar e resumir o estado da arte das técnicas de integração de modelos de \textit{features}. A análise dos dados foi realizada em ciclo de revisão, com o pesquisador e o auxílio dos professores orientadores para evitar falsos positivo/negativo e para cobrir as questões relevantes e ainda em aberto. 

\textbf{Notações de \textit{Features} (QP1).}  Os modelos de \textit{features} exibem um conjunto de diagramas, formando uma arvore conectando-se por meio das relações entre seus nós, isto é, \textit{features}.  A seguir são apresentadas as notações derivadas dos estudos investigados, incluindo\textit{ Feature-Oriented Domain Analysis (FODA), Feature-Oriented Reuse Method (FORM), Cardinality-based Feature Modeling (CBFM), Reuse-Driven Software Engineering Business Features (RSEB), Product Line Use Case Modeling for Systems and Software Engineering (PLUSS), Common Variability Language (CVL), Text-based Variability Language (TVL), AoURN-based Software Product Line (AoURN), Orthogonal Variability Model (OVM), the Model-Driven Product Lines Engineering (AMPLE), Generative Programming (GP) and Feature Oriented Product Line Software Engineering (FOPLE)}.

\textbf{Técnicas de Comparação (QP2).}   A literatura propõe um conjunto de operações para comparação e avaliação das diferenças encontradas entre os modelos de \textit{features}. Deste modo identificaram-se as seguintes técnicas recomendadas pela literatura: (1) sintática, para equiparar a sintaxe entre as \textit{features}; (2) semântica com a finalidade de comparar estrutura bem como o significado dos elementos presentes entre os modelos; (3) multi-estratégias, que abordam varias estratégias para melhorar o resultado final da similaridade entre os modelos comparados.

\textbf{Técnicas de Integração (QP3).} A integração entre dois modelos de \textit{features} é o resultado da combinação de ambos os elementos presentes nos diagramas de \textit{features }com a finalidade de produzir um novo modelo. Identificaram-se as seguintes operações responsáveis por executar a integração entre os modelos. (1) união, as diferenças entre os modelos de entrada são inseridos no modelo integrado e as semelhanças encontradas são adicionadas sem a repetição dos elementos comuns, por exemplo, MF$_{A}$ $\cup$ MF$_{B}$; (2) intersecção, o modelo integrado tem apenas os elementos comuns entre os diagramas de \textit{features}, desconsiderados os elementos incomuns entre os modelos comparados, por exemplo, MF$_{A}$ $\cap$ MF$_{B}$ ; (3) diferença, verifica a diferença entre os diagramas \textit{features} comparados, retornando os elementos pertencentes MF$_{A}$ incomuns a MF$_{B}$, isto é, MF$_{A}$ - MF$_{B}$. Observou-se que nos estudos investigados em alguns casos há pelo menos a aplicação de duas abordagens para melhorar a precisão das técnicas de integração. 

\textbf{Técnicas de Configuração e Validação (QP4).} A formalização bem como a validação dos modelos de \textit{features} nos estudos investigados assinalam as seguintes técnicas usadas na formalização dos modelos, (1) lógica proposicional, (2) lógica descritiva ou (3) programação restritiva assim como se utiliza de solucionadores e algoritmos para detectar possíveis inconsistências encontradas na formação do modelo em análise. Através dos estudos analisados identificaram-se os seguintes algoritmos aplicados na verificação dos modelos: (1) Famílias de Produtos da Álgebra (FPA); (2) Diagramas de Decisão Binária (BDD); (3) Problema de Satisfação Booleana (SAT); (4) Problemas de Satisfação de Restrições (CSP); (5) Forma Normal Conjuntiva (FNC), (6) Forma Normal Disjuntiva (FND); e (7) Multicritério, estudos que combinam diversas estratégias abordas para melhorar os resultados de integração.

\textbf{Suporte Ferramental (QP5).} A fim de conhecer os tipos de apoio ferramental fornecido para as equipes de desenvolvimento, investigou-se os trabalhos a partir de três perspectivas: 1) automático, que não requer qualquer interação humana; (2) semiautomático, ele requer que os desenvolvedores especifiquem os parâmetros de configuração diferenciando os elementos de entrada do modelo. Esta abordagem requer intervenção do desenvolvedor para o tratamento de procedimento de avaliação; e (3) etapas manuais para listar como as boas práticas conduzem o processo de integração dos elementos do modelo (por exemplo, \textit{features} e suas notações). 

\textbf{Métodos de Pesquisa (QP6).} Esta questão de pesquisa fornece uma visão geral sobre o direcionamento dos estudos atuais, ou seja, quais os tipos de estudos investigados foram produzidos ao longo destes anos. Devido a grande quantidade de estudos para serem classificados nos estudos primários a abrangência dos métodos de pesquisa é definida conforme as categorias propostas em \cite{kitchenham2010, wieringa2006}. Este estudo é classificado da seguinte forma: (1) \textbf{solução proposta}, propõe uma solução baseada em novos estudos ou estudos anteriores; (2) \textbf{pesquisa de avaliação}, avalia a aplicação de técnicas e os conceitos de estudos empíricos; (3) \textbf{pesquisa de validação}, concentra seus esforços em avaliar as técnicas desconhecidas pela indústria; (4) \textbf{artigos de opinião}, estudos que ampliam a discussão abordando o ponto de vista de seus autores sobre o problema de pesquisa e as possíveis soluções; (5) \textbf{artigos filosóficos}, aborda novas técnicas e estudos inéditos, propondo uma discussão sobre a metodologia aplicada; e (6) \textbf{artigos de experiência}, abordam estudos de experiência pessoal, investigações de ferramentas próprias aplicadas no contexto do autor, evidenciando seu aprendizado.  

\subsection{Seleção dos Estudos}\label{TR_SubSeleção}
Esta subseção apresenta o processo de execução dos filtros dos estudos selecionados, descrevendo os passos para alcançar os estudos primários. Aplicaram-se sete etapas para filtrar os estudos relevantes depois de fazer uma pesquisa inicial. A busca e filtragem foram executadas entre janeiro e fevereiro de 2016. A Figura 8 exibe as etapas do processo de seleção dos estudos relevantes, bem como da execução dos filtros, que são descritos da seguinte forma: 
\begin{itemize}
	\item Etapa 1: pesquisa inicial. Reúne uma lista ampla de estudos após a aplicação da estratégia de pesquisa, conduzida através da busca de palavras-chave, realizado nas seis bibliotecas digitais definidas de acordo com a Tabela 1. Ao todo foram identificados 775 estudos; 
	\item Etapa 2: remoção dos artigos duplicados.  Consiste na aplicação do critério de exclusão CE1, ou seja, estudos repetidos foram descartados em cada uma das bibliotecas pesquisadas; 
	\item Etapa 3: filtro por título: Este filtro procura estabelecer uma analise introdutória   através dos títulos das publicações investigadas verificando sua relação com a  linha de pesquisa em curso, ou seja, integração de modelos de \textit{features}, aplicaram-se os seguintes critérios de exclusão CE2 e CE3.
	\item Etapa 4: filtro por resumo: Este filtro é fundamental para  determinar os estudos relevantes  para esta pesquisas, pois apresenta um visão panorâmica dos estudos investigados.  Aplicaram-se os critérios de exclusão CE4 E CE5.
	\item Etapa 5: combinação: Todos os estudos filtrados a partir das fases anteriores foram agrupados em um novo diretório.  Aplicou-se novamente o critério de exclusão CE1, para remoção de publicações duplicadas;
	\item Etapa 6: filtro por texto completo: Selecionou-se os estudos através da leitura completa dos textos, a fim de verificar o conteúdo definido na seção de extração de dados, com ênfase em responder às questões de pesquisa inerente a este estudo, a aplicação do critério de exclusão CE 6; 
	\item Etapa 7:estudos representativos: A lista final dos estudos primários é definida após revisar todas as etapas anteriores e os critérios de inclusão e exclusão.
\end{itemize}

\begin{figure}[ht]
	\caption{Processo de seleção dos estudos relevantes.}
	\label{fig:filtro}
	\centering%
	\begin{minipage}{.9\textwidth}
		\includegraphics[width=\textwidth]{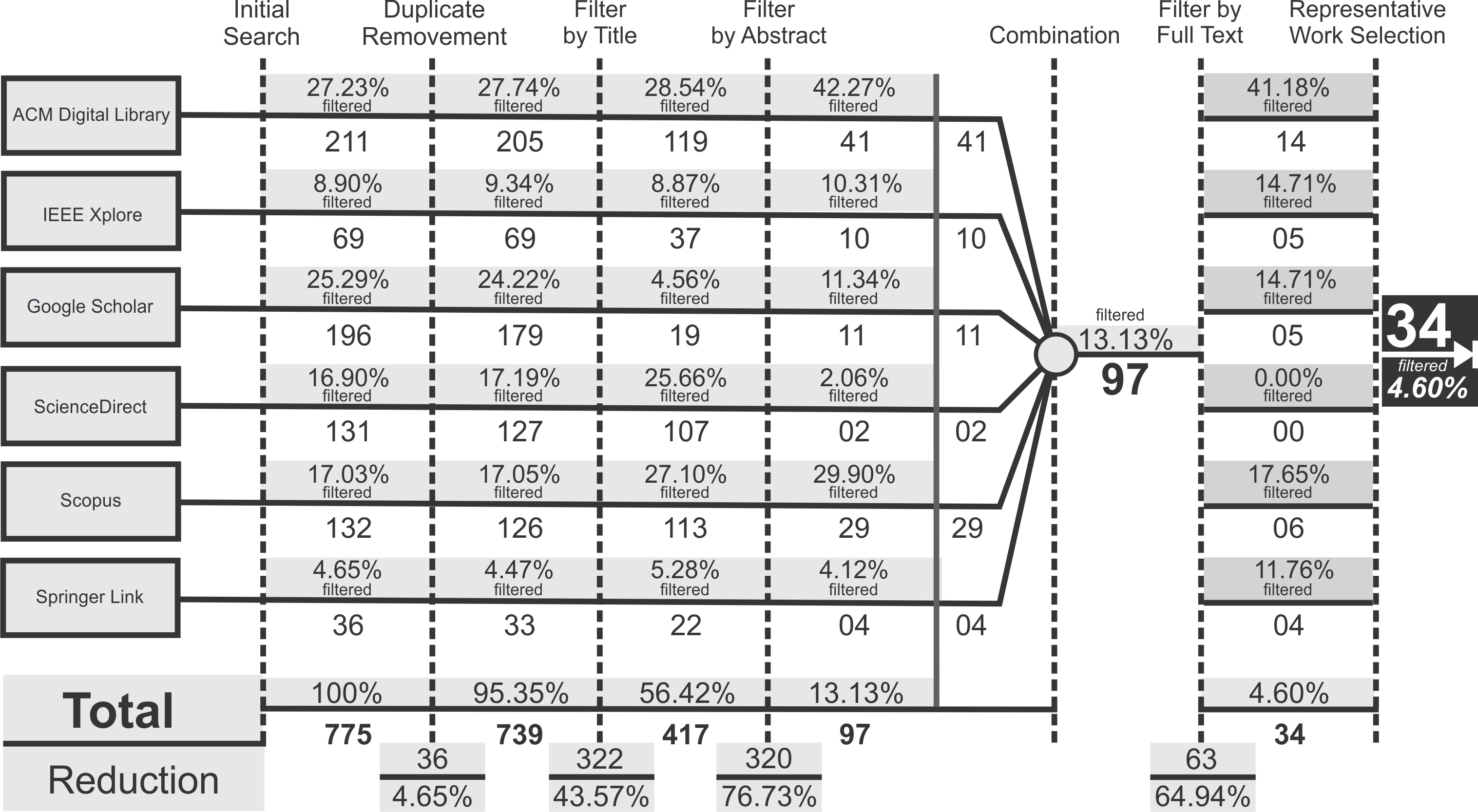}
		\fonte{Elaborado pelo autor.}
	\end{minipage}
\end{figure}

Após a aplicação das etapas inerentes a execução dos filtros, os resultados retornados na primeira etapa, (Etapa 1) totalizaram-se 775 (100\%) publicações, já a segunda etapa (Etapa 2), removem-se as publicações duplicadas obtendo 739 (95,35\%) publicações, isto é, reduzindo-se 36 publicações (4,65\%) a partir do total de 775. Seguindo a terceira etapa (Etapa 3), aplicaram-se os critérios de inclusão e exclusão aos demais estudos, filtrando 417 publicações (56,42\%), assim como, analisaram-se os títulos das publicações, reduzindo 322 (43,57\%) publicações de um total de 739 publicações e finalmente 97 (13,13\%) publicações após executar filtro do resumo (Etapa 4 e Etapa 5) das publicações analisadas, removendo 320 (76,73\%) publicações de 417. Na etapa consecutiva (Etapa 6) realizou-se uma análise criteriosa destas publicações de modo geral, sendo desconsideradas as publicações irrelevantes as questões de pesquisa formulada anteriormente, reduzindo 63 (64,94\%) publicações. Finalmente, selecionou-se 34 publicações, as publicações mais representativas (Etapa 7), constituindo 4,60\% a partir de 775 publicações iniciais localizadas. A lista completa dos estudos primários selecionados encontra-se no apêndice A deste trabalho.

\section{Resultados } \label{TR_Resultados}

Após a análise dos 34 estudos primários selecionados aplicando as etapas de filtragem anteriormente descritas, nesta seção são apresentados e interpretados os dados obtidos. O objetivo deste estudo procura entender, caracterizar e resumir o estado da arte sobre os assuntos relacionados à integração de modelos de \textit{features}. Assim, aborda-se cada questão de pesquisa conforme sua descrição na subseção \ref{TR_SubQuestões}.

\textbf{QP1: Notações de \textit{Features}.} Esta questão de pesquisa busca investigar as notações usadas para representar os modelos de \textit{features}. O objetivo da modelagem das \textit{features} é exibir as propriedades comuns e variáveis dos produtos possíveis de uma LPS.  A Tabela \ref{tab:tbNotacoes} apresenta os dados obtidos em relação ás expressões utilizados na representação dos diagramas de \textit{features}. Os dados indicam que a FODA é a notação preferida de modelagem para representar os diagramas de \textit{features}, 44\% (15/34) dos estudos primários. A notação FORML foi utilizada em apenas 3\% (1/34) dos estudos investigados. Além disso, as demais notações investigadas, FORM, CBFM, RSEB, PLUSS, CVL, TVL, entre outras são distribuídas em ambos os estudos, isto é, apresentam-se em mais de um estudo, representando 32\% (11/34) dos estudos primários. Finalmente, 21\% (7/34) dos artigos não mencionaram ou sugeriu qualquer notação. 

\begin{table}[ht]
\centering
\caption{Notações de \textit{features}}\label{tab:tbNotacoes}
\scalebox{0.8}{
\renewcommand{\arraystretch}{1.5}
\begin{tabular}{p{6cm}p{1.3cm}p{2cm}p{6cm}}
\hline
Notação         & Número & Percentual   & ID Artigos
\\ \hline
FODA             & 15     & 44\% & {[}S01{]},{[}S03{]},{[}S04{]},{[}S05{]},{[}S06{]}, {[}S09{]},{[}S17{]},{[}S19{]},{[}S20{]},{[}S22{]}, {[}S23{]},{[}S24{]},{[}S28{]},{[}S29{]},{[}S34{]} \\
FORML            & 1      & 3\%  & {[}S07{]}                                                                                                                                              
\\
Outras Notações & 11     & 32\% & {[}S02{]},{[}S10{]},{[}S11{]},{[}S14{]},{[}S16{]}, {[}S21{]},{[}S25{]},{[}S26{]},{[}S27{]},{[}S30{]}, {[}S32{]}                                        \\
Notações não especificadas   & 7      & 21\% & {[}S08{]},{[}S12{]},{[}S13{]},{[}S15{]},{[}S18{]}, {[}S30{]},{[}S32{]}                                                                                 
\\
Total            & 34     &100\%
\\ \hline
\fonte{Elaborada pelo autor.}
\end{tabular}}
\end{table}

\begin{figure}[!htb]
	\caption{Representação hierárquica das notações de \textit{features}.}
	\label{fig:hierarquiaca}
	\centering%
	\begin{minipage}{.6\textwidth}
		\includegraphics[width=\textwidth]{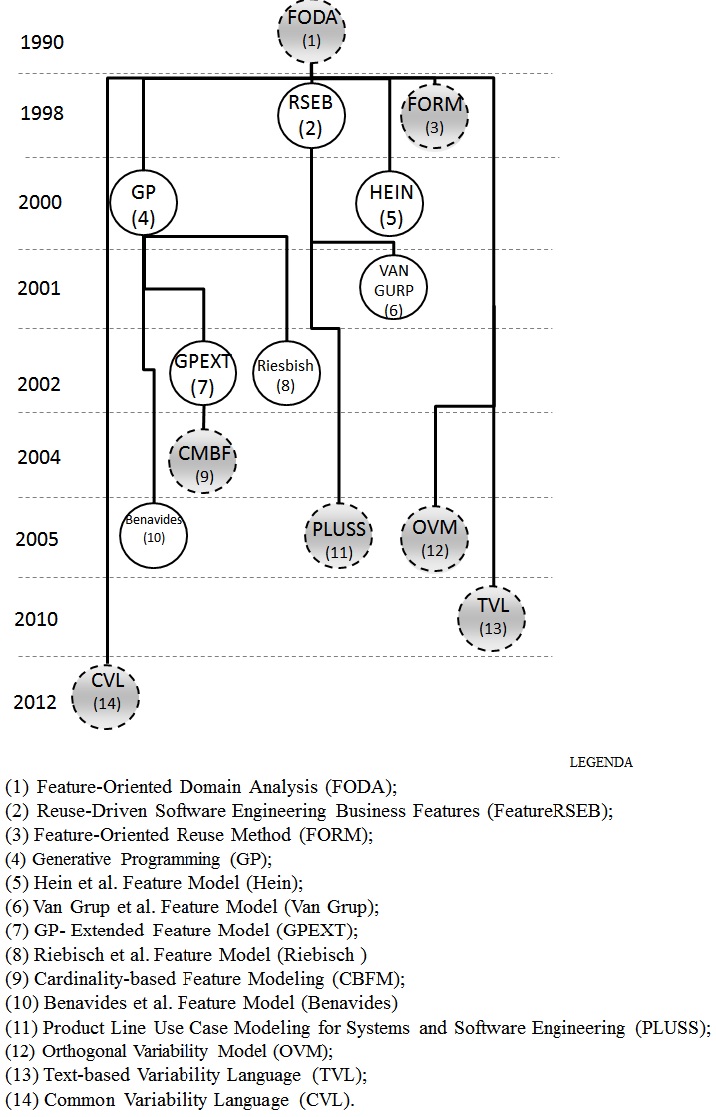}
		\fonte{Fonte: Elaborado pelo autor.}
	\end{minipage}
\end{figure}

A Figura \ref{fig:hierarquiaca} apresenta uma árvore com a evolução das técnicas empregadas para modelagem dos diagramas de \textit{features}, onde a notação FODA foi a primeira técnica proposta por \citetexto{kang1990}, sendo que as demais técnicas empregadas ao longo dos anos foi uma mutação desta, como por exemplo, RSEB \cite{griss1998}, FORM \cite{kang1998}, GP \cite{czarnecki2000}, Hein \cite{hein2000}, Gurp \cite{van2001}, GPEXT \cite{czarnecki2002}, Riebisch \cite{riebisch2002}, CMBF \cite{czarnecki2004}, Benavides \cite{benavides2005}, PLUSS \cite{eriksson2005}, OVM \cite{pohl2005}, TVL \cite{classen2011}, CVL \cite{reinhartz2014}, que sugere a necessidade de melhorias aplicadas à modelagem de \textit{features} nas LPS ao longo dos anos.

\textbf{QP2: Técnicas de Comparação.} Esta questão investiga as estratégias utilizadas na literatura para comparar e identificar a equivalência entre os dois modelos \textit{features} de entrada. A saída desta etapa serve como base para a próxima etapa de composição. Em seguida, os aspectos utilizados nesta etapa têm um forte impacto nos resultados da composição. A técnica de composição irá integrar elementos dos modelos de \textit{features} incorretamente se forem definidas relações de equivalência incorretas.

\begin{table}[ht]
\centering
\caption{Técnicas de comparação.}\label{tab:tbComparacao}
\scalebox{0.8}{
\renewcommand{\arraystretch}{1.5}
\begin{tabular}{p{6cm}p{1.3cm}p{2cm}p{6cm}}
\hline
Técnicas de Comparação         & Número & Percentual   & ID Artigos \\ \hline
Sintática              & 4      & 12\% & {[}S02{]},{[}S20{]},{[}S21{]},{[}S22{]}                               \\
Semântica              & 7      & 21\% & {[}S03{]},{[}S11{]},{[}S14{]},{[}S18{]},{[}S24{]}, {[}S29{]},{[}S31{]} \\
Multi-estratégia       & 4      & 12\% & {[}S09{]},{[}S17{]},{[}S32{]},{[}S34{]}                               \\
Não especificado       & 19     & 55\% & {[}S01{]},{[}S04{]},{[}S05{]},{[}S06{]},{[}S07{]}, {[}S08{]},{[}S10{]},{[}S12{]},{[}S13{]},{[}S15{]}, {[}S16{]},{[}S19{]},{[}S23{]},{[}S25{]},{[}S26{]}, {[}S27{]},{[}S28{]},{[}S30{]},{[}S33{]} \\																	   
Total        & 34  &100\%                                                                    \\ \hline
\fonte{Elaborada pelo autor.}
\end{tabular}}
\end{table}

A Tabela \ref{tab:tbComparacao} apresenta os dados obtidos  nesta pesquisa sobre as técnicas de comparação. Em primeiro lugar, poucas técnicas de comparação foram propostas para avaliar as equivalências entre os modelos de características de entrada. Em suma, apenas as estratégias semântica e sintática foram propostas. Os resultados mostram que 21\% (7/34) dos estudos primários se concentram na proposição de estratégia semântica, e 12\% (4/34) visam à estratégia sintática. Além disso, 12\% (4/34) dos artigos propõem ambas as técnicas denominadas de multi-estratégia, isto é, aplicam estratégias semânticas e sintáticas para definir a similaridade. Finalmente, a maioria dos estudos 55\% (19/34) não propõe qualquer técnica de comparação.

\textbf{QP3: Técnicas de Integração.} Esta questão investiga as técnicas atuais aplicadas à integração de modelos de recursos. Conforme mencionado anteriormente, a etapa de integração combina as \textit{features} equivalentes e, em seguida, produz o modelo de \textit{features} integrado (MF$_{I}$,). Em geral, existem heurísticas de composição bem conhecidas para a composição do modelo. Contudo, até onde se sabe as heurísticas de composição específicas para a integração de \textit{features} não têm sido amplamente discutidas até agora.

\begin{table}[ht]
\centering
\caption{Técnicas de integração.}\label{tab:tbIntegração}
\scalebox{0.8}{
\renewcommand{\arraystretch}{1.5}
\begin{tabular}{p{6cm}p{1.3cm}p{2cm}p{6cm}}
\hline
Técnicas de Integração         & Número & Percentual   & ID Artigos \\ \hline
União                 & 7      & 21\% & {[}S03{]},{[}S04{]},{[}S07{]},{[}S08{]},{[}S09{]}, {[}S15{]},{[}S23{]} \\
Interseção            & -      & - & {[}S03{]},{[}S04{]},{[}S07{]},{[}S08{]},{[}S09{]}, {[}S15{]},{[}S23{]} \\
Diferença             & -      & -  & {[}S04{]},{[}S09{]}                                                   \\
Não especificado      & 27     & 79\% & {[}S01{]},{[}S02{]},{[}S05{]},{[}S06{]},{[}S10{]}, {[}S11{]},{[}S12{]},{[}S13{]},{[}S14{]},{[}S16{]}, {[}S17{]},{[}S18{]},{[}S19{]},{[}S20{]},{[}S21{]}, {[}S22{]},{[}S24{]},{[}S25{]},{[}S26{]},{[}S27{]}, {[}S28{]},{[}S29{]},{[}S30{]},{[}S31{]},{[}S32{]}, {[}S33{]},{[}S34{]}\\																	   
Total        & 34 &100\%                                                                     \\ \hline
\fonte{Elaborada pelo autor.}
\end{tabular}}
\end{table}

A Tabela \ref{tab:tbIntegração} mostra os resultados obtidos nesta pesquisa sobre as técnicas de integração. Os resultados demonstram que três técnicas são comumente usadas em conjunto para integrar modelos de \textit{features}. As técnicas correspondentes são a união, a intersecção e diferença especificamente. Os resultados mostram que 21\% (7/34) dos estudos primários fazem uso da estratégia de união, estratégia de interseção, e estratégia de diferença como uma forma de compor os modelos de \textit{features}. Uma minoria de estudos (6\%, 2/34) adiciona especificamente a técnica de diferença. Finalmente, a maioria dos estudos (79\%, 27/34) não propõe nenhuma técnica de integração. Esse resultado sugere que trabalhos recentes não estão se concentrando em propor técnicas de integração para modelos de recursos.

\textbf{QP4: Técnicas de Configuração e Validação.} A etapa de validação é responsável pela identificação de inconsistências no modelo de feature integrado. Para resolver esse problema, os modelos de \textit{features} são analisados para verificar se as regras de formação são satisfeitas ou não. Técnicas como solucionador SAT e solucionador CSP são alguns exemplos.

A Tabela \ref{tab:tbConfVerVal} mostra as técnicas aplicadas na literatura para validação e verificação de modelos de \textit{features}. Pôde-se observar que 3\% (1/34) investigaram o algoritmo do problema de satisfação de restrições (CSP) e 6\% (2/34) de estudos primários aplicaram o algoritmo do problema de satisfação booleana (SAT). Estudos multicritérios são aqueles que se aplicam ou sugerem mais de uma estratégia, correspondendo a 24\% (8/34) dos estudos investigados.

\begin{table}[!ht]
\centering
\caption{Técnicas de configuração e validação.}\label{tab:tbConfVerVal}
\scalebox{0.8}{
\renewcommand{\arraystretch}{1.5}
\begin{tabular}{p{6cm}p{1.3cm}p{2cm}p{6cm}}
\hline
Estratégias de Validação         & Número & Percentual   & ID Artigos \\ \hline
CSP                   & 1      & 3\%  & {[}S14{]}\\ 
SAT                   & 2      & 6\%  & {[}S18{]},{[}S29{]}\\
Multicritério         & 8      & 24\% & {[}S02{]},{[}S09{]},{[}S20{]},{[}S24{]},{[}S25{]}, {[}S30{]},{[}S32{]},{[}S34{]}\\
Não especificado      & 23     & 68\% & {[}S01{]},{[}S03{]},{[}S04{]},{[}S05{]},{[}S06{]}, {[}S07{]},{[}S08{]},{[}S10{]},{[}S11{]},{[}S12{]}, {[}S13{]},{[}S15{]},{[}S16{]},{[}S17{]},{[}S19{]}, {[}S21{]},{[}S22{]},{[}S23{]},{[}S26{]},{[}S27{]}, {[}
S28{]},{[}S31{]},{[}S33{]} \\                                               																   
Total        & 34     &100\%                                                                 \\ \hline
\fonte{Elaborada pelo autor.}
\end{tabular}}
\end{table}

Essas técnicas geralmente combinam SAT, CSP e outras estratégias de validação para melhorar a análise de inconsistência. Pesquisas e revisões da literatura são também alguns dos estudos que são relatados aqui, e cobrem mais de uma técnica da validação. Finalmente, a maioria dos estudos (68\%, 23/34) não propõe nenhuma técnica de validação. Alguns estudos alegaram que o número de trabalhos sobre a abordagem de validação de modelos de \textit{features} aumentou \cite{benavides2010, andersen2012, benavides2013, eichelberger2013}. Embora tenha sido encontrado um número relevante de abordagens para validar os modelos neste estudo, à maioria deles (68\%, 23/34) não mencionou qualquer tipo de técnica de validação. Finalmente, também é importante ressaltar que uma execução eficiente da etapa de validação depende do tamanho do modelo analisado. Então, melhorar a eficiência na validação de grandes modelos é um importante desafio de pesquisa. 

\textbf{QP5: Suporte Ferramental.} Esta questão tem como objetivo investigar os tipos de ferramentas de suporte fornecidas para todo o processo de integração de modelos de \textit{features}. Identificaram-se três categorias: (1) automáticas, isto é, aquelas ferramentas que não suportam interferência humana; (2) semiautomáticas, as técnicas que permitem a interação entre o homem e a máquina; (3) o processo de integração é conduzido sem qualquer tipo de suporte ferramental, ou seja, é executado de forma manual.

\begin{table}[!ht]
\centering
\caption{Suporte ferramental.}\label{tab:tbSuporteFerramental}
\scalebox{0.8}{
\renewcommand{\arraystretch}{1.5}
\begin{tabular}{p{6cm}p{1.3cm}p{2cm}p{6cm}}
\hline
Suporte Ferramental         & Número & Percentual   & ID Artigos \\ \hline
Automático          & 20     & 59\% & {[}S01{]},{[}S02{]},{[}S04{]},{[}S06{]},{[}S07{]}, {[}S09{]},{[}S12{]},{[}S13{]},{[}S14{]},{[}S15{]}, {[}S18{]},{[}S19{]},{[}S20{]},{[}S21{]},{[}S22{]}, {[}S24{]},{[}S25{]},{[}S29{]},{[}S30{]}, {[}S34{]} \\
Semiautomático      & 2      & 6\%  & {[}S05{]},{[}S28{]}                                                                                                                                                                 
\\
Manual             & 0      & 0\%  &  \\
Não especificado      & 12     & 35\% & {[}S03{]},{[}S08{]},{[}S10{]},{[}S11{]},{[}S16{]}, {[}S17{]},{[}S23{]},{[}S26{]},{[}S27{]},{[}S31{]}, {[}S32{]},{[}S33{]} \\                                               																   
Total        & 34     &100\%                                                                 \\ \hline
\fonte{Elaborada pelo autor.}
\end{tabular}}
\end{table}

A Tabela \ref{tab:tbSuporteFerramental} apresenta as categorias de automação propostas pela literatura. Os resultados mostram que 59\% (20/34) dos estudos primários apresentam ferramentas que fornecem suporte automatizado, 6\% (2/34) dos estudos propostos apresentam ferramentas com apoio semiautomático. Além disso, não foi encontrado nenhum estudo sobre técnicas manuais. Finalmente, 35\% (12/34) dos estudos investigados não propõem suporte à ferramenta.

É importante ressaltar que não foi encontrado nenhum experimento para avaliar os desenvolvedores durante a aplicação das técnicas de integração manual. Esse tipo de experimento seria importante para demonstrar empiricamente a importância do auxílio computacional na integração de modelos de \textit{features}. Embora as ferramentas automáticas sejam as mais aplicadas não foi encontrada nenhuma evidência que comprove sua eficácia. O algoritmo aplicado para resolver modelos indesejados e erros propagados durante uma má integração pode ter impacto nos custos de produção. Por fim, as ferramentas semiautomáticas propostas na literatura não fizeram um diagnóstico de conflitos durante a comparação e também não permitem a edição de modelos de \textit{features} durante o processo de integração.

\textbf{QP6: Métodos de Pesquisa.} Classificam-se os estudos primários com base nos métodos de pesquisa definidos em \cite{wieringa2006, petersen2008} tais como: (1) Proposta de solução, propõe uma nova solução; (2) Pesquisa de avaliação, realiza estudos empíricos; (3) A pesquisa de validação, valida técnicas em um ambiente da indústria; (4) Artigos de opinião, uma opinião pessoal sobre um assunto específico em questão; (5) Artigo filosófico, propõe uma nova maneira de esboçar soluções; e (6) Artigos de Experiência, relata experiência ou lições aprendidas após usar uma técnica particular.

\begin{table}[!ht]
\centering
\caption{Métodos de pesquisa}\label{tab:tbmetodosPesquisa}
\scalebox{0.8}{
\renewcommand{\arraystretch}{1.5}
\begin{tabular}{p{6cm}p{1.3cm}p{2cm}p{6cm}}
\hline
Métodos         & Número & Percentual   & ID Artigos \\ \hline
Solução Proposta & 24     & 70\% & {[}S01{]},{[}S02{]},{[}S03{]},{[}S04{]},{[}S06{]}, {[}S07{]},{[}S08{]},{[}S09{]},{[}S11{]},{[}S12{]}, {[}S14{]},{[}S15{]},{[}S17{]},{[}S18{]},{[}S20{]}, {[}S23{]},{[}S24{]},{[}S25{]},{[}S26{]},{[}S27{]}, {[}
S29{]},{[}S30{]},{[}S31{]},{[}S34{]} \\
Pesquisa de Avaliação     & 6      & 18\% & {[}S05{]},{[}S16{]},{[}S19{]},{[}S21{]},{[}S28{]}, {[}S32{]} \\      
Pesquisa de Validação     & 1      & 3\%  & {[}S33{]} \\  
Artigos de Opinião        & 1      & 3\%  & {[}S10{]} \\ 
Artigos Filosóficos       & 1      & 3\%  & {[}S22{]} \\ 
Artigos de Experiência    & 1      & 3\%  & {[}S13{]} \\ 
                                             																   
Total        & 34     &100\%                                                                 \\ \hline
\fonte{Elaborada pelo autor.}
\end{tabular}}
\end{table}

A Tabela \ref{tab:tbmetodosPesquisa} mostra a classificação de estudos primários dos métodos de pesquisa. Os resultados indicam que a área de pesquisa com maior concentração de investigação dos estudos são as soluções propostas, com 70\% (24/34), dos trabalhos analisados. Além disso, pouco tem sido feito para avaliar empiricamente as atuais técnicas de integração.  As pesquisas e avaliação são apenas 18\% (6/34) dos estudos investigados. Finalmente, pesquisa de validação, artigo de opinião, artigo filosófico e artigo de experiência compreendem em 12\% (4/34) dos estudos investigados, evidenciando que a avaliação das técnicas tem sido amplamente baseada em reflexão e opinião de especialistas, e não em evidências empíricas.

\section{Discussão Adicional} \label{TR_Discussão}

Esta seção tem duplo objetivo: (1) revelar quando e onde os estudos primários foram publicados, revelando assim tendências de pesquisa e explorando como os estudos primários são distribuídos considerando as questões de pesquisa específicas nos últimos anos; e (2) esboçar uma visão panorâmica destacando alguns desafios adicionais, que podem ser tomados como base para construir uma agenda de pesquisa. Como tal, isso pode ajudar a aumentar a literatura, ilustrando um grande quadro sobre o estado da arte.

\begin{figure}[!htb]
	\caption{Distribuição dos estudos primários.}
	\label{fig:primarios}
	\centering%
	\begin{minipage}{.9\textwidth}
		\includegraphics[width=\textwidth]{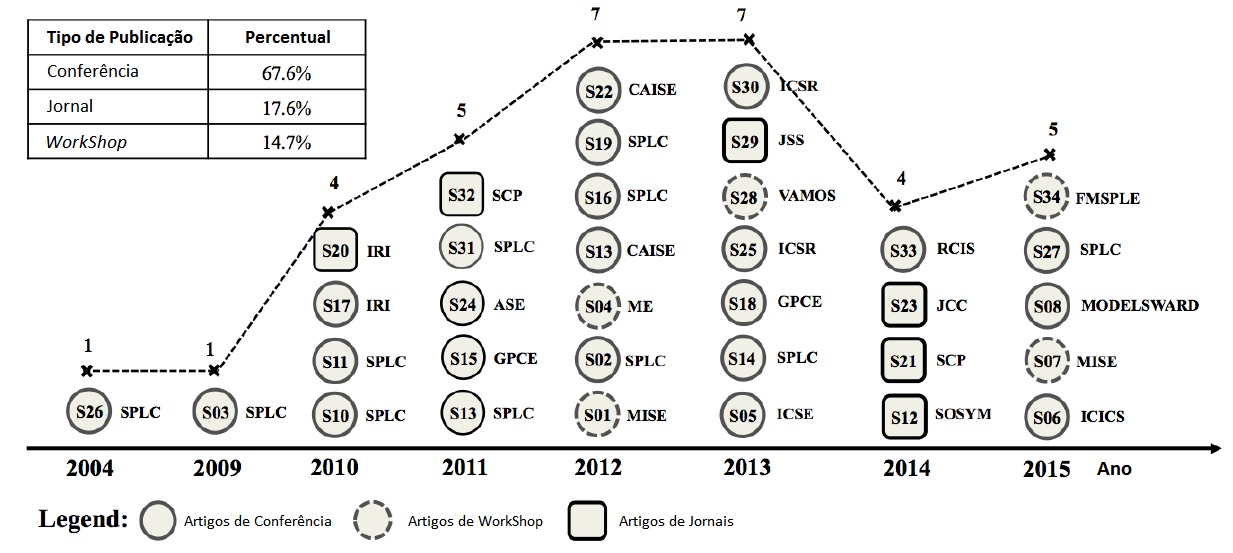}
		\fonte{Elaborado pelo autor.}
	\end{minipage}
\end{figure}

\textbf{Distribuição dos Estudos Primários.}  Para abordar o primeiro objetivo, obtivemos o ano de publicação e a fonte (ou seja, local) para cada estudo primário. Para isso, categorizamos todos os estudos primários para tipos de publicação: conferência, revista e \textit{workshop}. A Figura \ref{fig:primarios} mostra uma cronologia para todos os estudos primários, a quantidade de estudos publicados por ano e o local usado para publicar o artigo (por exemplo, conferência, revista e \textit{workshop}). Observe que o gráfico mostra a abreviatura da conferência, jornal e \textit{workshop }em que o artigo foi publicado ao lado da identificação do artigo. A abreviatura ajuda a simplificar o gráfico e a linha tracejada encontrada na Figura \ref{fig:primarios} resume o número de estudos primários publicados entre os anos de 2004 a 2015. 

Dado que em nenhum artigo foi identificado entre os anos de 2005 a 2008, o gráfico não descreve os resultados nesses anos. A partir de 2009, foram selecionados entre um a sete estudos primários por ano, demonstrando uma tendência ascendente gradual. De fato, de 2010 a 2015 pelo menos quatro artigos foram publicados. Isto significa que a academia investiu certo esforço para promover este campo de pesquisa, alcançando um pico de 7 artigos publicados em 2012 e 2013. Estes resultados ajudam a compreender o quanto a atividade de pesquisa sobre a integração de modelos de \textit{features} ocorreu nos últimos anos, bem como o local onde os pesquisadores têm procurado publicar seus resultados de pesquisa.

Considerando o local onde os estudos primários foram publicados, há uma predominância de publicação em conferência (67,6\%). Foram publicados onze artigos no SPLC (S02, S03, S10, S11, S13, S14, S16, S19, S26, S27, S31), sendo o local mais proeminente avaliado pelos pesquisadores para publicar seus artigos. O restante dos artigos foi publicado em diversas conferências. Dois estudos foram publicados em GPCE (S15, S18), CAISE (S13, S22), ICSR (S25, S30) e um estudo em ICSE (S05), RCIS (S33), MODELSWARD (S08), ICICS IRI (S20). Em seguida, uma menor quantidade de estudos primários foi publicada em Jornais (17,6\%). Dois artigos foram encontrados a partir de SCP (S21, S32), e um artigo de IS (S20), JSS (S29), JCC (S23) e SOSYM (S12). Finalmente, 14,7\% foram publicados em diferentes \textit{workshops}, sendo dois publicados em MISE (S01, S07) e uns publicados em ME (S04), VAMOS (S28) e FMSPLE (S34).

Com base nos dados coletados, podemos elaborar duas observações. Em primeiro lugar, a falta de estudos primários em locais de topo (por exemplo, TSE, TOSEM) pode ser motivada, por sua vez, devido à ausência de resultados com base em uma série de estudos empíricos controlados, em vez de baseados em exemplos-brinquedos e preliminares estudos casos. Em segundo lugar, a pesquisa sobre o tema da integração de modelos de \textit{features} tem mostrado uma tendência ascendente; No entanto, o número de publicações é ainda pequeno, por vezes inexistente em determinado período de tempo. Isto pode indicar duas direções possíveis: (1) uma vez que a integração de artefatos de software (por exemplo, código fonte, modelos conceituais, entre outros) é amplamente conhecida como um problema severo \cite{schaefer2012}, e o número de lacunas abertas é enorme, mais trabalhos de pesquisa devem estar em andamento. Portanto, novas publicações devem ocorrer nas próximas edições de conferências, \textit{workshops} e revistas; (2) a saída de potenciais técnicas eficazes e revolucionárias deve ocorrer nos próximos anos, derivadas de técnicas teóricas, que não são sólidas. Hoje, não existe uma ferramenta fácil de usar e pronta para produção, enquanto a viabilidade comercial não está comprovada.

\emph{Vista panorâmica e outros desafios.} Abordando o segundo objetivo, a Figura \ref{fig:bolha} apresenta um gráfico de bolhas, que é uma variação de um gráfico de dispersão. O eixo-x representa as questões de pesquisa consideradas, ou seja, a amplitude do método de pesquisa utilizado (à esquerda) e o tipo de técnicas propostas (à direita). O eixo-y representa o ano de publicação. Uma dimensão adicional dos dados é representada no tamanho das bolhas, o que representa o número de estudos primários publicados.
A característica principal observada neste gráfico é a presença de um padrão de distribuição considerando o uso de métodos de pesquisa ao longo dos anos. Os trabalhos de "proposta de solução" têm sido os métodos de pesquisa mais adotados, sendo encontrados em 70\% (24/34) dos estudos primários, enquanto que os demais métodos de pesquisa utilizados registraram os 30\% dos casos restantes (10/34), distribuídos da seguinte forma: (3\%, 1/34), artigos de experiência, (3\%, 1/34), artigos de opinião, (3\%, 1/34), pesquisa de validação (3\%, 1/34) e pesquisa de avaliação (17\%, 6/34). Esta superioridade pode significar que este campo de pesquisa ainda está em uma fase inicial, em que o número de trabalhos é predominantemente maior do que um relacionado à avaliação empírica. Além disso, observou-se que os estudos primários se concentraram principalmente na proposição de técnicas de verificação e validação (47\%, 16/34), técnicas de comparação (33\%, 11/34) e técnicas de integração representando (20\%, 7/34 ).

\begin{figure}[ht]
	\caption{Evolução das publicações em termos de temas de pesquisa ao longo dos anos.}
	\label{fig:bolha}
	\centering%
	\begin{minipage}{.8\textwidth}
		\includegraphics[width=\textwidth]{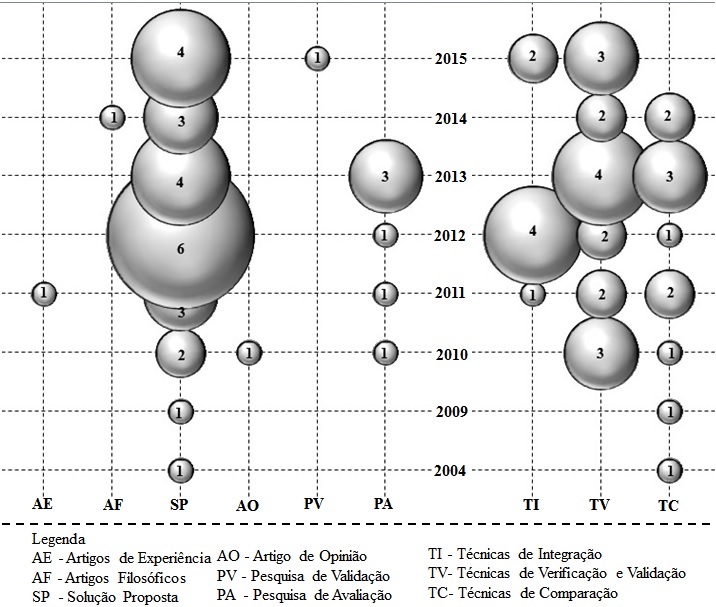}
		\fonte{Elaborado pelo autor.}
	\end{minipage}
\end{figure}

Em 2012, apenas um dos sete artigos relacionados à execução da pesquisa de avaliação foi proposto, enquanto seis artigos publicados foram "proposta de solução", sendo quatro estudos sobre técnicas de integração, dois estudos explorando técnicas de verificação e validação e um estudo sobre técnicas de comparação. Além disso, sete estudos propõem técnicas de integração de modelos de \textit{features}, abordando as técnicas de união, a interseção e diferença. Ressaltamos também que os estudos sobre integração de modelos de \textit{features} são recentes, uma vez que foram publicados predominantemente nos últimos anos. Isso reforça que os estudos primários selecionados são pioneiros. De fato, apenas técnicas de conceito de projeção foram propostas, ao invés de técnicas conviviais e práticas, apoiadas em dados empíricos derivados de projetos de desenvolvimento de software convencionais. Esses dados são essenciais para apoiar os desenvolvedores a tomar decisões sobre a adoção (ou não) de tais técnicas em projetos de software convencionais, onde o tempo e o custo são apertados.

Analisando mais de perto as técnicas de integração, a maioria delas fornece abordagem primitiva. Mesmo que eles possam lidar com um conjunto de casos de integração elementar, suportando questões estruturais, sintáticas e semânticas, eles ainda são propensos a erros como casos de integração mais crítica não são devidamente apoiados. Na verdade, geralmente eles são incapazes de identificar a similaridade, ou diferenças, causada por mudanças de reestruturação. Este tipo de modificações é tipicamente encontrado em casos mais severos de evolução de modelos de \textit{features}, que podem aparecer a partir das tarefas de refatoração ou do refinamento de \textit{features} principais no SPL (por exemplo, incluindo, alterando ou excluindo \textit{features} devido à mudança de requisitos do projeto). Uma direção de pesquisa muito interessante é sobre como melhorar a comparação técnica entre os modelos de \textit{features}, que estão cientes das mudanças estruturais, sintáticas e semânticas na evolução de novos modelos.

Observa-se que a integração de modelos de \textit{features} depende de uma etapa crítica, a saber, a comparação entre os elementos do modelo seguindo uma abordagem única e multi-estratégica, onde se podem contemplar questões estruturais, sintáticas, semânticas, léxicas e de layout. Se o nível de detalhe para comparar modelos de \textit{features} puder ser ajustado, os desenvolvedores comparariam modelos de \textit{fetaures} em diferentes níveis de abstração. Isso permitiria ter um melhor controle sobre os tipos de alterações conflitantes que poderiam ocorrer.

Além disso, outra direção de pesquisa interessante seria a falta de conhecimento empírico sobre a aplicação das técnicas atuais. A avaliação tem sido largamente baseada na reflexão e na opinião de especialistas, em vez de provas derivadas de estudos empíricos (por exemplo, estudos de caso, experiência controlada e quase experimental). Em \citetexto{kleinner2012, farias2015}, os autores apontam que as técnicas de integração têm sido largamente baseadas na opinião de especialistas (evangelista), em vez de experiências de conhecimento prático. Infelizmente, as opiniões dos especialistas geralmente divergem. Como tal, questões mais práticas sobre a integração de modelos de \textit{features} devem ser exploradas em ambientes reais. Ou seja, pouco se sabe sobre o esforço que os desenvolvedores investem para integrar modelos de \textit{features}, bem como eles lidam com conflitos entre os elementos dos modelos de \textit{features} a serem integrados.

Se os conflitos entre os elementos de entrada dos modelos de \textit{features} forem resolvidos de forma inadequada, os desenvolvedores serão capazes de lidar com defeitos (ou inconsistências). Dado que as técnicas atuais de modelagem não são para detectar e resolver um amplo espectro de conflitos, os desenvolvedores acabam tendo que lidar com modelos de \textit{features} com problemas. Com isso em mente, uma lacuna importante no conhecimento sobre o uso de modelos de \textit{features} está relacionada aos efeitos na produção de modelos de má qualidade (por exemplo, esforço, compreensibilidade e reutilização) na prática.

\section{Ameaças à Validade} \label{TR_Ameaças}

Esta seção discute as estratégias utilizadas para mitigar algumas ameaças à validade, ou seja, a confiabilidade dos resultados obtidos no processo de investigação, sendo considerada a validade de construção, validade externa e interna, e por fim, a validade de conclusão \cite{wohlin2000, kitchenham2010, kitchenham2011}.  

\textbf{Validade de construção.} Considera os relacionamentos entre a teoria e a observação, consiste em evidenciar os resultados obtidos durante a investigação com os resultados pretendidos em sua elaboração. Dado que a execução de um estudo de mapeamento sistemático retorna uma extensa lista de artigos publicados, um risco comum é não incluir todos os estudos relevantes. Para atenuar esta ameaça, seguiram-se as diretrizes estabelecidas para definir palavras-chave e a \textit{string} busca, aplicando estas às bases de dados eletrônicas (Biblioteca Digital ACM, IEEE, o Google Scholar, Scopus, Springer Link e Science Direct).

\textbf{Validade interna.} Poderá ser influenciada por medições incorretas ou instrumentos inadequados prospectando possíveis diferenças entre os resultados. A validação esta relacionada aos filtros de investigação que determinam a análise desta pesquisa, para evitar esta ameaça, aplicou-se seis etapas conforme segue: (1) pesquisa inicial, através da \textit{string} de busca (2) remoção dos artigos duplicados, (3) filtro por título das publicações conforme palavras-chave, (4) filtro por abstrato das publicações, realizando uma leitura previa destas verificando já as questões de pesquisa investigadas, (5) combinação das publicações, ou seja, agrupamentos após a execução filtros e finalmente (6) filtros por texto, isto é, uma densa leitura dos artigos selecionados, para após obter o numero de artigos selecionados, dos quais as questões de pesquisa investigadas estão relacionadas diretamente a estas.

\textbf{Validade externa.} Considera até que ponto os resultados são generalizáveis, limitando os resultados do estudo para contextos dentro do ambiente avaliado.  As ameaças mais comuns à validade externa são selecionar amostras não representativas quantitativamente ou qualitativamente e adotar matérias ou instrumentos distantes da realidade. Para evitar estas ameaças seguiram-se as etapas descritas na literatura sobre revisão sistemática, definindo um escopo apresentado neste trabalho, no contexto de integração entre modelos de \textit{features}, desta forma minimizando as dificuldades em reproduzir as etapas de investigação próprias deste trabalho, ou seja, dentro deste universo restrito.

\textbf{Validade de conclusão.} Por fim, realizou-se um diagnóstico dos resultados da investigação, apurando a confiabilidade destes, intensificando sua veracidade, garantido a consistência e a precisão desta investigação. Para isso, seguiram-se rigorosamente os passos descritos na literatura \cite{petersen2008, kitchenham2010, kitchenham2011} para uma boa condução deste estudo de mapeamento sistemático, assim como a correta análise e interpretação das questões de pesquisa investigadas.  

\section{Novas Direções de Pesquisa} \label{TR_Conclusões}

Este trabalho procurou compreender, caracterizar e resumir o estado da arte sobre técnicas de integração de modelos de \textit{features}. Para isso, realizamos um estudo de mapeamento sistemático para investigar seis questões de pesquisa. Selecionou-se 34 estudos primários aplicando um cuidadoso processo de filtragem em uma amostra de 775 estudos pesquisados em 6 bases de dados eletrônicas (escopo de pesquisa).

Resumisse os principais achados da seguinte forma. Em relação às notações utilizadas, a maioria (44\%, 15 de 34) dos estudos escolheu a Análise de Domínios Orientada a \textit{Features} (\textit{Feature-Oriented Domain Analysis – FODA}) para modelagem de diagramas de \textit{features}. Este resultado mostra uma forte preferência do meio acadêmico em usar esta notação. Portanto, ainda não existem investigações que considerem tanto as etapas de comparação como as de integração. A comparação das técnicas dos modelos propostos na literatura abrangeu estratégias sintáticas (12\%, 4 de 34 estudos) e semânticas (21\%, 7 de 34 estudos). Em relação às técnicas de integração de modelos de \textit{features}, verificou-se que a maioria dos estudos se concentra na proposição de estratégias de união (21\%, 7 de 34) e interseção (21\%, 7 de 34). Além disso, uma pequena parte dos trabalhos 6\% (2 de 34 estudos) propôs a estratégia de diferenciação.

Os resultados referentes às técnicas de validação mostraram que existem diversos tipos de estratégias de validação, isto é, técnicas como SAT, CSP, BDD, CNF e DNF foram discutidas em 24\% (8 de 34) dos estudos em conjunto. Além disso, há também artigos argumentando sobre SAT (6\%, 2 de 34 estudos) e CSP (3\%, 1 de 34 estudos) exclusivamente. Os resultados relacionados ao suporte de ferramentas apontam que a maioria (56\%, 19 de 34 estudos) de técnicas de integração é automática. Além disso, o apoio semiautomático é coberto por uma minoria (9\%, 3 de 34) dos estudos. No entanto, uma grande parte (35\%, 12 de 34) de obras não propõe qualquer ferramenta. Além disso, o suporte de ferramentas para todas as etapas de integração propostas não foi encontrado na literatura atual. Quanto aos métodos de pesquisa, os resultados mostraram que a maioria (67\%, 23 de 34) dos estudos investigados foi à proposta de soluções, ou seja, é importante notar a falta de estudos empíricos em condições reais. Assim, mais experimentos e estudos de caso devem ser validados no contexto industrial.

Como trabalho futuro, é necessária realizar estudos experimentais, ou seja, experiência controlada para verificar o esforço empregado dos pelos desenvolvedores na detecção de conflitos e o grau de precisão das integrações, bem como o requisito necessário para integrar modelos de \textit{features} em configurações do mundo real. Além disso, é necessário implementar e propor uma técnica de integração de modelos de \textit{features} aplicando todos os passos de integração propostos neste trabalho. Finalmente, espera-se que os comentários discutidos ao longo do estudo possam encorajar pesquisadores e profissionais a explorar as informações relatadas. Além disso, este estudo pode ser visto como um primeiro passo para uma agenda mais ambiciosa sobre como caracterizar e melhorar as técnicas de integração de modelos de \textit{features}.

\chapter{Técnica de Integração de Modelos de \textit{Features}} \label{TécnicaDeIntegração}
Para explorar as oportunidades de pesquisa elencadas na Capítulo \ref{TrabalhosRelacionados}, assim como atender aos objetivos descritos na Seção \ref{I_Objetivos}, propõe-se uma técnica de integração de modelos de \textit{features}. A técnica proposta busca identificar elementos equivalentes entre dois modelos de \textit{features}, bem como integrá-los. Este Capítulo encontra-se organizado da seguinte forma. A Seção \ref{TiProcesso} apresenta um processo genérico de integração de modelos de \textit{features}, o qual é utilizado como um guia para integrar os modelos. A Seção \ref{TiEstrategiasComparacao} descreve as estratégias de comparação de modelos definidas para identificar equivalência entre os modelos de \textit{features}. Por fim, a Seção \ref{TiEstratégiasdeIntegração} descreve as estratégias de integração dos modelos.

\section{Processo de Integração do Modelos de \textit{Features}} \label{TiProcesso}

A composição de modelos desempenha um papel fundamental no desenvolvimento de \textit{software} dirigido a modelos, para adicionar novos elementos ou reconciliar modelos desenvolvidos em paralelo por diferentes equipes \cite{farias2015}.
A integração e a automação de processos estão entre os fatores mais significativos dirigidos a indústria de \textit{software}, devido aos multiplos desafios tecnológicos voltados a integração e automatização de modelos, concentrando seus esforços na obtenção de melhorias no processo \cite{koehler2005}. Conforme \citetexto{bezivin2006}, não há um consenso sobre as atividades e suas transições, bem como os artefatos gerados a partir da execução das mesmas para dar suporte a integração de modelos. 

Em \cite{oliveira2007}, os autores reconhecem essa necessidade e propõem um guia para integração de perfis UML. Entretanto os autores em \cite{oliveira2007} apresentam um guia para auxiliar na composição de modelos, orientando as atividades, assim como os procedimentos que devem ser adotados a fim de obter uma completa iteração e alinhamento entre as atividades desenvolvidas

Há peculiaridades inerentes à integração de modelos, bem como uma metodologia que auxilia na elaboração dos processos decorrentes para executar e integrar especificamente modelos de \textit{features}. Isso pode ser motivado pela área de integração de modelos de \textit{features} não ter ainda sido explorada.

O modelo de composição proposto por \cite{oliveira2007} pode ser aplicado em diferentes domínios e plataformas, sendo assim  o modelo é aplicado à composição  de modelos de  \textit{features}, o qual passa por pequenos ajustes exibido na Figura \ref{fig:propostaProcessoIntegracao}. O processo proposto é organizado em quatro fases, as quais são apresentadas a seguir:

\begin{enumerate} 
   \item \textbf{Análise dos modelos de entrada.} Esta primeira fase consiste em identificar e analisar os modelos de \textit{features} de entrada, MF$_{A}$ e MF$_{B}$, buscando garantir a compatibilidade para sua integração, bem com prevenir a entrada de modelos de \textit{features} inconsistentes \cite{oliveira2008c}. Esta fase procura verificar os seguintes critérios: (1) os modelos de \textit{features} de entrada são do mesmo formato, isto é, um arquivo XML; e (2) os modelos de \textit{features} de entrada apresentam inconsistências. Caso os modelos de entrada não atendam aos critérios estabelecidos, a integração é considerada invalidada, finalizando o processo de integração.
   \item \textbf{Comparação dos modelos de entrada.} A segunda fase incide em comparar os modelos de \textit{features} de entrada para determinar a semelhança entre seus elementos. Para isso, têm-se como entrada as (1) estratégias de comparação, sejam estas sintáticas ou semânticas; e o (2) limiar. A comparação gera os seguintes resultados: 1)  similaridade, especificando o grau de similaridade entre os elementos dos diagramas de \textit{features} de entrada com um valor de 0 (zero) a 1 (um);  (2) descrição de equivalência , umas descrição dos elementos dos modelos comparados, MF$_{A}$ e MF$_{B}$, sendo considerados equivalentes; (3) descrição dos elementos não correspondentes, isto é, a descrição dos elementos dos modelos, MF$_{A}$ e MF$_{B}$ que não são equivalentes. Os elementos dos modelos MF$_{A}$ e MF$_{B}$ que tiverem um grau de similaridade acima do limiar definido, eles serão considerados equivalentes. Dessa forma, o limiar pode ser considerado como um mecanismo para estabelecer uma relação de equivalência entre os elementos de entrada. Além disso, ele permite que os usuários da técnica atue diretamente na definição da equivalência. 

\begin{figure}[ht]
	\caption{Processo de integração de modelos de \textit{features} proposto.}
	\label{fig:propostaProcessoIntegracao}
	\centering%
	\begin{minipage}{.6\textwidth}
		\includegraphics[width=\textwidth]{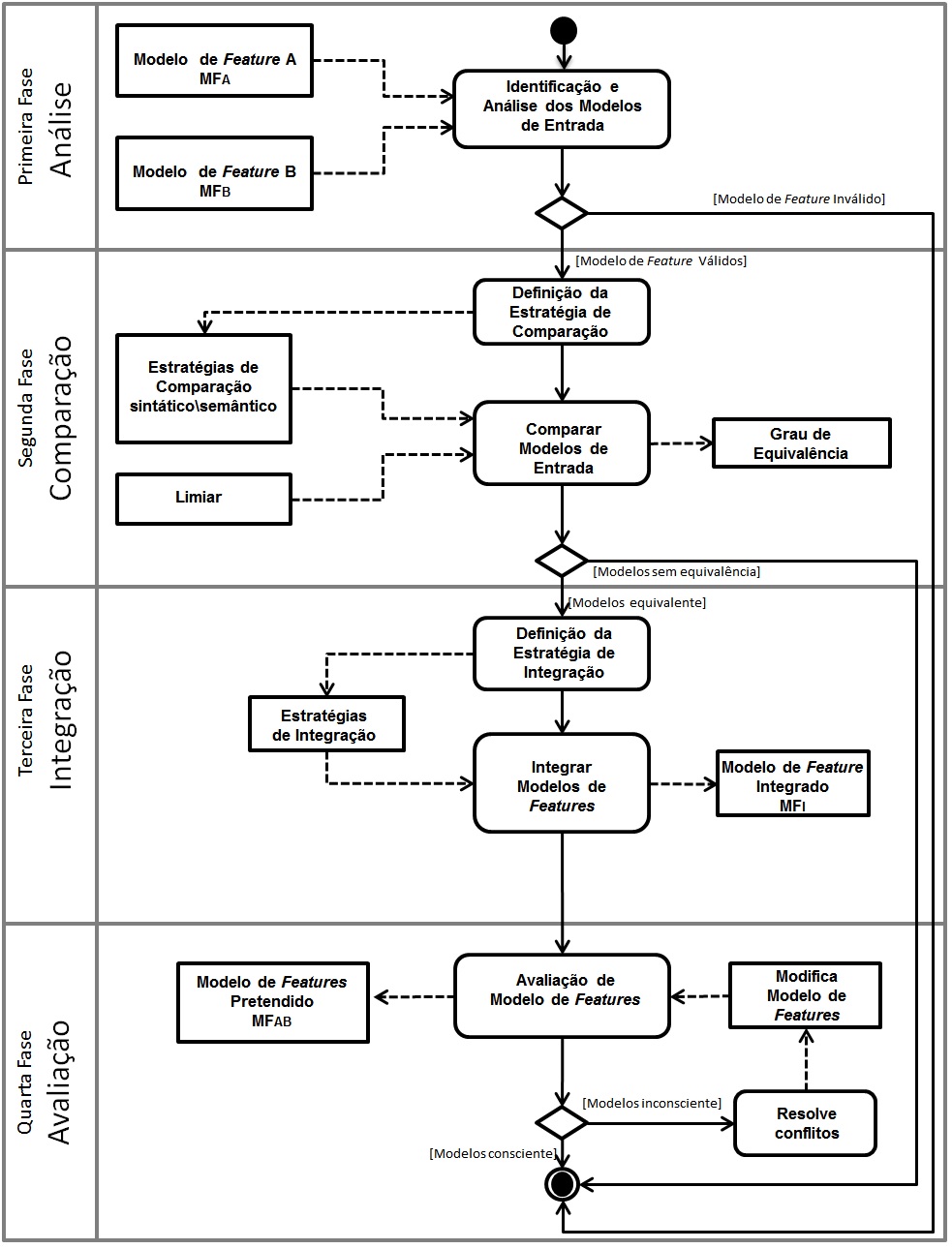}
		\fonte{Adaptado de Oliveira (2007).}
	\end{minipage}
\end{figure}

   \item \textbf{Integração dos modelos.} A terceira fase consiste em integrar os elementos dos modelos de entrada, considerando a descrição de equivalência e a similaridade produzida na fase anterior. Essa integração irá gerar um modelo integrado, MF$_{I}$. Se dois elementos dos modelos de entrada não forem definidos como equivalentes, então uma abordagem semiautomática de integração será executa, a qual permite o usuário da técnica definir se os dois modelos serão integrados ou não. Caso contrário, os elementos definidos na fase anterior serão integrados automaticamente. Sendo assim, baseado na equivalência entre os elementos de entrada, uma abordagem automática ou semiautomática será seguida. Em outras palavras, esta fase decorre  principalmente de quais estratégias de integração será seguida: (1) automática, sendo os modelos de \textit{features} totalmente correspondentes, ou seja, grau de equivalência é igual a 1, ou sendo os modelos não correspondentes, isto é, o grau de equivalência igual a 0, aplica-se múltiplas estratégia de integração a qual acomoda em sua saída quatro prospecções de integração (adicional,  formal, parcial e complementar) \cite{acher2010, acher2009, thum2009}; (2) semiautomático, decorre dos modelos comparados não correspondentes possuírem um grau de equivalência  que varia entre 0.2 a 0.9 \cite{farias2015, farias2015b}, sendo que esta fase requer intervenção humana, devido aos conflitos detectados entre os modelos. A fase seguinte é responsável por gerenciar os modelos de saída,  MF$_{I}$ e  MF$_{AB}$, sejam estes equivalentes ou não. 
   \item \textbf{Avaliação da integração realizada.} A quarta fase analisa a saída do modelo produzido, ou seja, o modelo de integrado é igual ao modelo  pretendido, MF$_{I}$ = MF$_{AB}$. Inconsistências inerentes de uma má integração podem surgir, isto é, o modelo de \textit{feature} é diferente do pretendido, MF$_{I}$ $\neq$ MF$_{AB}$. Assim o modelo , MF$_{I}$, precisa ser manipulado para que os conflitos identificados possam ser resolvidos. Para isso, o processo passa a ser realizado de forma semiautomática, verificando as inconformidades na fase anterior, ou seja, durante o processo de integração retomando o processo decisório para que analistas ou desenvolvedores atuem pontualmente nos conflitos, deste modo, transformando  MF$_{I}$ em MF$_{AB}$. 
\end{enumerate}
Em síntese, a técnica lê e avalia dois modelos de entrada, realizando os procedimentos de comparação e integração dos elementos que compõe os diagramas, produzindo um modelo \textit{feature} composto. Na próxima Seção são apresentadas as propostas de comparação dos modelos: (1) sintática, (2) semântica e (3) estrutural, os quais definem e estabelecem os critérios de similaridade entre os modelos, assim como os procedimentos integração.

\section{Estratégias de Comparação} \label{TiEstrategiasComparacao}

As estratégias de comparação definem os métodos aplicados para verificar o grau de equivalência entre os modelos de \textit{features}. No Capítulo \ref{FundamentaçãoTeórica} é citado os elementos necessários para desenvolver os diagramas de \textit{features}, nesta Seção são relembrados alguns deste fundamentos, para uma melhor compreensão das técnicas que virão a ser propostas. 

 O modelo \textit{features} é arquitetado visualmente através de notações gráficas, sendo suas \textit{features} representadas por retângulos e seus elementos gráficos através círculos (preenchido ou vazado), triângulos (preenchido ou vazado) e setas (simples, tracejada ou bidirecional), sua variabilidade é representada pelos relacionamentos entre as \textit{features}, conforme demostrado na Figura \ref{fig:mapLogicofeature}. 
 
 O conjunto de \textit{features} forma uma hierarquia contendo um elemento principal (\textit{feature}-pai) e elementos secundários (\textit{features}-filhos) conectados através de relacionamentos (opcional, obrigatório, alternativo, dependência e exclusão), sendo possível formalizar sua configuração, a qual pode ser representada através da semântica formal \cite{oliveira2009b}. Considerando que os modelos de \textit{features} seguem uma linguagem natural, diferentes interpretações são aceitáveis, ocasionando ambiguidade. Buscando verificar esta equidade entre os significados são aplicadas estratégias de comparação. Embora ontologias possam ser utilizadas na comparação de modelos, eles precisam ser atualizados frequentemente, inviabilizando seu uso em ambientes reais de desenvolvimento de \textit{software}.

Em \cite{farias2015} os autores relatam que a principal deficiências é a adoção de técnicas generalizadas na composição de modelos e a falta de conhecimento empírico sobre seus efeitos no esforço dos desenvolvedores na resolução de conflitos. A comparação tem como parte de seus objetivos verificar a existência de sobreposição semântica e sintática nos modelos de entrada \cite{oliveira2008}. Tais sobreposições devem ser evitadas, devido a necessidade do modelo final produzido representar seus  conceito, por exemplo o modelos de \textit{features} descritos anteriormente, a fim de evitar conflitos e transformações de modelos ineficientes.

\subsection{Estratégia Semântica} \label{TiSubEstrategisSemantica}

\begin{figure}[ht]
	\caption{Mapeamento de definições: modelo de \textit{features} e lógica proposicional.}
	\label{fig:mapLogicofeature}
	\centering%
	\begin{minipage}{1\textwidth}
		\includegraphics[width=\textwidth]{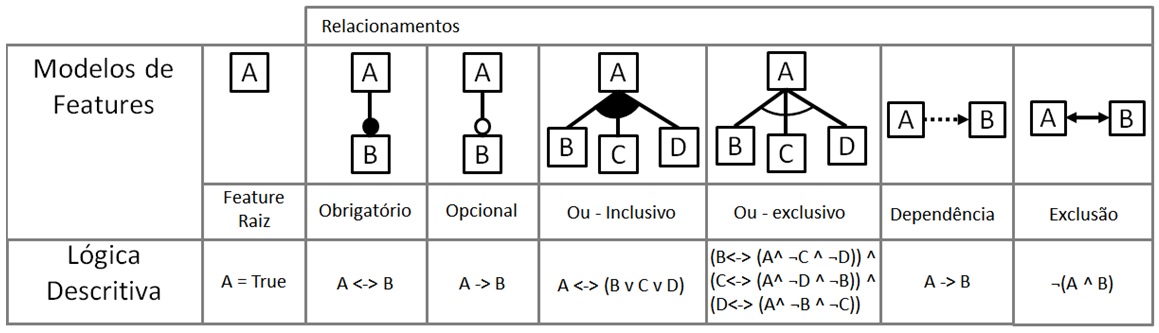}
		\fonte{Elaborado pelo Aultor.}
	\end{minipage}
\end{figure}

A estrutura para representar o modelo de \textit{feature} ocorre através de um conjunto de símbolos e regras de formação, constituindo assim um conjunto de configurações. Na Figura \ref{fig:mapLogicofeature} é possível visualizar as regras de derivação do modelo de\textit{features} estabelecendo as fórmulas de conectividade. Cada \textit{feature} corresponde a uma variável booleana (0 ou 1) capturada através da lógica proposicional ($\vee$, $\wedge$,  $\rightarrow$, $\leftrightarrow$ e negação $\neg$).

O modelo de \textit{features} é definido como segue:\\
Modelo de \textit{Features}  MF = (G, r, Eobr, FXOR, FOR, Depen, Excl)

\begin{itemize}
   \item G = (F,E) é a formação de uma árvore que contém uma raiz, onde  F é um conjunto finito de \textit{features}, e E $\subseteq$ F x F é um conjunto finito de arestas (\textit{subfeatures}), ou seja, é a decomposição hierárquica entre as \textit{features}, pai-filhos e seus relacionamentos;
   \item r $\in$ F é \textit{feature} raiz;
   \item \textit{Fetures} que não são obrigatórias e que não fazem parte de um grupo de \textit{features} são \textit{features} opcionais; 
   \item Uma \textit{feature} pai poderá ter vários grupos de \textit{feature}, mas uma \textit{feature} deve pertencer apenas a um grupo \textit{feature};
   \item FXOR $\subseteq$ P (F) x F e FOR $\subseteq$ P (F) x F define os grupos de \textit{features}, os quais são um conjunto de pares filhos junto com feature pai. As \textit{features} filhas são grupos exclusivos (XOR) ou inclusivos (OR).
   \item Conjunto de restrições que implica Depen (uma feature depende de outra feature) e  Excl (exclusão entre \textit{features}) em que A $\in$ F e B $\in$ F.
 \end{itemize}   
  
\begin{figure}[ht]
	\caption{Modelo de \textit{features} representação gráfica para transformação lógica proposicional.}
	\label{fig:modeloLogicoProp}
	\centering%
	\begin{minipage}{.8\textwidth}
		\includegraphics[width=\textwidth]{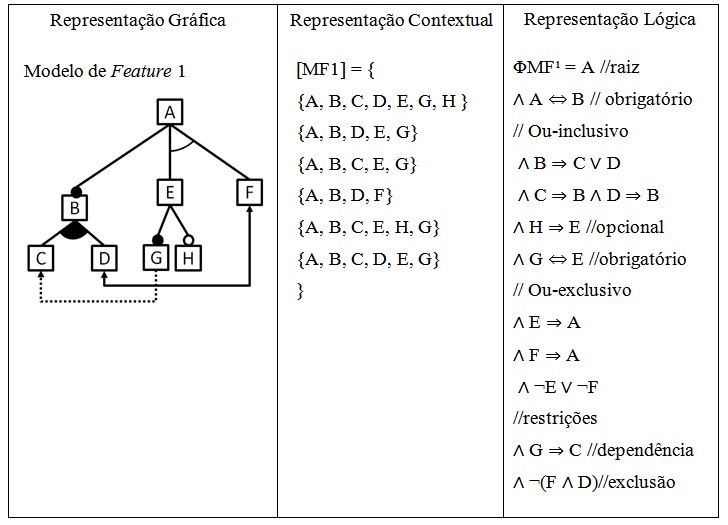}
		\fonte{Elaborado pelo Aultor.}
	\end{minipage}
\end{figure}

   A partir desta estrutura se estabelece um conjunto de configurações válidas para o modelo de \textit{features}, ou seja, são estabelecidas as regras de boa formação. A Figura \ref{fig:modeloLogicoProp} apresenta um modelo de \textit{features}, sendo o mesmo mapeado para a fórmula proposicional. Cada \textit{feature} corresponde a uma variável da lógica proposicional e suas possíveis configurações derivadas. O modelo de \textit{features} 1 apresenta três níveis hierárquicos, sete relacionamentos  e oito \textit{features}. Sua representação contextual, [MF1],  demostra seis possíveis configurações de uma linha de produto. Por fim, o modelo lógico proposicional é $\Phi$MF$^{1}$.

A estratégia semântica apresenta seis tipos de relacionamento, obrigatória, opcional, alternativa inclusiva e exclusiva, e os relacionamentos transversais que são exclusão e dependência, os quais são aplicados às fórmulas de conectividade, cada relacionamento corresponde a uma variável booleana (0 ou 1). A fórmula para identificar a similaridade entre conectividade das \textit{features} é subdividida em duas etapas:

\begin{itemize}
   \item A primeira etapa consiste em verificar a diferença entres os relacionamentos dos modelos de \textit{features}.
   \item A segunda etapa consiste na aplicação da fórmula da estratégia semântica para se obter a similaridade que determina a equidade entre os modelo de \textit{features}.
 \end{itemize}

A estratégia semântica para verificar a diferença é representada por “difSem” o qual recebe a diferença entre conectividade da \textit{Feature} Referência  representada por “ frdif” e a conectividade da \textit{Feature} Comparada representada “fcdif”, ou seja:

	\begin{center}
	\textbf{Equação 1.}
		\\difSem = frdif – fcdif, onde:
	\end{center}
\textbf{difSem} = 1.00 se ambos os relacionamentos  apresentarem a mesma conectividade;\\
\textbf{difSem} = 0.00 caso os relacionamentos entre as \textit{features}  não apresentem a mesma conectividade.\\

\begin{figure}[ht]
	\caption{Comparação entre relacionamentos de modelos de \textit{features}.}
	\label{fig:compModelos}
	\centering%
	\begin{minipage}{.6\textwidth}
		\includegraphics[width=\textwidth]{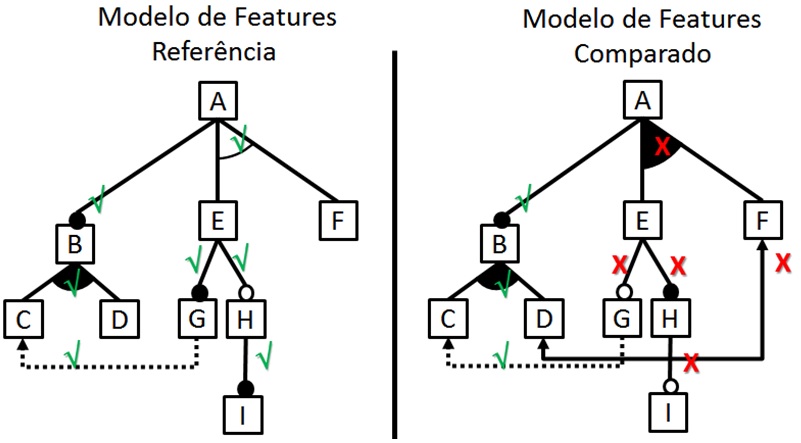}
		\fonte{Elaborado pelo Aultor.}
	\end{minipage}
\end{figure}

Na Figura \ref{fig:compModelos}, ilustra-se a comparação entre dois modelos de \textit{features}, o referência, MF$_{R}$, e  comparado, MF$_{C}$. É possível observar no exemplo, o relacionamento em ambos os modelos de \textit{features}, seja este opcional, obrigatório, alternativo inclusivo ou exclusivo, dependência e exclusão. Os modelos exibem o número de relacionamentos, onde o modelo referência possui 7 notações e o modelo comparado exibe 8 notações, conforme a Tabela \ref{tab:compModelos}. A comparação entre os elementos correspondentes conforme demostra a Tabela 7 expõe o escore de pontuação sendo similar igual a 1 e diferente igual 0. Por exemplo, na Figura 14, ao comparar a conectividade entre os pares {A, B} de ambos os modelos, verifica-se que a similaridade é igual a 1, entretanto a conectividade entre {A, E, F} não é similar, isto é, implica em 0 e assim sucessivamente verifica-se todos os relacionamentos conforme demostra-se na figura abaixo, o modelo comparado, MF$_{C}$,  exibe com um “x” os relacionamentos que não são similares e com “v” os relacionamentos similares.  Aplicando a fórmula da diferença semântica, difSem, obtêm-se o resultado demostrado na Tabela \ref{tab:compModelos}, onde o escore da pontuação é obtido após a comparação entre os modelos de \textit{features}. 

\begin{table}[htbp]
  \centering
  \caption{Escore de pontuação relacionamentos entre as \textit{features}.}
    \begin{tabular}{rrrrrrrrr}
    \toprule
    \multicolumn{9}{l}{\textbf{Conectividades}} \\
    \midrule
    \multicolumn{1}{l}{\textbf{MF Referência }} & \multicolumn{1}{c}{Obr}                & \multicolumn{1}{c}{Aex}                & \multicolumn{1}{c}{Ain}                & \multicolumn{1}{c}{Obr}                & \multicolumn{1}{c}{Opc}                & \multicolumn{1}{c}{Obr}                & \multicolumn{1}{c}{Dep}                & \multicolumn{1}{c}{-} \\
    \multicolumn{1}{l}{\textbf{MF Comparado}} & \multicolumn{1}{c}{Obr}                & \multicolumn{1}{c}{Ain}                & \multicolumn{1}{c}{Ain}                & \multicolumn{1}{c}{Opc}                & \multicolumn{1}{c}{Obr}                & \multicolumn{1}{c}{Opc}                & \multicolumn{1}{c}{Dep}                & \multicolumn{1}{c}{Exc} \\
    \multicolumn{1}{l}{\textbf{Escore}}    & \multicolumn{1}{c}{1}                  & \multicolumn{1}{c}{0}                  & \multicolumn{1}{c}{1}                  & \multicolumn{1}{c}{0}                  & \multicolumn{1}{c}{0}                  & \multicolumn{1}{c}{0}                  & \multicolumn{1}{c}{1}                  & \multicolumn{1}{c}{0} \\
    \midrule
    \multicolumn{1}{l}{\textbf{Lengenda}}  &                                        &                                        &                                        &                                        &                                        &                                        &                                        &  \\
    \multicolumn{4}{r}{Obrigatório = Obr}                                                                                                                             & \multicolumn{5}{r}{Alternativo Inclusivo = Ain } \\
    \multicolumn{4}{r}{Opcional = Opc}                                                                                                                                & \multicolumn{5}{r}{Alternativo Exclusivo = Aex} \\
    \multicolumn{4}{r}{Exclusão = Exc }                                                                                                                               & \multicolumn{5}{r}{Dependência = Dep } \\
    \end{tabular}%
  \label{tab:compModelos}%
  \fonte{Elaborado pelo Autor.}
\end{table}%

A segunda etapa consiste na aplicação da fórmula para calcular a estratégia semântica retornando o resultado da equivalência entre os modelos de \textit{features} comparados:

\begin{center}
\textbf{Equação 2.}
  \\$ESTSEM = \frac{\Sigma(difSem)}{c}$~onde:
\end{center}
\textbf{difSem }é o somatório total do escore obtido, ou seja, somatório das diferenças contidas entre os relacionamentos; \\
\textbf{c} é numero de conectividade existente.\\

 Após aplicar a fórmula para calcular a estratégia semântica, ESTSEM, verifica-se uma parte do conjunto das estratégias para formular o cálculo de efetividade estratégica, CEE, ou seja, o escore de pontuação exibido a partir dos dados contidos na Tabela \ref{tab:compModelos} entre os modelos de \textit{features}, MF$_{R}$ e MF$_{C}$, no exemplo acima ESTSEM é representado por difSem igual 3,  dividido pela número de conectividades existentes. Neste exemplo, c é igual a 8,  retornando como resultado o grau de equivalência entre as conectividades das \textit{features}, ESTSEM  é igual 0.37. 
 \begin{center}
  $\Sigma$ = 1+0+1+0+0+0+1+0 
  $ESTSEM = \frac{\Sigma(3)}{8}$~logo,
  ESTSEM = 0,37
\end{center}

\subsection{Estratégia Estrutural}\label{TiSubEstrategiaEstrutural}

A similaridade estrutural procura analisar a hierarquia entre os elementos dos modelos de \textit{features}, examinando o nível hierárquico em que as \textit{features} se encontram \cite{becan2015, oliveira2009}.  Com a aplicação desta técnica, investiga-se o grau de disponibilidade de configuração de possíveis produtos que podem sofrer alterações conforme o nível hierárquico em que se encontra. A fórmula para identificar a similaridade entre os níveis hierárquicos dos modelos de \textit{features} é subdividida em duas etapas. A primeira etapa consiste em verificar a diferença entres os níveis hierárquicos dos modelos de \textit{features}. A segunda etapa consiste na aplicação da fórmula para calcular estratégia estrutural para se obter a similaridade que determina a equidade entre os modelo de \textit{features}.

O Cálculo da Diferença Estrutural (difEst) é calculado de acordo com a equação ESEST:
Entre as hierarquias recebe no máximo 1 ponto,  estando os dois elementos (\textit{fetaures}) no mesmo nível hierárquico. Considera-se 0.5 ponto, para elementos (\textit{features}) que se encontram em um nível hierárquico abaixo ou acima do seu similar, aplica-se 0.25 pontos para elementos (\textit{features}) que se encontram dois ou mais níveis hierárquicos acima ou abaixo ao seu similar. Por fim, o valor 0 (zero) é atribuído, caso a \textit{feature} não encontre outra \textit{feature} similar nos níveis hierárquicos. Apresenta-se a fórmula para identificar o cálculo estrutural de diferença hierárquica entre as \textit{features}, sendo a diferença estrutural representada por “difEst”, a qual recebe a diferença entre nível hierárquico da \textit{feature} referência  representada por “ frn” e o nível hierárquico da \textit{feature} comparada representada “fcn” ou seja:\\

\fbox{
  \parbox {0.9\textwidth}{
  \begin{center}
  \textbf{Equação 3.}\\
	difEst = frn – fc\\
	------------------------------------------------------------------------------------------------------ 
\end{center}

Onde:

\textbf{difEst} = 1,00, se ambas as \textit{features} estiverem no mesmo nível hierárquico;\\
\textbf{difEst} = 0,50, caso as \textit{features} encontram-se  um nível hierárquico abaixo ou acima;\\
\textbf{difEst} = 0.25, caso as \textit{features} encontram-se  dois ou mais níveis hierárquico abaixo ou acima;\\
\textbf{difEst} = 0,00, caso não encontre \textit{features} similar relacionadas aos níveis hierárquico abaixo ou acima

 }
}\\

\begin{figure}[ht]
	\caption{Comparação entre o nível hierárquico de \textit{features}.}
	\label{fig:nivel}
	\centering%
	\begin{minipage}{.6\textwidth}
		\includegraphics[width=\textwidth]{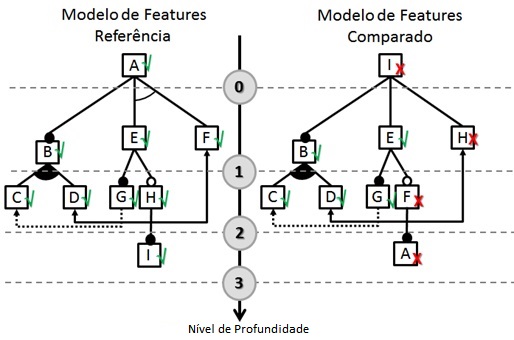}
		\fonte{Elaborado pelo autor.}
	\end{minipage}
\end{figure}

Na Figura \ref{fig:nivel}, ilustra-se a comparação entre dois modelos de\textit{features}, o modelo referência, MF$_{R}$, e modelo comparado, MF$_{C}$.   O nível hierárquico entre os modelos  comparados, neste exemplo, é de 0 a 3, isto é, quatro níveis hierárquicos. O exemplo ilustra abaixo, a \textit{feature} $\lbrace$A$\rbrace$ exibida no MF$_{R}$, encontra-se no nível zero, porém no modelo comparado, MF$_{C}$, a \textit{feature} $\lbrace$A$\rbrace$ encontra-se no terceiro nível, conforme demostrado na Tabela \ref{tab:nívelHierárquico}. O escore de pontuação, difEst é igual a 0.25, devido a diferença dos níveis. Assim, sucessivamente verifica-se em que nível hierárquico, encontram-se as \textit{features} conforme demostra-se na figura abaixo, o modelo comparado, MF$_{C}$, exibe com um “x” os níveis hierárquicos que não são similares e com “v” os níveis hierárquicos similares. Aplicando a fórmula da diferença estrutural, difEst, obtêm-se o resultado demostrado na Tabela \ref{tab:nívelHierárquico}, cujo escore da pontuação é obtido após a comparação entre os modelos de \textit{features}.

\begin{table}[htbp]
  \centering
  \caption{Escore de pontuação nível hierárquico.}
    \begin{tabular}{lccccccccc}
    \toprule
    \textit{\textbf{Features}}             & \textbf{A}                             & \textbf{B}                             & \textbf{C}                             & \textbf{D}                             & \textbf{E}                             & \textbf{F}                             & \textbf{G}                             & \textbf{H}                             & \textbf{I} \\
    \midrule
    \textbf{Escore}                        & 0,25                                   & 1                                      & 1                                      & 1                                      & 1                                      & 0,5                                    & 1                                      & 0,5                                    & 0,25 \\
    \bottomrule
    \end{tabular}%
  \label{tab:nívelHierárquico}%
  \fonte{Elaborado pelo autor.}
\end{table}%

A segunda etapa consiste na aplicação da fórmula da estratégia estrutural para calcular o nível da estrutura hierárquica, retornando o resultado da equivalência dos níveis hierárquicos entre os modelos de \textit{features} comparados: 

\begin{center}
\textbf{Equação 4.}\\
  $ESTEST = \frac{\Sigma(difEst)}{f}$~onde:
\end{center}
\textbf{difEst} é o somatório total do escore obtido através das diferenças estruturais;\\
\textbf{f }é número de \textit{features} existente. \\

 O grau máximo de similaridade obtido na estratégia estrutural é 1, para os casos em que as \textit{features} f estejam  todas contidas no mesmo nível hierárquico n.
 
Após aplicar a fórmula para calcular a estratégia estrutural, ESTEST, verifica-se uma parte do conjunto das estratégias para formular o Cálculo de Efetividade Estratégica, CEE. O escore de pontuação é exibido a partir dos dados contidos na Tabela \ref{tab:nívelHierárquico} entre os modelos de \textit{features}, MF$_{R}$, e MF$_{C}$. No exemplo acima ESTEST é representado por difEst igual 6.5,  dividido pela número de \textit{features} existentes e neste exemplo, f é igual a 9,  retornando como resultado o grau de equivalência entre os níveis hierárquicos das \textit{features}, onde ESTEST  é igual 0.72. 

\begin{center}
  $\Sigma$ = 0,25+1+1+1+1+0,5+1+0,5+0,25 
  $ESTEST = \frac{\Sigma(6.5)}{9}$~logo,
  ESTEST = 0,72
\end{center}

\subsection{Estratégia Sintática} \label{TiEstrategisSintatica}

Considerando que os modelos de \textit{features} seguem uma linguagem natural é aceitável diferentes interpretações ocasionando ambiguidade na comparação entre modelos, buscando verificar esta equidade entre os significados das palavras procura-se aplicar um dicionário de sinônimos em conjunto com uma técnica para comparação de \textit{string}. 

O algoritmo \textit{Jaro distance} é formado através de uma matriz, sua aplicação foi desenvolvida para comparação de pequenas strings \cite{cohen2003}, e baseia-se no número de caracteres comuns entre duas strings e sua ordem sequencial. A métrica da distância projetada se adapta melhor a distâncias curtas, e a pontuação normalizada para o seu algoritmo é 0, quando não apresenta nenhuma similaridade e  1 quando o mesmo apresenta  similaridade.  
A fórmula para identificar a similaridade entre as \textit{features} é subdividida em duas etapas:
\begin{itemize}
   \item A primeira etapa consiste em verificar a diferença gramatical entres as \textit{features} que compõem os modelos.
   \item A segunda etapa consiste na aplicação da fórmula estratégia sintática para se obter a similaridade que determina a equidade entre os modelo de \textit{features}.
 \end{itemize}
 
 A distância de \textit{Jaro dj} entre duas \textit{strings} é produzida através da comparação da \textit{string} 1  representada por |s1| e da  \textit{string} 2 representada por |s2|, conforme segue sua fórmula:
 
\begin{center}
	$$ d_{j} = \begin{cases} 0 \left( \frac{m}{|s1|}+\frac{m}{|s2|}+\frac{m}{|s3|} \right) \\ \frac{1}{3} \end{cases} $$
\end{center}
 
 \begin{description}
   \item[m] é o número de caracteres correspondente;
   \item[t] é o número de transposições suportadas.
 \end{description} 
 
 Quando o valor de m for igual a 0, não existe similaridade entre as strings. As strings não correspondentes, ou seja, aquelas que têm sua ordem de sequencia encontra-se diferente entre s1 e s2, onde aplica-se a fórmula de transposição, contendo o valor máximo dos seus caracteres entre |s1| e |s2| que é a divisão por 2, definindo, desta forma, o número de transposições suportadas. 
 
 \begin{center}
	$$ t = \left[ \frac{max(|s1|,|s2|)}{2}\right] $$
\end{center}

\begin{figure}[ht]
	\caption{Comparação gramatical entre \textit{features}.}
	\label{fig:compGramatical}
	\centering%
	\begin{minipage}{.6\textwidth}
		\includegraphics[width=\textwidth]{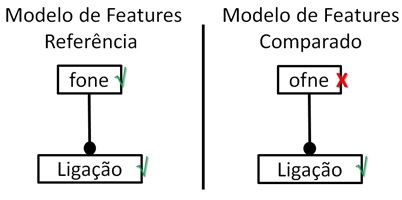}
		\fonte{Elaborado pelo autor.}
	\end{minipage}
\end{figure}

Na Figura \ref{fig:compGramatical}, é apresentado um exemplo de comparação entre dois modelos de \textit{features}, o modelo referência, MF$_{R}$, e o modelo comparado, MF$_{C}$. A similaridade entre a feature {fone} no modelo MFR, e a feature {ofne} no modelo MF$_{C}$, demostra que não há uma similaridade, entretanto a feature $\lbrace$ofne$\rbrace$ tem o mesmo significado contento um erro de grafia, conforme demostrado na Tabela \ref{tab:escoreSintatico}, o escore de pontuação, \textit{dj} é igual a 0.91, devido a diferença entre as strings. Assim, sucessivamente verifica-se a sintaxe entre as \textit{features }conforme demostra-se na figura abaixo, o modelo comparado,  MF$_{C}$, exibe com um “x” a sintaxe que não são similares e com “v” a sintaxe similares. Aplicando a fórmula da diferença entre as \textit{features}, isto é, a fórmula de \textit{Jaro Distance}, com o objetivo de verificar a ortografia entre os modelos, obtêm-se o resultado demostrado na Tabela \ref{tab:escoreSintatico}, onde se tem o escore da pontuação obtido após a comparação entre os modelos de \textit{features}. 

A aplicação da fórmula de \textit{Jaro Distance} para comparar as strings “fone” e “ofen”, é apresentada abaixo. Para obter os dados correspondentes aplicam-se as \textit{strinsg}, |s1|, e |s2| na matriz para verificar sua similaridade, onde m = 4, S1 = 4,  S2 = 4 e t = 1. O resultado retornado através da aplicação da fórmula corresponde a 91\% de similaridade entre as strings |s1| e |s2|.

\begin{center}
	$$ d_{j} = \begin{cases} \frac{1}{3} \left( \frac{4}{4}+\frac{4}{|4|}+\frac{4-1}{|4|} \right),logo  \end{cases} $$
	d$_{j}$ = 0,916
\end{center}

\begin{table}[htbp]
  \centering
  \caption{Escore de pontuação de similaridade entre \textit{strings}.}
    \begin{tabular}{lcc}
    \toprule
    \textit{\textbf{Features}}             & \textbf{Fone}                          & \textbf{Ligação} \\
    \midrule
    Escore                                 & 0,91                                   & 1 \\
    \bottomrule
    \end{tabular}%
  \label{tab:escoreSintatico}%
  \fonte{Elaborado pelo autor.}
\end{table}%

A segunda etapa consiste na aplicação da fórmula da estratégia sintática para calcular a similaridade entre as strings que compõem o modelo de \textit{feature}, retornando o resultado da equivalência entre os modelos comparados: 

\begin{center}
\textbf{Equação 5.}\\
  $ESTSIN = \frac{\Sigma(dj)}{f}$~onde:
\end{center}

\textbf{dj} é o somatório total do escore obtido através da diferença entre as \textit{strings};
\textbf{f} é número de \textit{features} existente. \\

 O grau máximo de similaridade obtido na estratégia sintática é 1, para os casos em que o conjuntos de \textit{string} de  dj é igual a número de \textit{features} de f, isto é, todos as \textit{features} sejam iguais.  
Após a aplicar a fórmula para calcular a estratégia sintática, ESTSIN, verifica-se uma parte do conjunto das estratégias para formular o cálculo de efetividade estratégica, CEE, ou seja, o escore de pontuação exibido a partir dos dados contidos na Tabela \ref{tab:escoreSintatico} entre os modelos de \textit{features}, MF$_{R}$ e MF$_{C}$, no exemplo acima ESTSIN é representado por dj igual 1,91,  dividido pela número de \textit{features} existentes, neste exemplo, f é igual a 2,  retornando como resultado o grau de equivalência entre as \textit{features}, ESTSIN  é igual 0,95.

\begin{center}
  $\Sigma$ = 0,91+1  
  $ESTSIN = \frac{\Sigma(1,91)}{2}$~logo,
  ESTSIN = 0,95
\end{center}

\subsection{Cálculo de Efetividade Estratégica}\label{TiSubCEE}

Após a definição parcial das fórmulas para comparação entre os modelos de \textit{features} por meio dos cálculos de estratégia estrutural representada por ESTEST, estratégia semântica representada por ESTSEM, e a estratégia sintática representada por ESTSIN, define-se o calculo de efetividade das estratégias, obtendo-se o resultado de similaridade entre os modelos de \textit{features} comparados.
O grau de equivalência é representado por CEE, sendo que sua variação encontra-se entre os valores de 0 a 1, o seu calculo é executado através de uma média aritmética simples, sendo sua pontuação calculada por meio de uma divisão das estratégias somadas por sua quantidade, conforme segue sua fórmula: 

\begin{center}
\textbf{Equação 6.}\\
  $CEE = \frac{ste\Sigma(ESTEST+ESTSEM+ESTSIN)}{n}$~$\rightarrow$[0..1]
\end{center}
\textbf{ste} é o somatório das estratégias de comparação;\\
\textbf{n} é o somatórios do número de estratégias aplicados. \\

A aplicação para formular o cálculo de efetividade estratégica, CEE, após a comparação modelos, MF$_{R}$, F$_{C}$, conforme os exemplos aplicados a este trabalho apresentam o resultado do somatório das estratégias de acordo com a fórmula CEE, sendo o seu resultado igual a 0.68, isto é, a equivalência entre os modelos é de 68 por cento.

\begin{center}
  $CEE = \frac{2,04}{3}$~$\rightarrow$0,68
\end{center}

Sendo assim, para realizar a integração entre os modelos de \textit{features} é necessário analisar o cálculo de efetividade estratégica, o qual define o parâmetro de corte, isto é, o limiar que define qual estratégia de integração deverá seguir: automática ou semiautomática. 

\section{Estratégias de Integração} \label{TiEstratégiasdeIntegração}

Nesta seção são definidas as estratégias para integração entre os modelos de \textit{features}, o qual segmenta o entendimento dos processos estabelecidos na Seção \ref{TiProcesso}, bem como as diretrizes que determinam a equivalência entre a comparação dos modelos de \textit{features} fundamentados na Seção \ref{TiEstrategiasComparacao}.
As estratégias de integração consistem em:
\begin{itemize}
   \item Especificar dois modelos de \textit{features} como entrada;
   \item Determinar as estratégias de comparação a ser aplicada, estrutural, semântica e sintática;
   \item Definir as estratégias de integração a ser seguida, automática ou semiautomática;
   \item Obter o resultado da comparação entre os modelos de \textit{features} integrados, ou seja, o modelo pretendido como saída.
 \end{itemize}
 
 A partir da entrada dos modelos \textit{features} a serem comparados, é possível obter uma evolução do modelo de \textit{features}, isto é, o modelo pretendido, MF$_{AB}$, independente da estratégia de comparação a ser seguida sempre originará um modelo como saída. Sendo assim, a comparação entre dois modelos de \textit{features} pode originar com a saída múltiplas estratégias operacionais, as quais serão acopladas conforme sua especificidade \cite{farias2015}.
Retornado o cálculo de efetividade estratégica, CEE, obtém-se o parâmetro de corte, ou seja, limiar que define qual estratégia deverá ser seguida. A similaridade entre os modelos de \textit{features} é indicada pelo o grau de equivalência. Quanto mais próximo de 1 menor será sua diferença, sendo assim define-se qual estratégia de integração deverá seguir. Os índices estabelecidos para o corte de similaridade variam entre 0.7 e 0.8, conforme \citetexto{farias2015, farias2015b}, este trabalho aplica um limiar igual ou superior 0.95, tomando por verdade melhorar a precisão da técnica de integração.

\begin{itemize}
   \item \textbf{CEE = 1.} Elementos que compõem os diagramas de \textit{features} 100\% similares aplica-se a estratégia automática, sem a interferência humana.
   \item \textbf{CEE = 0.} Elementos que compõem os diagramas de \textit{features} que não possuem nenhuma similaridade aplica-se a estratégia automática, sem a interferência humana.
   \item \textbf{CEE $\geq$ 0.95 e < 1.} Os elementos que compõem os diagramas de \textit{features} que possuírem um alto índice de similaridade aplica-se a estratégia automática, sem a interferência humana.
   \item \textbf{CEE > 0 e < 0.95.} Os elementos que compõem os diagramas de \textit{features} que possuírem alguma similaridade aplica-se a estratégia semiautomática, o qual requer intervenção humana ou automática, sem a interferência humana.
 \end{itemize}
 
\subsection{Múltipla Estratégia de Integração}\label{TiSubMuliEstrategia}

A definição das estratégias sejam estas automáticas ou semiautomáticas serão vinculadas as técnicas de múltipla estratégia de integração operacional. A entrada ocorre por meio de dois modelos de \textit{features }definidos como: o modelo referência, MF$_{R}$, e modelo comparado, MF$_{C}$, após realizar a comparação entre os modelos e execução das técnicas de integração obtêm-se como resultado uma saída, o modelo pretendido, MF$_{AB}$. Procurando obter o melhor precisão possível entre os modelos de \textit{features} integrados, produzindo um resultado desejado, classifica-se as seguintes categorias de estratégias operacionais:

\begin{description}
   \item[ Estratégia Comum:] Ao analisar a comparação entre os dois modelos de \textit{features} obtêm-se como resultado a igualdade entre os modelos, isto é, tudo que contém em um modelo também está em outro, implicando em: 
   
\begin{center}
	MF$_{R}$ = MF$_{C}$ $\Leftrightarrow$ MF$_{R}$ $\subset$ MF$_{C}$ $\wedge$ MF$_{C}$ $\subset$ MF$_{R}$
\end{center}

   \item[ Estratégia Adicional:] Ao analisar a comparação entre os dois modelos de \textit{features}, tudo que contém em um modelo passa a estar contido em outro, implicando em:
   
\begin{center}
   MF$_{R}$ $\cup$ MF$_{C}$ = \{ $\chi$ $\mid$ $\chi$ $\in$ MF$_{R}$ $\vee$ $\chi$ $\in$ MF$_{R}$ \}
\end{center}    
      
   \item[ Estratégia Formal:] Ao analisar a comparação entre os dois modelos de \textit{features} verificam-se as \textit{features} comuns que estão contidas em ambos os modelos, implicando em: 

\begin{center}
   MF$_{R}$ $\cap$ MF$_{C}$ = \{ $\chi$ $\mid$ $\chi$ $\in$ MF$_{R}$ $\wedge$ $\chi$ $\in$ MF$_{R}$ \}
\end{center}    
   
   \item[ Estratégia Parcial:] Ao analisar a comparação entre os dois modelos de \textit{features} verifica-se a diferença entre os modelos, isto é, retorna a diferença que contém o modelo referência, implicando em: 

\begin{center}
   MF$_{R}$ $-$ MF$_{C}$ = \{ $\chi$ $\mid$ $\chi$ $\in$ MF$_{R}$ $\wedge$ $\chi$ $\notin$ MF$_{R}$ \}
\end{center}    
   
   \item[ Estratégia Complementar:] Ao analisar a comparação entre os dois modelos de \textit{features} verifica-se a diferença entre ambos os modelos, implicando em:

\begin{center}
   MF$_{R}$ $\bigtriangleup$ MF$_{C}$ = \{ $\chi$ $\mid$ $\chi$ $\in$ MF$_{R}$ $-$ MF$_{C}$ $\vee$ $\chi$ $\in$ MF$_{C}$ $-$ MF$_{R}$ \}
\end{center} 
    
   \item[ Estratégia Nula:] Ao analisar a comparação entre os dois modelos de \textit{features} estes não possuem \textit{fetuares} comuns, isto é são disjuntos, implicando em:

\begin{center}
 MF$_{R}$ $\cap$ MF$_{C}$ = 0
\end{center}   
    
 \end{description}

\subsection{Estratégia Semiautomática}\label{TiSubEstrategiaSemiautomatica}

O modelo semiautomático será introduzido para os desenvolvedores e analistas, após retornar o cálculo de efetividade estratégia, CEE, ou seja, se o grau de equivalência for menor que 0.95.  A estratégia semiautomática requer a intervenção humana para garantir que seja realizada a integração entre os modelos de \textit{features} através da aplicação das estratégias descritas nas Seções anteriores, tendo como fundamento exibir os conflitos existentes no processo de comparação, para que os desenvolvedores ou analistas tendam a decidir qual elemento conflitante será integrado em conformidade com as funcionalidades ou requisitos inerentes ao projeto. 

\begin{figure}[ht]
	\caption{Exemplo de integração técnicas - semiautomático e automático}
	\label{fig:exemploTecSemiAuto}
	\centering%
	\begin{minipage}{.8\textwidth}
		\includegraphics[width=\textwidth]{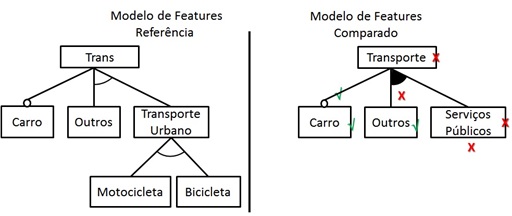}
		\fonte{Elaborado pelo autor.}
	\end{minipage}
\end{figure}

A aplicação da técnica envolve na varredura de ambos os modelos de \textit{features}, conforme sua entrada, ou seja, modelo referência, MF$_{R}$, e modelo comparado, MF$_{C}$, visando comparar as métricas definidas entre os modelos de \textit{features} \cite{kleinner2008},sintaxe, semântica, notações, número de elementos, sua ordem e seu nível hierárquico. Após realizar o processo de comparação entre os dois modelos de \textit{features}, são aplicadas as estratégias; (1) estrutural, (2) semântica e (3) sintática, exibindo aos usuários os conflitos existentes para sua alteração. Por fim, são executadas as técnicas de múltipla estratégia de integração operacional para retornar o modelo integrado,ou seja, a saída que deseja, MF$_{AB}$. A Figura \ref{fig:exemploTecSemiAuto} ilustra dois modelos de \textit{features} que servirão de exemplo para aplicação da técnica de múltipla estratégica de integração operacional, aplicando a técnica semiautomática e automática.

Nesse exemplo são visualizados os conflitos existentes entre o modelo referência, MF$_{R}$, e o modelo comparado, MF$_{C}$ a \textit{feature} $\lbrace$Trans$\rbrace$ exibida no MF$_{R}$, apresenta diferenças (1) sintática quando comparada ao MF$_{C}$, a \textit{feature} $\lbrace$Transporte, Serviços Públicos$\rbrace$, exibem com um “x” as \textit{features} que não são similares e com “v” a sintaxe similar $\lbrace$Carro, Outros$\rbrace$; (2) Semântica quando comparada ao MF$_{C}$, as notações que compõem o diagrama \textit{feature} $\lbrace$ou Inclusivo, Nulo$\rbrace$ e por fim (3) estrutural, quando avalia a ordem das \textit{features}, bem como o número de elementos que compõem o diagrama quando comparada ao MF$_{C}$, a \textit{feature} $\lbrace$Motocicleta, Bicicleta$\rbrace$, no que refere ao número de elementos contém diferenças. Após a comparação é apresentado o grau de equivalência para compor o cálculo de efetividade estratégico.

O escore para identificar a equivalência entre os modelos é calculado aplicando as estratégias: sintática, semântica e estrutural entre os modelos de \textit{features}, o qual possibilita calcular a diferença encontrada entre os dois modelos, obtendo como resultado o percentual de diferença entre os modelos comparados.

\begin{center}
  $CEE = \frac{0,3+0,6+0,4}{3}$
\end{center}

Por conseguinte, o grau de equivalência é 0,43. Isto é, os modelos de \textit{features} apresentam baixa similaridade. Os desenvolvedores e analistas podem optar, neste caso, em aplicar uma das estratégias, pois o limiar é menor que 0.95. A opção indicada neste trabalho para o limiar que varia entre 0.25 a 0.95 é a opção semiautomática, onde a interação com o usuário acontece para que se possa tomar a melhor decisão de acordo com as funcionalidades ou requisitos estabelecidos no projeto de \textit{software}.  O modelo semiautomático permitirá ao usuário que o mesmo altere os conflitos localizados no modelo, comparado ou não. Permanecendo o modelo o mais próximo de desejado, por exemplo, entre o conflito $\lbrace$Tran$\rbrace$ e $\lbrace$Transporte$\rbrace$ a técnica questiona-se junto ao usuário qual das duas opções é a considerada ideal, percorrendo todos os elementos que constituem o diagrama até não encontrar mais conflitos. Assim, o próximo passo é a integração modelo, gerando o modelo pretendido, exibindo um novo grau de equivalência entre o modelo de referência, MF$_{R}$, e o modelo pretendido, MF$_{AB}$.

\subsection{Estratégia Automática}\label{TiSubEstrategiaAutomática}

O cálculo de efetividade estratégico, retornado como resultado 1, implica que os dois modelos de \textit{features} possuem igualdade, aplicando a técnica de múltipla estratégia de integração operacional que melhor ajustar, neste caso a estratégia comum retornando como saída o mesmo modelo, MF$_{R}$. O cálculo de efetividade estratégico, retornado como resultado 0, implicando que não há qualquer similaridade entre os modelos de \textit{features}, neste caso, aplicada a técnicas de múltipla estratégia adicional, retorna como saída o modelo integrado, MF$_{I}$.

\begin{figure}[ht]
	\caption{Exemplo de integração com o uso da técnicas automático}
	\label{fig:tecnica2}
	\centering%
	\begin{minipage}{.8\textwidth}
		\includegraphics[width=\textwidth]{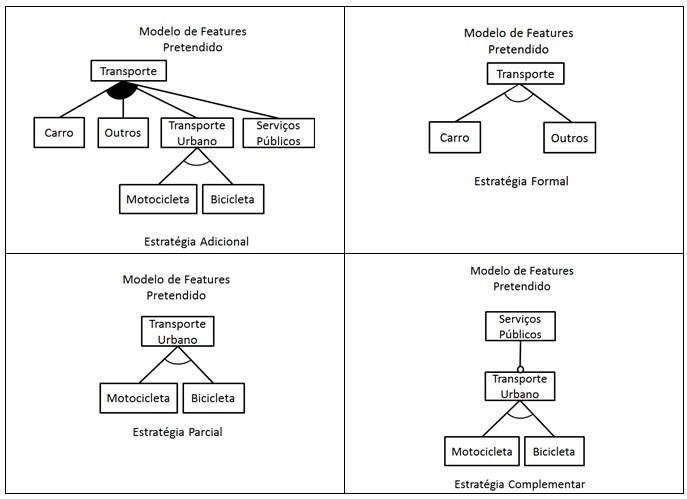}
		\fonte{Elaborado pelo autor.}
	\end{minipage}
\end{figure}

A Figura \ref{fig:tecnica2} exibe as estratégias geradas quando selecionada pelos analistas ou desenvolvedores a estratégia automática, cuja técnica gera quatro modelos, onde o mesmo poderá ter como parâmetro também quatro modelos, verificando qual o melhor deles para se adaptar às suas necessidades para então executar as alterações. 

\chapter{Aspectos de Implementação} \label{AspectosDeImplementação}
Com a finalidade de colocar em prática as técnicas  de integração de modelos apresentadas no Capítulo 4, assim como atender os objetivos descritos na Seção \ref{I_Objetivos} deste trabalho, este Capítulo apresenta a FMIT (\textit{Feature Model Integration Tool}), uma ferramenta para integração de modelos de \textit{features}. Desenvolvedores se beneficiam com o uso da FMIT ao reduzir o esforço de integração, e podem diminuir a propensão a erros, os quais podem ser convertidos em inconsistências nos modelos compostos. 

Este Capítulo é organizado da seguinte forma. A Seção \ref{AspVisaoGeral} apresenta uma visão geral do protótipo, FMIT, bem como uma descrição de sua aplicação. A Seção \ref{AspArquitetura} exibe a arquitetura do protótipo juntamente com uma descrição do mesmo. A Seção \ref{AspInterface} ilustra sua interface e suas funcionalidades com o objetivo de demostrar a interação do protótipo com os analistas e desenvolvedores. A Seção \ref{AspAlgoritmos} apresenta alguns dos principais algoritmos de implementação (interpretação do modelo, identificação de equivalência e processo de integração) do protótipo, FMIT. Por fim a Seção \ref{AspTecnologias} apresenta uma descrição das tecnologias empregadas para o desenvolvimento do protótipo.

\section{FMIT - Visão Geral} \label{AspVisaoGeral}

A integração manual de modelos de \textit{features} trata-se de uma tarefa árdua e propensa a erros \cite{becan2015, farias2015, farias2015b}. A modelagem de \textit{features} é uma atividade complexa quando introduzida em grandes LPS \cite{teixeira2013, apel2009}, em consequência da vasta variabilidade que emerge da composição de modelos. A execução de uma integração segura é fundamental na qualidade dos produtos gerados. Deve-se levar em consideração, a quantidade de \textit{features}, as notações, ou seja,  seus relacionamentos, suas dependências e observar a ordem em que os elementos são configurados \cite{becan2015, teixeira2013}. Além disso, o grande número de técnicas, por exemplo, FODA \cite{kang1990}, RSEB \cite{griss1998}, FORM \cite{kang1998}, CMBF \cite{czarnecki2004}, TVL \cite{classen2011}, CVL \cite{reinhartz2014},torna-o um dificultador, o que eleva o risco de más integrações por  parte das equipes de desenvolvimento.

\begin{figure}[!ht]
	\caption{Fluxo de atividades para execução do protótipo FMIT.}
	\label{fig:fluxo}
	\centering%
	\begin{minipage}{.8\textwidth}
		\includegraphics[width=\textwidth]{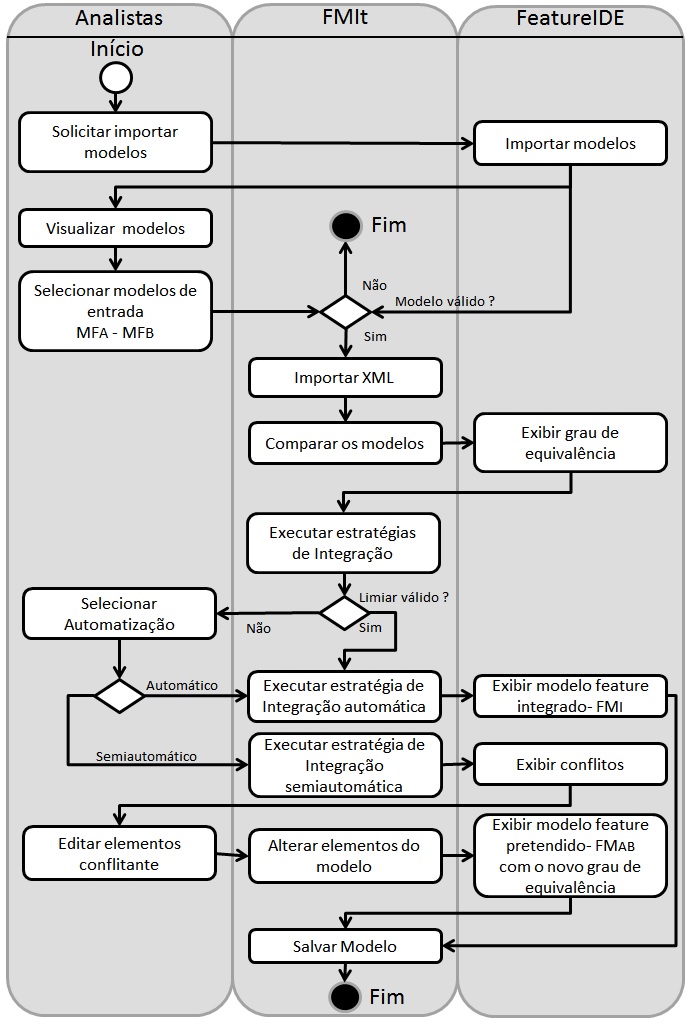}
		\fonte{Elaborado pelo autor.}
	\end{minipage}
\end{figure}

Sendo assim, a utilização de uma ferramenta que automatize parcial ou completamente a integração pode trazer melhorias. Neste contexto, a ferramenta FMIT foi desenvolvida com o objetivo de introduzir melhorias no processo de automatização de integração de modelos de \textit{features}, assim como para auxiliar as equipes de desenvolvimento na atividade de integração. As atividades para a integração de modelos, proposto pelo protótipo, FMIT, compreende uma sequência de passos, conforme ilustra a Figura \ref{fig:fluxo}.

Os passos descritos abaixo visam conduzir as atividades inerentes à equipe de desenvolvimento, especificamente aos analistas e desenvolvedores, visto a interação entre o \textit{framework}, FeatureIDE, que atua como uma ferramenta intermediária, isto é,  facilitando a visualização dos modelos \textit{features}, sua configuração, bem como a criação ou a edição de modelos. 

Finalmente, tem-se o protótipo, FMIT, que age como um condutor no processo de integração de modelos, procurando identificar conflitos entre os modelos comparados, sendo capaz de exibir a similaridade destes, pois, em suma, sua diretriz é aperfeiçoar o processo de composição.  Resumidamente, os passos incluem importação, seleção, leitura, validação e visualização dos modelos de entrada, a interação com o protótipo, FMIT, tendo em vista aceitar ou rejeitar conflitos identificados e, por fim, a visualização do modelo integrado. Os passos apresentados na Figura \ref{fig:fluxo} serão descritos a seguir:

\begin{itemize}
   \item \textbf{Solicitar importação de modelos.} O analista importa os modelos de entrada MF$_{A}$ e MF$_{B}$, ou seja, importa os arquivos no formato XML para o \textit{framework} FetureIDE.
   \item \textbf{Importar modelos.} O \textit{framework} analisa os arquivos para importação, executando uma validação interna; os arquivos que se encontram conforme os padrões estabelecidos são importados, caso contrário o processo é finalizado.
   \item \textbf{Importar XML.} O protótipo, FMIT, executa uma rotina de leitura e importação dos arquivos XML para que esses sejam validados e estruturados para uso do mesmo. 
   \item \textbf{Visualizar modelos.} O analista pode visualizar os modelos importados no \textit{framework} graficamente ou textualmente, bem como fazer uso de todas as funcionalidades que a mesma permita, por exemplo, alterar alguma \textit{feature}.
   \item \textbf{Selecionar modelos de entrada.} O analista executa o método principal do protótipo, onde o mesmo informa o modelo de origem, MF$_{A}$, isto é, modelo referência, bem como o modelo de destino, MF$_{B}$, isto é, o modelo comparado. O sistema executa um método de verificação e análise dos arquivos XML. Se os arquivos de importação não forem localizados o mesmo é finalizado.
   \item \textbf{Comparar modelos.} Essa funcionalidade consiste em uma das principais atividades do processo de comparação descrita no Capítulo 4.2. Este método processa uma série de informações referentes aos elementos que constituem o modelo de \textit{features}, formando a matriz similaridade. É considerado método base para processar a integração de modelos
   \item \textbf{Exibir grau de equivalência.} O \textit{framework} exibe uma serie de informações textuais em seu console para o analista ou desenvolvedor, advindas do protótipo.  As informações são inerentes à contextualização da comparação dos elementos dos diagramas de \textit{features}, por exemplo, o grau similaridade sintática e semântica. Conforme descrito na Seção \ref{TiProcesso}, o sistema apresenta o cálculo de efetividade estratégia, CEE, em síntese é o grau de equivalência entre os dois modelos comparados.  
   \item \textbf{Executar estratégia de integração.}  Este método retém as informações advindas da equivalência dos modelos, que consolidam o processo de integração descrito na Seção \ref{TiEstratégiasdeIntegração} que irá delinear a estratégia a ser seguida de acordo com o limiar estabelecido. Enfim, caso o cálculo de efetividade estratégica, CEE, seja igual a 0 ou 1, executa-se a estratégia automática, sendo menor que 0.95 aplica-se a estratégia semiautomática.
   \item \textbf{Selecionar Automatização.} O analista ou desenvolvedor define que tipo de automatização pretende escolher, automático onde o sistema sugere quatro saídas de composição ou semiautomático que necessita de sua intervenção para solucionar os conflitos existentes.
   \item \textbf{Executar estratégia de integração semiautomática.} Este método executa as estratégias de composição descritas na Seção \ref{TiEstratégiasdeIntegração}, conforme as regras pré-estabelecidas, ou seja, necessita de intervenção humana. 
   \item \textbf{Executar estratégia de integração automática.} Este método executa as estratégias de composição descritas na subseção Seção \ref{TiEstratégiasdeIntegração}, conforme as regras pré-estabelecidas, ou seja, sem a intervenção humana.  
   \item \textbf{Exibir modelos de features integrado.} O \textit{framework} exibe no console o modelo \textit{features} integrado, MF$_{I}$, após a conclusão da composição que compreende a integração dos modelos MF$_{A}$ e MF$_{B}$.
   \item \textbf{Exibir conflitos.} O \textit{framework} exibe no console os modelos conflitantes, após a comparação dos modelos de entrada, MF$_{A}$ e MF$_{B}$.
   \item \textbf{Editar elementos conflitantes.} O protótipo identifica os elementos conflitantes, sendo necessária a intervenção do analista no processo decisório em aceitar ou rejeitar as mudanças dos elementos em conformidade com os requisitos de projeto.
   \item \textbf{Alterar elementos do modelo.} O protótipo alterar os elementos do modelo de \textit{features} comparado, isto é MF$_{B}$, conforme a decisão do analista, formando um novo modelo.
   \item \textbf{Exibir modelo de \textit{features} pretendido.} O \textit{framework} exibe no console o modelo feature pretendido, MF$_{AB}$, após a conclusão da composição que compreende a integração dos modelos MF$_{A}$ e MF$_{B}$, exibindo o novo grau de equivalência.
   \item \textbf{Salvar modelo.} O sistema salva todas as informações exibidas no console em um arquivo textual, modelos de entrada, MFA e MFB, grau de equivalência dos modelos comparados, processos de edição e alterações de elementos do modelo, bem como apresenta os elementos integrados, MF$_{I}$ ou MF$_{AB}$ oriundos da composição dos modelos de entrada. 
 \end{itemize}

\section{Arquitetura do Protótipo FMIT} \label{AspArquitetura}

Análise, comparação, integração, avaliação e persistência, desta forma produzindo uma técnica de integração específica, para a situação proposta. Através da manipulação dos modelos de entrada e aplicação das estratégias de comparação seguidos das técnicas de múltipla estratégia operacional pretende-se obter como saída o modelo mais perto do idealizado entre os analistas e desenvolvedores.

A integração entre os modelos de \textit{features} é definida através de um conjunto de funcionalidades que tem por finalidade produzir um novo modelo, isto é, o modelo de \textit{features} pretendido, MF$_{AB.}$ \cite{farias2013}. As atividades propostas para o desenvolvimento serão realizadas de forma sequencial, entretanto ocorrerá de forma independente onde cada atividade terá sua função isolada, ou seja, separadamente. Contudo, para a execução de uma atividade, um conjunto de informações deve ser indicado, por exemplo, na análise do modelo de \textit{features}. Na comparação devem-se informar quais são as \textit{features}, notações e estrutura equivalentes e qual estratégia integração deverá ser aplicada.

\subsection{Diagrama de \textit{Features} da FMIT}\label{AspSubDiagrama}

O suporte à implementação de funcionalidades da arquitetura FMIT, é proposto devido a várias razões e requisitos identificados em trabalhos anteriores \cite{farias2015, oliveira2009}. Utilizável, portanto, quando identificanda a necessidade de uma arquitetura para a integração de modelos de \textit{features} \cite{bischoff2016}, reutilizável para suportar e orientar o desenvolvimento de novas ferramentas de integração. É representativo o domínio de integração de modelos, uma vez que seu \textit{design }decompõe as principais preocupações com \textit{features} bem modularizados. Por último, permite avaliar os modelos gerados e persistir os resultados. Assim, a arquitetura proposta fornece um conjunto de características fundamentais, incluindo análise dos modelos de entrada, comparação dos modelos de entrada, integração do modelo de entrada equivalente, persistência do modelo de saída gerado e avaliação do modelo de saída.

\begin{figure}[!ht]
	\caption{Funcionalidades da ferramenta de integração.}
	\label{fig:funciopnalidade}
	\centering%
	\begin{minipage}{.8\textwidth}
		\includegraphics[width=\textwidth]{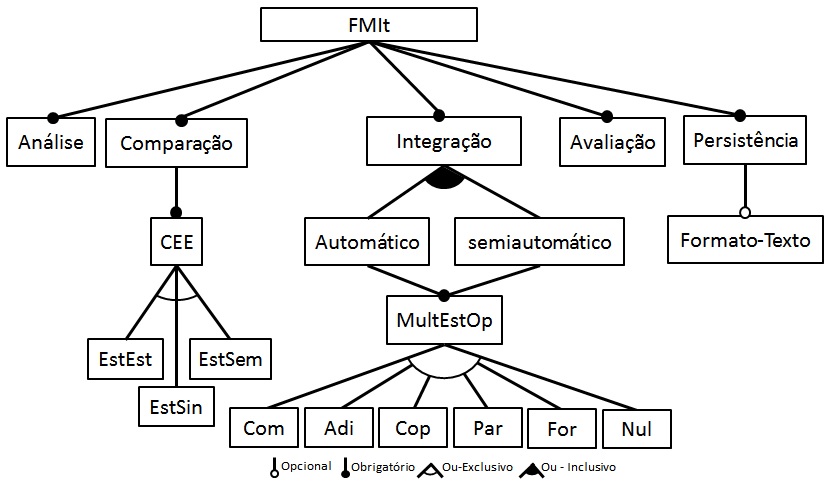}
		\fonte{Bischoff, 2016.}
	\end{minipage}
\end{figure} 

A Figura \ref{fig:funciopnalidade} mostra uma visão simplificada do modelo de \textit{features} da FMIT. Assim, para desenvolver ferramentas de integração, os desenvolvedores devem primeiro implementar as \textit{features } obrigatórias, incluindo análise, comparação, integração, avaliação e persistência. Além de identificar um conjunto de funcionalidades essenciais, as funcionalidades obrigatórias especificam suas dependências de forma fácil e compreensível. Uma preocupação presente em toda a arquitetura do FMIT foi assegurar que os recursos obrigatórios cumpram o processo de integração do modelo descrito na Figura \ref{fig:funciopnalidade}. Por exemplo, a \textit{feature} de análise implementa o primeiro passo e a \textit{feature}  de persistência fornece a funcionalidade necessária para manter o modelo integrado de saída gerado no final do processo de integração.

A \textit{feature }opcional é referente ao formato do arquivo, isto é, um arquivo texto, gerado após finalizar o modelo integrado ou desejado em sua saída. As principais \textit{features} são representadas pelas estratégias de Comparação que formaliza o Cálculo de Efetividade Estratégica, CEE, definindo a equivalência entre os modelos, descrita no Capítulo \ref{TécnicaDeIntegração} e as estratégias de integração, Automática e Semiautomática que segue umas das Múltiplas Estratégias de Operação para conduzir a composição entre os modelos.

\subsection{Diagrama de Casos de Uso}\label{AspSubDiagramas}

Esta Seção tem como objetivo descrever os requisitos do protótipo, definindo de que forma deverá funcionar. O diagrama de casos de uso tem o objetivo de auxiliar a comunicação, assim como, descrever os cenários que exibe as funcionalidades do sistema \cite{eriksson2003}.

\begin{figure}[!ht]
	\caption{Atividade realizada pelo usuário.}
	\label{fig:atividadeUser}
	\centering%
	\begin{minipage}{.7\textwidth}
		\includegraphics[width=\textwidth]{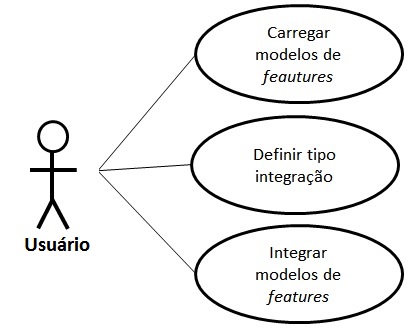}
		\fonte{Elaborado pelo Autor.}
	\end{minipage}
\end{figure}

Apresenta-se as funcionalidades básicas do protótipo, FMIT, tendo como referência a visão do usuário. Sendo assim, o protótipo pode ser dividido em três funcionalidades: (1) carregar modelos de \textit{features}, (2) definir tipo integração e (3) integrar os modelos de \textit{features}. A Figura \ref{fig:atividadeUser} ilustra estas funcionalidades.

\begin{figure}[!ht]
	\caption{Caso de uso carregar modelos de entrada.}
	\label{fig:atividadeCarregarModelo}
	\centering%
	\begin{minipage}{.7\textwidth}
		\includegraphics[width=\textwidth]{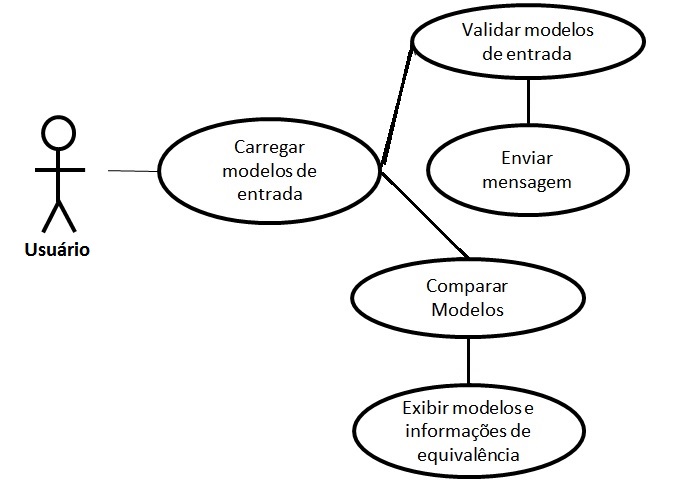}
		\fonte{Elaborado pelo Autor.}
	\end{minipage}
\end{figure}

A próxima interação consiste em (1) importar os modelos de entrada, (2) validar o formato dos modelos de entrada, (2.1) enviar mensagem caso os modelos sejam inválidos, (3) comparar os modelos de entrada e (4) exibir os modelos importados e informações sobre a equivalência entre os modelos. Na figura \ref{fig:atividadeCarregarModelo} é ilustrado o caso de uso para a atividade de configuração dos parâmetros de automatização.

\begin{figure}[!ht]
	\caption{Caso de uso definir tipo de integração.}
	\label{fig:atividadeDefiniIntegração}
	\centering%
	\begin{minipage}{.7\textwidth}
		\includegraphics[width=\textwidth]{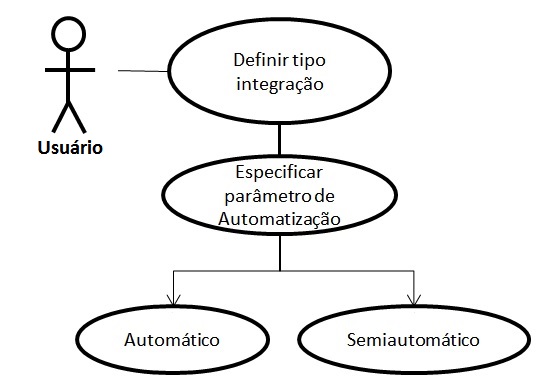}
		\fonte{Elaborado pelo Autor.}
	\end{minipage}
\end{figure}

Além disso, será necessário realizar a integração entre os modelos, definindo como a mesma será executada. O usuário poderá optar em executar uma análise mais profunda no caso a opção (1) semiautomática para produzir o modelo pretendido, ou seja, há uma intervenção do usuário em aceitar ou rejeitar um conflito, entretanto ao optar pelo modelo (2) automático, este não requer intervenção do usuário. Na Figura \ref{fig:atividadeDefiniIntegração} é ilustrado o caso de uso para definir o tipo de integração.

\begin{figure}[!ht]
	\caption{Caso de uso de integração do modelo de \textit{features}}
	\label{fig:atividadeIntegraçãoSalvar}
	\centering%
	\begin{minipage}{.7\textwidth}
		\includegraphics[width=\textwidth]{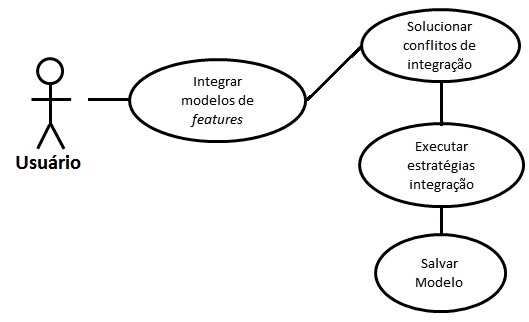}
		\fonte{Elaborado pelo Autor.}
	\end{minipage}
\end{figure}

Uma vez definido o parâmetro de integração, será possível integrar os modelos de \textit{features} de entrada. Para integrar será necessário (1) solucionar os conflitos de integração, (2) Executar as estratégias de Integração e finalmente (3) Salvar as informações do modelo gerado. Na figura \ref{fig:atividadeIntegraçãoSalvar} é ilustrado o caso de uso para a atividade de integração dos modelos de \textit{features} e persistir na sua saída. 

\subsection{Componentes Arquiteturais}\label{AspSubComponentes}

Os componentes são responsáveis pela implementação de cada atividade ilustrada na  Figura \ref{fig:funciopnalidade}. Os componentes permitem ser utilizados de forma isolada, ou seja, independentemente, de modo a vir modularizar os elementos do protótipo, provendo o reuso dos componentes que virão a ser produzidos, bem como permitem o desenvolvimento e manutenção pontual da ferramenta proposta. 

\begin{figure}[!ht]
	\caption{Arquitetura de componentes.}
	\label{fig:arqComponentes}
	\centering%
	\begin{minipage}{.8\textwidth}
		\includegraphics[width=\textwidth]{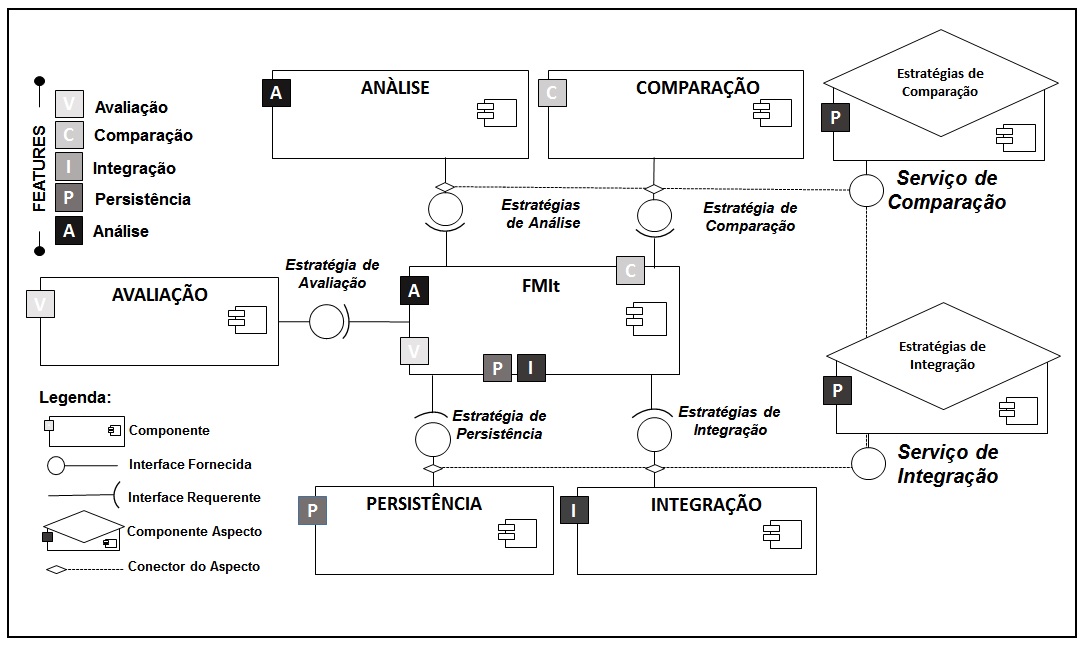}
		\fonte{ Adaptado de Farias (2015b).}
	\end{minipage}
\end{figure}

Cada componente será tratado como um módulo independente, que encapsula seu comportamento através de um conjunto de atividades internas. O papel desempenhado pelos componentes ocorre por meio dos módulos e sua interação entre os elementos demostrando o comportamento esperado.  Deste modo, a Figura \ref{fig:arqComponentes} ilustra a arquitetura de componentes a ser aplicada, o qual se concentra em apresentar os componentes como um grupo de elementos responsável em desempenhar cada atividade especificamente, ou seja, módulos independentes que desempenham um fator determinante na integração de modelos de \textit{features}. 

\begin{itemize}
   \item \textbf{FMIT.} O motor principal do protótipo se concentra no componente FMIT, e é responsável por administrar todos os processos do protótipo, iniciar a execução de todas as atividades, bem como exibir e avaliar o retorno dos métodos determinados.
   \item \textbf{Análise.} Este componente implica em autenticar os modelos entrada, avaliando se ambos os modelos de \textit{features} são do mesmo formato e verifica se o processo de comparação poderá ser aplicado sobre os modelos. O componente realiza a leitura dos modelos entrada, ou seja, o Modelo \textit{Features} Referência, MF$_{R}$  e Modelo \textit{Features} Comparado, MF$_{C}$, executa uma análise das especificações das notações da modelagem utilizada, especificamente a modelagem \textit{FODA- Feature-Oriented Domain Analysis }\cite{kang1990}, identificando os elementos que compõem o modelo, bem como seus relacionamentos, após agregando os dados no formato XML para o \textit{framework} FeatureIDE \cite{bischoff2016, farias2015b, thum2014}, responsável por projetar a visualização graficamente dos modelos. Este será o primeiro componente executado pelo motor do FMIT. 
   \item \textbf{Comparação.} Este componente consiste em analisar as estratégias de similaridade entre os modelos, verificando todos os elementos que compõem o modelo de \textit{features}, especificando individualmente uma checagem entre ambos os modelos, sobre cada categoria, para verificar o grau de equivalência. As estratégias aplicadas para comparar os modelos de \textit{features} são aplicadas em três níveis: (1) estratégia estrutural, (2) estratégia sintática e (3) estratégia semântica; são responsáveis por demostrar os conflitos surgidos durante a composição dos modelos para os desenvolvedores e analistas, e por fim, é retornado o Cálculo de Efetividade Estratégico (CEE), que determina o grau de equivalência entre ambos os modelos de \textit{features} determinando qual estratégia de integração melhor se adapta. Infere, também, no limite a ser aplicado, ou seja, no corte para considerar um modelo de \textit{features} equivalente ou não.
   \item \textbf{Integração.} O componente de integração é invocado após a avaliação das equivalências encontradas e atua na integração do  MF$_{I}$, ou MF$_{AB}$. Este componente implica em duas estratégias, (1) Automática e (2) Semiautomática, ambas segmentadas por seis técnicas de Múltipla Estratégia de Integração Operacional descritas na Seção \ref{TiEstratégiasdeIntegração}, sendo sua finalidade formar um novo modelo de \textit{features}, ou seja, o MF$_{I}$ como saída. A Estratégia Automática (1) ocorrerá de duas formas: a primeira, sendo ambos os modelos diagnosticados como sendo idênticos é aplicada à Estratégia de Integração Operacional Comum, a segunda etapa consiste em verificar se ambos as modelos serão totalmente diferentes, isto é, não existe nenhuma equivalência entre eles aplicando assim à Estratégia de Integração Operacional Adicional. Por fim, a Estratégia Semiautomática (2), será aplicada, quando o grau de equivalência entre os modelos de \textit{features} estiver abaixo do limiar, pré-estabelecido, conforme descrito na Seção \ref{TiEstratégiasdeIntegração}. Possibilita ao analista ou desenvolvedor escolher entre uma de duas opções, automático e semiautomático. Caso a escolha seja pelo modo automático o protótipo retorna quatro possíveis soluções integradas em sua saída, MF$_{I}$1... MF$_{I}$4. As estratégias, aplicadas são: Adicional, Formal, Parcial ou Complementar. O modelo semiautomático conecta o componente de comparação e avaliação para produzir o modelo pretendido, MF$_{AB}$.
   \item \textbf{Avaliação.} O componente de avaliação retorna visualmente os conflitos localizados no componente de comparação, exibindo os mesmos para tomada de decisão do analista ou desenvolvedor, na compreensão de retornar o modelo pretendido, MF$_{AB}$., com base na Múltipla Estratégia de Integração Operacional.
   \item \textbf{Persistência.} É a camada que armazena os modelos carregados pelo componente de análise e os modelos produzidos durante os processos de comparação, integração e avaliação. Os dados são armazenados em um arquivo texto no disco rígido. 
 \end{itemize}

\subsection{Arquitetura em Camadas}\label{AspSubArquitetura}

\begin{figure}[!ht]
	\caption{Arquitetura de camadas do protótipo.}
	\label{fig:arqCamadas}
	\centering%
	\begin{minipage}{.5\textwidth}
		\includegraphics[width=\textwidth]{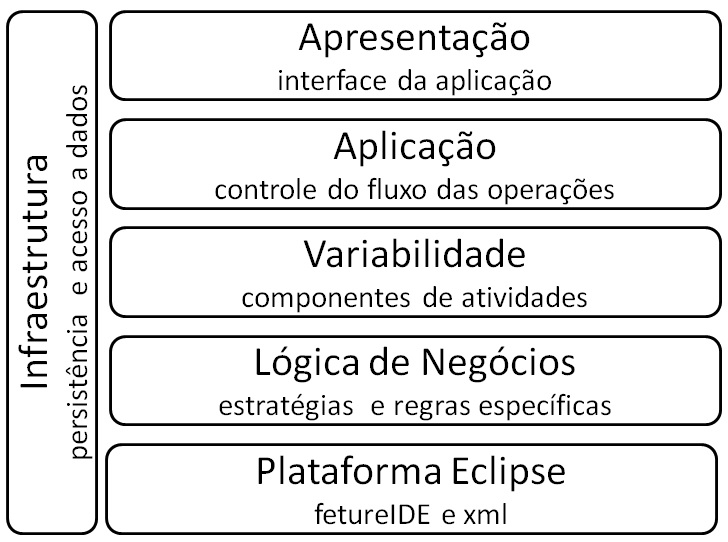}
		\fonte{ Adaptado de Farias (2015b).}
	\end{minipage}
\end{figure}

Para atender a funcionalidade especificada \cite{farias2015b, oliveira2007} através das estratégias para integração de modelos de \textit{features}, bem como as características do protótipo propõe-se um design projeto arquitetural modularizado em seis camadas: (1) Infraestrutura, (2) Apresentação, (3) Aplicação, (4) variabilidade, (5) Lógica de Negócios e  (6) Plataforma Eclipse conforme a Figura \ref{fig:arqCamadas}.

\begin{itemize}
\item \textbf{Camada de Apresentação.} Representa a camada mais próxima do usuário, ou seja, interface da aplicação. Interage diretamente com os analistas e desenvolvedores onde serão coletados os parâmetros de configuração, bem como as entradas necessárias para a execução das funcionalidades. Além disso, tem como responsabilidade retornar as informações da aplicação: conflitos encontrados na composição e propor um modelo de \textit{features} como saída, ou seja, o resultado final. 
\item \textbf{Camada da Aplicação.} Esta camada abrangerá fluxo de controle das operações, tendo como responsabilidade coordenar e controlar a execução das atividades. Análise, Comparação, Integração, Avaliação e Persistência, onde determinará as chamadas, bem como as repostas das atividades, demandadas em sua execução.
\item \textbf{Camada de Variabilidade.} O objetivo desta camada está em aplicar o comportamento das operações propostas (estratégias de análise, comparação e integração), isto é, executar as operações na camada de aplicação, obtendo o procedimento esperado da camada de negócios.
\item \textbf{Camada da Lógica de Negócio.}  A implementação da lógica de negócios será empregadas através dos padrões de projeto \cite{gamma1995}, evitando que um único objeto seja responsável por manter todas as estratégias de operação .
\item \textbf{Camada de Infraestrutura.} Sua principal responsabilidade será a de fornecer o controle das funcionalidades entre as camadas, bem como realizar o acesso aos dados, sua persistência e a manipulação das exceções surgidas no decorrer da aplicação.  
\end{itemize}

\section{Interface do Protótipo} \label{AspInterface}

O protótipo apresenta uma interface simples e intuitiva, tendo por finalidade verificar a aplicação da técnica proposta bem como torná-la mais prática possível para os analistas e desenvolvedores.  O protótipo, FMIT, faz uso de outras tecnologias para dar suporte às atividades necessárias descritas no processo de integração. O FMIT associa a implementação destas tecnologias,  por exemplo, o Eclipse e FetureIDE de tal forma que facilite o uso,  para usuários com pouca experiência em codificação. O protótipo lê e filtra as informações das \textit{tags} de arquivos escritos em XML e transforma-os em um modelo de \textit{features} abstrato, no qual os elementos do modelo de entrada podem ser manipulados.

O termo integração de modelos de \textit{features} pode ser definido como um conjunto de atividades que devem ser executadas com dois (ou mais) modelos de \textit{features} de entrada, ou seja, o Modelo de \textit{Features} Referencia, MF$_{R}$, e Modelo de \textit{Features} Comparado, MF$_{C}$.  O objetivo é a união destes modelos para a produção de um novo modelo, ou seja, o Modelo de \textit{Features} Pretendido, MF$_{AB}$. O principal desafio é resolver os conflitos que surgem na composição destes modelos. A Figura \ref{fig:toolFMIT} ilustra o protótipo desenvolvido. Cada parte da interface será descrita a seguir. 

\textbf{Definição do ferramental.} Para execução do protótipo se faz necessário três ferramentas: o (1) Eclipse, que é parte integrante do \textit{framework } utilizado, o (2) FeatureIDE para modelagem, e por fim, o protótipo(3)FMIT, para dar suporte a integração dos modelos.  
 
\textbf{Definição dos modelos.} Esta atividade consiste primeiramente na criação do modelo de \textit{features} ou na importação de um modelo já existente. Estes dois processos podem ser realizados no \textit{framework} FeatureIDE. 

\textbf{Execução do Protótipo}. Esta atividade será usada para construir o novo modelo. Exibindo os resultados do protótipo, FMIT, inerente à integração de dois modelos de \textit{features}, de acordo com as estratégias aplicadas, retornando como saída o modelo integrado ou pretendido. A abordagem automática ou semiautomática é estabelecida pelo limiar de acordo com as regras de negócio. A Figura \ref{fig:toolFMIT} mostra uma visão geral do protótipo, FMIT, juntamente com alguns dos componentes da estrutura destacados com letras, por exemplo, (A), (B), (C) e (D-d). O protótipo apresenta uma visão inicial para a integração de modelos específicos dentro da estrutura, e é discutida brevemente abaixo:

\begin{figure}[!ht]
	\caption{Protótipo \textit{feature model integration tool} - FMIT.}
	\label{fig:toolFMIT}
	\centering%
	\begin{minipage}{1\textwidth}
		\includegraphics[width=\textwidth]{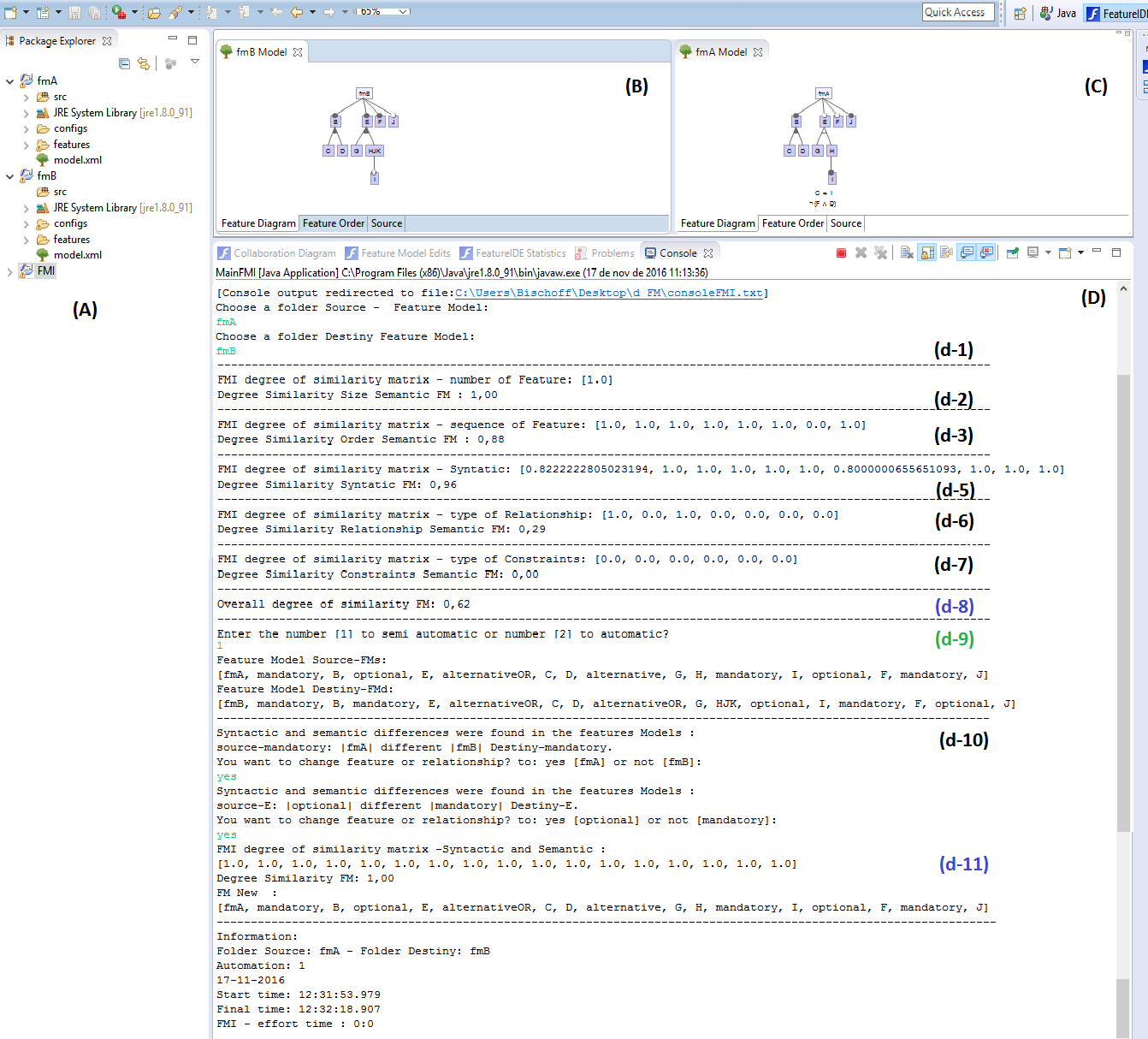}
		\fonte{Elaborado pelo Autor.}
	\end{minipage}
\end{figure}

\begin{itemize}
\item \textbf{Package Explorer (A).}  Para cada novo projeto de modelagem, é necessário criar ou importar arquivos que são usados durante o processo de modelagem. O explorador de pacotes tem como funcionalidade central permitir a organização dos projetos dos modelos de \textit{features}, bem como acomodar o protótipo de integração.
\item \textbf{Visualização (B-C).} Os modelos criados, necessariamente devem ser visualizados com o objetivo de atender dois requisitos básicos ao usar os modelos: compreensão e comunicação.  Diante desta necessidade, a visualização dos modelos permite aos analistas ou desenvolvedores analisar e editar o modelo interativamente, entre o protótipo e \textit{framework}. Os modelos de \textit{features }são acomodados em conjunto com o protótipo, FMIT,  para ajudar a explorar os conflitos quando existir. Nesse exemplo temos dois modelos de \textit{features}, o fmA e fmB a serem integrados.
\item\textbf{ Visualização do Console (D).} Uma vez que os modelos foram criados ou importados, uma visão geral da distribuição dos elementos das \textit{features} presentes é exibida e pode ser executada através do console, constituindo na usabilidade do protótipo. 
\begin{description}
   \item[d-1, persistência:] Consiste na persistência dos dados em um arquivo no formato de texto salvo no disco rígido, bem como define a entrada dos modelos, isto é, a importação dos arquivos de origem e destino no formato XML, no exemplo fmA e fmB. Após o protótipo executa uma validação dos arquivos, basicamente para verificar se não há uma inconstância em seu formato, sendo o passo seguinte a comparação dos modelos mencionados e exibe os resultados obtidos. 

   \item[d-2 a d-8, resultados:] Exibe os resultados oriundos da comparação entre os modelos fmA e fmB, individualmente, em um vetor que contém os valores  da similaridade, que variam entre 0 a 1. Por fim, é exibido o seu grau de equivalência das estruturas analisadas.  Este cenário atribui-se como um filtro, em virtude de diagnosticar e avaliar diversos aspectos na contribuição de uma integração de boa qualidade tais como, número de \textit{features}, estrutura sintática e semântica entre outros. Pode-se verificar (d-8) que a similaridade geral entre os modelos comparados no exemplo é de 0.62, ou seja, o modelo destino, fmB é 62\% equivalente ao modelo origem, fmA. Neste caso, o limiar foi inferior ao esperado. Há muitos conflitos que surgiram durante a fase de comparação.
   
      \item[d-9, automatização:] Esta etapa consiste em auxiliar os analistas e desenvolvedores na condução de integração segura dos modelos. O protótipo disponibiliza duas opções de integração para escolher, a automática que atua sem a intervenção do analista ou desenvolvedor, prospectando quatro modelos de saída conforme as estratégias de integração (união, interseção, diferença e complemento) para uma futura edição, ou a semiautomática, que atua com a intervenção humana, recomendada neste caso.
   
      \item[d-10,] Após selecionar a opção semiautomática os conflitos identificados são exibidos no console para sua escolha, incumbindo ao analista ou desenvolvedor tomar a decisão que melhor direcionar os requisitos estabelecidos em seu projeto.
   
      \item[d-11,] Finalmente o \textit{console} exibe a integração do novo modelo de \textit{features}, bem como o seu grau de equivalência. Após a composição dos modelos neste exemplo é 100\% similar, indicando que posteriormente à sua edição os conflitos existentes foram solucionados de acordo com os requisitos estabelecidos no projeto.

 \end{description}
\end{itemize}

\section{Algoritmos do Protótipo} \label{AspAlgoritmos}

Esta Seção apresenta os algoritmos elaborados para implementar as funcionalidades de (1) carregar modelos, (2) comparar modelos, (3)definir o tipo de integração, e (4)integrar os modelos. Tais atividades foram anteriormente ilustradas na Figura \ref{fig:funciopnalidade}. Os algoritmos são apresentados a seguir.

 O framework FeatureIDE exibe  o modelo de \textit{features} de duas formas, graficamente e textualmente, o formato adotado para representar os diagramas textualmente segue o padrão XML. Dessa forma, o seu conteúdo e sua estrutura são constituídos por instruções de marcações, ou seja, \textit{tags}. 
 
 Para realizar a leitura dos modelos de \textit{features }de entrada no protótipo, FMIT, o componente de análise executa o método de conversão das instruções de marcação para capturar e armazenar em memória, todas as propriedades e atributos  dos elementos contidos no arquivo em análise. Cada tipo de notação (opcional, alternativo, obrigatório, etc) possui sua própria características, que são representados por atributos e propriedades nas \textit{tags} XML, de modo que o interpretador precisa reconhecer cada uma delas para identificar corretamente as informações contidas na \textit{tag}.
 
 \begin{figure}[!ht]
	\caption{Protótipo \textit{tags} XML de modelos de \textit{features}}
	\label{fig:tags}
	\centering%
	\begin{minipage}{.7\textwidth}
		\includegraphics[width=\textwidth]{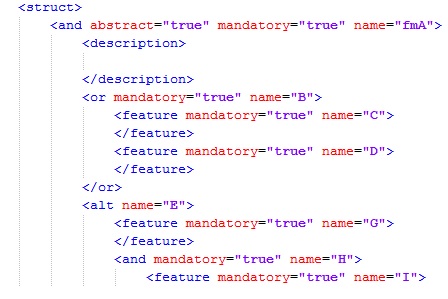}
		\fonte{Elaborado pelo Autor.}
	\end{minipage}
\end{figure}

A Figura \ref{fig:tags} exibe um exemplo em que a \textit{tag} struct apresenta as propriedades \textit{abstract}, \textit{mandatory}, e \textit{name} e os atributos \textit{and}, \textit{or},\textit{ alt}, \textit{description} e \textit{feature}. O modelo segue a representação proposta por \cite{kang1990},\textit{Feature-Oriented Domain Analysis (FODA)}.

A Tabela \ref{tab:alg1} exibe o Algoritmo 1, que executa o método de exploração do conteúdo do modelo de \textit{features} de entrada , que realiza um laço de repetição no arquivo (modelo de  \textit{feature}) para capturar todas as tags que representam os elementos e atributos. Assim que uma \textit{tag} (elemento) é capturada, a mesma é identificada e seus atributos são armazenados em memória. Como suporte ao ferramental aplicou-se o uso da biblioteca  JDom \footnote{http://www.jdom.org/downloads/docs.html}, que é uma API para facilitar a leitura, criação e  atualização de documentos.

\begin{table}[!htbp]
\scalefont{0.9}
  \centering
  \caption{Algoritmo 1.}
    \begin{tabular}{l}
    \toprule
    Algoritmo 1:  Leitura e importação de modelo  \\
    \midrule
    Entrada: Modelo de \textit{Features} \\
    Saída: Elementos do modelo carregado em memória \\
    1:   Importa biblioteca jdom2.org; \\
    2:   arquivoXML  $\leftarrow$   modeloFeature.xml;   \\
    3:   Enquanto  arquivoXML Faça // realiza a leitura até o fim do arquivo \\
    4:      tipoElemento $\leftarrow$   tagXML ;  \\
    5:      tipo $\leftarrow$ tipoElemento.identificaElemento(); \\
    6:      Para i = 0 até tipo.qtdAtributos() Faça \\
    7:         objeto$\leftarrow$  tipo.Elemento.analisaXML(tipo); \\
    8:         persiste (objeto); \\
    9:      Fim para \\
    10: Fim enquanto \\
    \bottomrule
    \end{tabular}%
  \label{tab:alg1}%
  \fonte{Elaborado pelo Autor.}
\end{table}%

A identificação de equivalência entre os elementos dos modelos de \textit{features} entrada (MF$_{R}$ e MF$_{C}$) e a respectiva composição desses elementos para a produção de um Modelo de \textit{Features} Integrado, MF$_{I}$, é realizada pelos algoritmos abaixo. Os algoritmos são responsáveis pela implementação das técnicas de comparação e integração descritos no Capitulo \ref{TécnicaDeIntegração}. A Tabela \ref{tab:alg2} apresenta o Algoritmo 2 que, executa um laço entre elementos comparados (MF$_{R}$ e MF$_{C}$), verifica a sequência, ou seja, analisa se as \textit{features }estão no mesmo índice e atribui um escore entre 0 e 1, para  calcular o Grau de Equivalência Estrutural dos modelos comparados.

\begin{table}[htbp]
  \scalefont{0.8}
  \centering
  \caption{Algoritmo 2.}
    \begin{tabular}{lrrrrrrrr}
    \toprule
    Algoritmo 2:  Equivalência Estrutural  &                                        &                                        &                                        &                                        &                                        &                                        &                                        &  \\
    \midrule
    Entrada: Modelo de \textit{Features} – MFR e MFC &                                        &                                        &                                        &                                        &                                        &                                        &                                        &  \\
    Saída: Equivalência Estrutural carregado em memória &                                        &                                        &                                        &                                        &                                        &                                        &                                        &  \\
    1:   Algoritmo1  (Modelo de \textit{Features} Origem - MFR) &                                        &                                        &                                        &                                        &                                        &                                        &                                        &  \\
    2:   vetorOrigem [] $\leftarrow$ objetoOrigem     &                                        &                                        &                                        &                                        &                                        &                                        &                                        &  \\
    3:   Algoritmo1 (Modelo de \textit{Features} Destino – MFC) &                                        &                                        &                                        &                                        &                                        &                                        &                                        &  \\
    4:   vetorDestino [] $\leftarrow$ objetoDestino   &                                        &                                        &                                        &                                        &                                        &                                        &                                        &  \\
    5:   Para i = 0 até vetorOriem() Faça  &                                        &                                        &                                        &                                        &                                        &                                        &                                        &  \\
    6:      Para j = 0 até vetorDestino() Faça     &                                        &                                        &                                        &                                        &                                        &                                        &                                        &  \\
    7:         Se    elemento.vetorDestino().índice[i] $==$ elemento.vetorOrigem().índice [j] &                                        &                                        &                                        &                                        &                                        &                                        &                                        &  \\
    8:             vetorEstrutura[i] $\leftarrow$ 1;  &                                        &                                        &                                        &                                        &                                        &                                        &                                        &  \\
    9:               Senão Se elemento.vetorDestino().índice[i] != elemento.vetorOrigem().índice [j] &                                        &                                        &                                        &                                        &                                        &                                        &                                        &  \\
    10:                vetorEstrutura[i] $\leftarrow$ 0; &                                        &                                        &                                        &                                        &                                        &                                        &                                        &  \\
    11:             Fim Se                 &                                        &                                        &                                        &                                        &                                        &                                        &                                        &  \\
    12:        Fim Se                      &                                        &                                        &                                        &                                        &                                        &                                        &                                        &  \\
    13:     Fim para                       &                                        &                                        &                                        &                                        &                                        &                                        &                                        &  \\
    14: Fim para                           &                                        &                                        &                                        &                                        &                                        &                                        &                                        &  \\
    15: Para i = 0 até vetorEstrutura() faça &                                        &                                        &                                        &                                        &                                        &                                        &                                        &  \\
    16:      equivalênciaEstrutural + $\leftarrow$  vetorEstrutura[i]/ vetorEstrutura .tamanhoÌndice; &                                        &                                        &                                        &                                        &                                        &                                        &                                        &  \\
    17: Fim para                           &                                        &                                        &                                        &                                        &                                        &                                        &                                        &  \\
    18: Imprimir (“FMI – Vetor de Comparação Estrutural:”) + vetorEstrutura[i] ;  &                                        &                                        &                                        &                                        &                                        &                                        &                                        &  \\
    19: Imprimir (“FMI – Grau de Equivalência Estrutural :”) + equivalênciaEstrutural;    &                                        &                                        &                                        &                                        &                                        &                                        &                                        &  \\
    \bottomrule
    \end{tabular}%
  \label{tab:alg2}%
  \fonte{Elaborado pelo Autor.}
\end{table}%

A Tabela \ref{tab:alg3} demostra o Algoritmo 3, que exibe em sua saída o Grau de Equivalência Sintática, aplica-se a função de Jaro Winkler (WINKLER, 1990), esta métrica é indicada à comparação de pequenas strings \cite{cohen2003, kleinner2008}, ou seja a comparação de palavras curtas, neste caso adaptado à \textit{features}.  O algoritmo executa um laço entre \textit{features} comparado seu rótulo (MF$_{R}$ e MF$_{C}$), e atribui um escore que vária entre 0 a 1.

\begin{table}[htbp]
  \centering
  \scalefont{0.9}
  \caption{Algoritmo 3.}
    \begin{tabular}{lrrrrrr}
    \toprule
    Algoritmo 3:  Equivalência Sintática   &                                        &                                        &                                        &                                        &                                        &  \\
    \midrule
    Entrada: Modelo de \textit{Features}            &                                        &                                        &                                        &                                        &                                        &  \\
    Saída: Elementos do modelo carregado em memória &                                        &                                        &                                        &                                        &                                        &  \\
    1:   Algoritmo1 (Modelo de \textit{Features} Origem - MFR) &                                        &                                        &                                        &                                        &                                        &  \\
    2:   vetorOrigem [] $\leftarrow$ objetoOrigem     &                                        &                                        &                                        &                                        &                                        &  \\
    3:   Algoritmo1 (Modelo de \textit{Features} Destino – MFC) &                                        &                                        &                                        &                                        &                                        &  \\
    4:   vetorDestino [] $\leftarrow$ objetoDestino   &                                        &                                        &                                        &                                        &                                        &  \\
    5:   Para i = 0 até vetorOriem() Faça  &                                        &                                        &                                        &                                        &                                        &  \\
    6:      Para j = 0 até vetorDestino() Faça     &                                        &                                        &                                        &                                        &                                        &  \\
    8:             vetorJW [i] $\leftarrow$ jaroWinkler (elemento de MFR, elemento MFC) ; &                                        &                                        &                                        &                                        &                                        &  \\
    9:     Fim para                        &                                        &                                        &                                        &                                        &                                        &  \\
    10: Fim para                           &                                        &                                        &                                        &                                        &                                        &  \\
    11: Para i = 0 até vetorJW() Faça      &                                        &                                        &                                        &                                        &                                        &  \\
    12:      equivalênciaSintática + $\leftarrow$  vetorJW [i]/ vetorJW .tamanhoÌndice; &                                        &                                        &                                        &                                        &                                        &  \\
    13: Fim para                           &                                        &                                        &                                        &                                        &                                        &  \\
    14: Imprimir (“FMI – Vetor de Comparação Sintática:”) + vetorJW[i] ;  &                                        &                                        &                                        &                                        &                                        &  \\
    15: Imprimir (“FMI – Grau de Equivalência Sintática:”) + equivalênciaSintática;  &                                        &                                        &                                        &                                        &                                        &  \\
    \bottomrule
    \end{tabular}%
  \label{tab:alg3}%
  \fonte{Elaborado pelo Autor.}
\end{table}%

A Tabela \ref{tab:alg4} exibe o Algoritmo 4, que executa um laço entre elementos comparados (MF$_{R}$ e MF$_{C}$), verificando o relacionamento entre as \textit{features} (opcional, obrigatório, alternativo inclusivo, alternativo exclusivo, dependências, e exclusão) e atribui um escore entre 0 e 1, para  calcular o Grau de Equivalência Semântica dos modelos comparados.

\begin{table}[htbp]
  \centering
  \scalefont{0.9}
  \caption{Algoritmo 4}
    \begin{tabular}{lrrrrrr}
    \toprule
    Algoritmo 4:  Equivalência Semântica   &                                        &                                        &                                        &                                        &                                        &  \\
    \midrule
    Entrada: Modelo de \textit{Features}            &                                        &                                        &                                        &                                        &                                        &  \\
    Saída: Elementos do modelo carregado em memória &                                        &                                        &                                        &                                        &                                        &  \\
    1:   Algoritmo1 (Modelo de \textit{Features} Origem - MFR) &                                        &                                        &                                        &                                        &                                        &  \\
    2:   vetorOrigem [] $\leftarrow$  objetoOrigem     &                                        &                                        &                                        &                                        &                                        &  \\
    3:   Algoritmo1 (Modelo de \textit{Features} Destino – MFC) &                                        &                                        &                                        &                                        &                                        &  \\
    4:   vetorDestino [] $\leftarrow$  objetoDestino   &                                        &                                        &                                        &                                        &                                        &  \\
    5:   Para i = 0 até vetorOriem() Faça  &                                        &                                        &                                        &                                        &                                        &  \\
    6:      Para j = 0 até vetorDestino() Faça     &                                        &                                        &                                        &                                        &                                        &  \\
    7:         Se    elemento.vetorDestino().índice[i] == elemento.vetorOrigem().índice [j] &                                        &                                        &                                        &                                        &                                        &  \\
    8:             vetorSemântico[i] $\leftarrow$  1;  &                                        &                                        &                                        &                                        &                                        &  \\
    9:               Senão Se elemento.vetorDestino().índice[i] != elemento.vetorOrigem().índice [j]                   &                                        &                                        &                                        &                                        &                                        &  \\
    10:                vetorSemântico [i] $\leftarrow$  0; &                                        &                                        &                                        &                                        &                                        &  \\
    11:             Fim Se                 &                                        &                                        &                                        &                                        &                                        &  \\
    12:        Fim Se                      &                                        &                                        &                                        &                                        &                                        &  \\
    13:     Fim para                       &                                        &                                        &                                        &                                        &                                        &  \\
    14: Fim para                           &                                        &                                        &                                        &                                        &                                        &  \\
    15: Para i = 0 até vetorSemântico () Faça &                                        &                                        &                                        &                                        &                                        &  \\
    16:      equivalênciaSemântica + $\leftarrow$   vetorSemântico [i]/ vetorSemântico .tamanhoÌndice; &                                        &                                        &                                        &                                        &                                        &  \\
    17: Fim para                           &                                        &                                        &                                        &                                        &                                        &  \\
    18: Imprimir (“FMI – Vetor de Comparação Semântico:”) + vetorSemântico;  &                                        &                                        &                                        &                                        &                                        &  \\
    19: Imprimir (“FMI – Grau de Equivalência Semântica:”) + equivalênciaSemântica;  &                                        &                                        &                                        &                                        &                                        &  \\
    \bottomrule
    \end{tabular}%
  \label{tab:alg4}%
  \fonte{Elaborado pelo Autor.}
\end{table}%

Finalmente, a Tabela \ref{tab:alg5} demonstra o Algoritmo 5, que representa o grau de equivalência global entre os modelos comprados (MF$_{R}$ e MF$_{C}$), ou seja, o Calculo de Efetividade Estratégica que representa as táticas adotadas na verificação de similaridade dos modelos, obtidos através somatório dos Algoritmos 2, 3 e 4 divididos pelo número de estratégias de comparação aplicadas.

\begin{table}[htbp]
\scalefont{0.9}
  \centering
  \caption{Algoritmo 5}
    \begin{tabular}{lrrrrrrr}
    \toprule
    Algoritmo 5:  Equivalência  Global     &                                        &                                        &                                        &                                        &                                        &                                        &  \\
    \midrule
    Entrada: Algoritmos 2, 3 e 4.          &                                        &                                        &                                        &                                        &                                        &                                        &  \\
    Saída: Cálculo de Efetividade Estratégica, CEE, carregado em memória. &                                        &                                        &                                        &                                        &                                        &                                        &  \\
    1:   CEE$\leftarrow$ 0;                           &                                        &                                        &                                        &                                        &                                        &                                        &  \\
    2:   CEE$\leftarrow$ (equivalênciaEstrutural+ equivalênciaSintática+ equivalênciaSemântica)/3; &                                        &                                        &                                        &                                        &                                        &                                        &  \\
    3:   Imprimir (“FMI – Cálculo de Equivalência Global :”) + CEE; &                                        &                                        &                                        &                                        &                                        &                                        &  \\
    \bottomrule
    \end{tabular}%
  \label{tab:alg5}%
  \fonte{Elaborado pelo Autor.}
\end{table}%

Após a obtenção da equivalência global dos modelos comparados, aplicam-se os algoritmos para implementação das estratégias de integração: estratégia automática (1), exibida a partir do Algoritmo 6, na Tabela \ref{tab:alg6} e por fim a estratégia semiautomática (2), demostrada na Tabela \ref{tab:alg7} o Algoritmo 7. Os algoritmos implementados estendem classes e interfaces oriundos da linguagem \textit{Java}, como por exemplo, o\textit{ Java Utilities}, para trabalhar com estruturas de coleções e \textit{interface}, para trabalhar com, \textit{Set} e \textit{List}. A adoção destes métodos advém da facilidade de uso de recursos para operações de comparação entre elementos de conjuntos ou listas.

\begin{table}[htbp]
  \centering
  \caption{Algoritmo 6}
    \begin{tabular}{lrrrrr}
    \toprule
    Algoritmo 6:  Integração Automática    &                                        &                                        &                                        &                                        &  \\
    \midrule
    Entrada: MFR e  MFC                    &                                        &                                        &                                        &                                        &  \\
    Saída:   MFI, carregados em memória    &                                        &                                        &                                        &                                        &  \\
    1:   Algoritmo1 (Modelo de Feature Origem - MFR) &                                        &                                        &                                        &                                        &  \\
    2:   vetorOrigem [] $\leftarrow$  objetoOrigem     &                                        &                                        &                                        &                                        &  \\
    3:   Algoritmo1 (Modelo de Feature Destino – MFC) &                                        &                                        &                                        &                                        &  \\
    4:   vetorDestino [] $\leftarrow$  objetoDestino   &                                        &                                        &                                        &                                        &  \\
    //união dos elementos                  &                                        &                                        &                                        &                                        &  \\
    5:   Lista <\textit{features}> adicional $\leftarrow$  novaLista<features>();    &                                        &                                        &                                        &                                        &  \\
    6:   adicional.adicionarTodos(vetorOrigem), &                                        &                                        &                                        &                                        &  \\
    7:   adicional.adicionarTodos(vetorDestino), &                                        &                                        &                                        &                                        &  \\
    //intersecção dos elementos            &                                        &                                        &                                        &                                        &  \\
    8:   Lista <\textit{features}> formal $\leftarrow$  novaLista<\textit{features}>();  &                                        &                                        &                                        &                                        &  \\
    9:   formal.adicionarTodos(vetorOrigem), &                                        &                                        &                                        &                                        &  \\
    10: formal.manterTodos(vetorDestino),  &                                        &                                        &                                        &                                        &  \\
    //diferença dos elementos              &                                        &                                        &                                        &                                        &  \\
    11: Lista <\textit{features}> parcial $\leftarrow$  novaLista<\textit{features}>();  &                                        &                                        &                                        &                                        &  \\
    12: parcial.adicionarTodos(vetorOrigem), &                                        &                                        &                                        &                                        &  \\
    13: parcial.removerTodos(vetorDestino), &                                        &                                        &                                        &                                        &  \\
    //complemento dos elementos            &                                        &                                        &                                        &                                        &  \\
    14: Lista <\textit{features}> complementar $\leftarrow$  novaLista<\textit{features}>();  &                                        &                                        &                                        &                                        &  \\
    15: complementar.adicionarTodos(vetorDestino), &                                        &                                        &                                        &                                        &  \\
    16: complementar.removerTodos(vetorOrigem), &                                        &                                        &                                        &                                        &  \\
    17: Imprimir (“FMI – União:”) + adicional;  &                                        &                                        &                                        &                                        &  \\
    18: Imprimir (“FMI – Intersecção:”) + formal;  &                                        &                                        &                                        &                                        &  \\
    19: Imprimir (“FMI – Diferença:”) + parcial; &                                        &                                        &                                        &                                        &  \\
    20: Imprimir (“FMI – Complemento:”) + complementar; &                                        &                                        &                                        &                                        &  \\
    \bottomrule
    \end{tabular}%
   \label{tab:alg6}%
  \fonte{Elaborado pelo Autor.}
\end{table}%

O Algoritmo 6, apresenta as técnicas operacionais de integração descritas na Seção \ref{TiSubMuliEstrategia} (Comum, Adicional, Formal, Parcial, Complementar, Nula), as demais técnicas a Estratégia Nula é chamada quando não há qualquer similaridade entre os elementos comparados aplicando o método adicional, descrito na linha 5,  já a Estratégia Comum é, repito, o vetor de origem, pois não existe diferenças entre os modelos comparados, descrito na linha 2. O algoritmo acima contém os métodos para integrar os modelos após sua comparação, a linha 5 chama o método adicional, responsável por adicionar novos elementos aos modelo integrado, MF$_{I}$;  a linha 8 chama o método formal,  que exclui os elementos incomuns de ambos os modelos comparados;  a linha 11 chama o método parcial que exibe as diferenças contidas no modelo origem, MF$_{R}$,  comparadas ao modelo destino, MF$_{C}$, gerando o modelo integrado, MF$_{I}$;  por fim, o método complementar é executado na linha 14 exibindo as diferenças contidas no modelo destino, MF$_{C}$, compradas ao modelo origem, MF$_{R}$, gerando o modelo integrado, MF$_{I}$.

\begin{figure}[ht]
	\caption{Exemplo de integração automática.} 
	\label{exemploIntAut}
	\centering%
	\begin{minipage}{0.9\textwidth}
		\includegraphics[width=\textwidth]{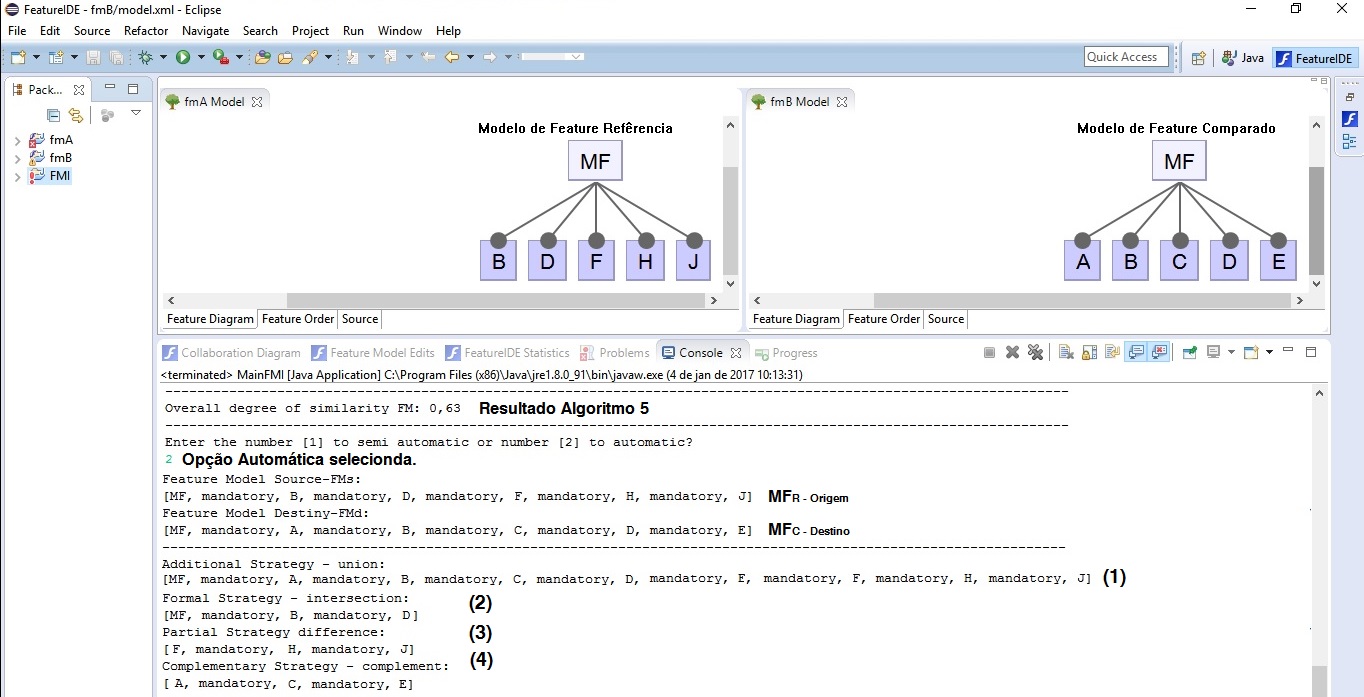}
		\fonte{Elaborado pelo Autor.}
	\end{minipage}
\end{figure}

A Figura \ref{exemploIntAut} ilustra a integração entre o MF$_{R}$, e MF$_{C}$,, exibindo a estratégia de integração automática bem como sua respectiva saída conforme o algoritmo 6 propõe.  A Figura também exibe em sua saída os resultados oriundos dos algoritmos 2, 3, 4 e 5.  O cálculo global de equivalência entre os modelos é de 63\%, originado do algoritmo 5. Ao selecionar a opção automática o mesmo integra ambos os modelos produzindo as seguintes saídas: (1) Estratégia Adicional-União: [MF, A, B, C, D, E, F, H, J], (2) Estratégia Formal-Intersecção: [MF, B, D], (3) Estratégia Parcial-Diferença: [F, H, J], e por fim, a (4) Estratégia Complementar-Complemento: [A, C, E].

A Tabela \ref{tab:alg7} e a Tabela \ref{tab:alg71} exibem o Algoritmo 7, que ilustra a Estratégia de Integração Semiautomática que tem como objetivo auxiliar na produção final do modelo  pretendido, MF$_{AB}$, ou seja, sua função é localizar os conflitos inerentes aos modelos que virão a ser integrados, o modelo referência, MF$_{R}$, e o modelo comparado, MF$_{C}$, exibindo aos analistas ou desenvolvedores para que os mesmos tomem a decisão de qual elemento deve-se integrar ao modelo conforme os requisitos de projeto estabelecidos.

\begin{table}[htbp]
\scalefont{0.8}
  \centering
  \caption{Algoritmo 7}
    \begin{tabular}{lrrrrrrrr}
    \toprule
    Algoritmo 7:  Integração Semiautomática &                                        &                                        &                                        &                                        &                                        &                                        &                                        &  \\
    \midrule
    Entrada: Modelo de \textit{Features} Origem - MFR e Modelo de \textit{Features} Destino – MFC &                                        &                                        &                                        &                                        &                                        &                                        &                                        &  \\
    Saída: Modelo de \textit{Features} Pretendido - MFAB, carregados em memória &                                        &                                        &                                        &                                        &                                        &                                        &                                        &  \\
    1:   Algoritmo1 (Modelo de Feature Origem - MFR) &                                        &                                        &                                        &                                        &                                        &                                        &                                        &  \\
    2:   vetorOrigem [] $\leftarrow$ objetoOrigem     &                                        &                                        &                                        &                                        &                                        &                                        &                                        &  \\
    3:   Algoritmo1 (Modelo de Feature Destino – MFC) &                                        &                                        &                                        &                                        &                                        &                                        &                                        &  \\
    4:   vetorDestino [] $\leftarrow$ objetoDestino   &                                        &                                        &                                        &                                        &                                        &                                        &                                        &  \\
    5:   Para i = 0 até vetorOriem() faça  &                                        &                                        &                                        &                                        &                                        &                                        &                                        &  \\
    6:      Para j = 0 até vetorDestino() faça     &                                        &                                        &                                        &                                        &                                        &                                        &                                        &  \\
    7:         Se elemento.vetorDestino().índice[i] == elemento.vetorOrigem().índice [j] &                                        &                                        &                                        &                                        &                                        &                                        &                                        &  \\
    8:              verificaEquivalencia[i] $\leftarrow$ 1; &                                        &                                        &                                        &                                        &                                        &                                        &                                        &  \\
    9:              Senão Se elemento.vetorDestino().índice[i] != elemento.vetorOrigem().índice [j]  &                                        &                                        &                                        &                                        &                                        &                                        &                                        &  \\
    10:  Imprimir (“Existem diferenças sintáticas ou semânticas entre os modelos comparados.”);  &                                        &                                        &                                        &                                        &                                        &                                        &                                        &  \\
    11:  Imprimir (“Origem – [+vetorOrigem[i]]+ é diferente do Destino – [+vetorDestino[j]]”);               &                                        &                                        &                                        &                                        &                                        &                                        &                                        &  \\
    12:  Imprimir (“Deseja alterar o relacionamento ou feature? sim para: [+vetorOrigem[i]]  ou &                                        &                                        &                                        &                                        &                                        &                                        &                                        &  \\
                                                  não para: [+vetorDestino[j]].”) &                                        &                                        &                                        &                                        &                                        &                                        &                                        &  \\
    13:                Ler (retornoQuestão);   &                                        &                                        &                                        &                                        &                                        &                                        &                                        &  \\
    14:                Faça                &                                        &                                        &                                        &                                        &                                        &                                        &                                        &  \\
    15:                    Se retornoQuestão == “sim” ou retornoQuestão== “não” &                                        &                                        &                                        &                                        &                                        &                                        &                                        &  \\
    16:                        Se retornoQuestão == “sim”  &                                        &                                        &                                        &                                        &                                        &                                        &                                        &  \\
    17:                             vetorDestino().índice[i] $\leftarrow$ vetorOrigem().índice [j]; &                                        &                                        &                                        &                                        &                                        &                                        &                                        &  \\
    18:                        Senão Se retornoQuestão == “não”  &                                        &                                        &                                        &                                        &                                        &                                        &                                        &  \\
    19:                            vetorOrigem().índice [j] $\leftarrow$ vetorDestino().índice[i]; &                                        &                                        &                                        &                                        &                                        &                                        &                                        &  \\
    20:                          Fim Se    &                                        &                                        &                                        &                                        &                                        &                                        &                                        &  \\
    21:                        Fim Se      &                                        &                                        &                                        &                                        &                                        &                                        &                                        &  \\
    22:                     Fim Se         &                                        &                                        &                                        &                                        &                                        &                                        &                                        &  \\
    23:                     verificaQuestão $\leftarrow$ verdadeiro &                                        &                                        &                                        &                                        &                                        &                                        &                                        &  \\
    24:                     Se             &                                        &                                        &                                        &                                        &                                        &                                        &                                        &  \\
    25:                       Imprimir (“Escolha invalida !!!”); &                                        &                                        &                                        &                                        &                                        &                                        &                                        &  \\
    26:                       Imprimir (“Deseja alterar o relacionamento ou feature?  &                                        &                                        &                                        &                                        &                                        &                                        &                                        &  \\
                                               sim para: [+vetorOrigem[i]]  ou   não para: [+vetorDestino[j]].”); &                                        &                                        &                                        &                                        &                                        &                                        &                                        &  \\
    27:                       Ler (retornoQuestão);   &                                        &                                        &                                        &                                        &                                        &                                        &                                        &  \\
    28:                       verificaQuestão $\leftarrow$ falso &                                        &                                        &                                        &                                        &                                        &                                        &                                        &  \\
    29:                     Fim Se         &                                        &                                        &                                        &                                        &                                        &                                        &                                        &  \\
    30:                Fim Enquanto(verificaQuestão $\leftarrow$ falso); &                                        &                                        &                                        &                                        &                                        &                                        &                                        &  \\
    31:                verificaEquivalencia[i]$\leftarrow$ 0; &                                        &                                        &                                        &                                        &                                        &                                        &                                        &  \\
    32:             Fim Se                 &                                        &                                        &                                        &                                        &                                        &                                        &                                        &  \\
    33:        Fim Se                      &                                        &                                        &                                        &                                        &                                        &                                        &                                        &  \\
    34:     Fim para                       &                                        &                                        &                                        &                                        &                                        &                                        &                                        &  \\
&...continua                                        &  \\
    \bottomrule
    \end{tabular}%
  \label{tab:alg7}%
  \fonte{Elaborado pelo Autor.}
\end{table}%

\begin{table}[htbp]
\scalefont{0.8}
  \centering
  \caption{Algoritmo 7 complemento}
    \begin{tabular}{lrrrrrrrr}
    ...continuação                            &                                        &                                        &                                        &                                        &                                        &                                        &                                        &  \\
    35: Fim para                           &                                        &                                        &                                        &                                        &                                        &                                        &                                        &  \\
    \midrule
    36: Para i = 0 até vetorOrigem () faça \textbackslash{}\textbackslash{} armazena valores do vetor antes de sua alteração &                                        &                                        &                                        &                                        &                                        &                                        &                                        &  \\
    37:      vetorOrigemInalterado[i] $\leftarrow$  vetorOrigem [i]; &                                        &                                        &                                        &                                        &                                        &                                        &                                        &  \\
    38: Fim para                           &                                        &                                        &                                        &                                        &                                        &                                        &                                        &  \\
    39: Para i = 0 até vetorOrigemInalterado () faça &                                        &                                        &                                        &                                        &                                        &                                        &                                        &  \\
    40:      Para j = 0 até vetorDestino() faça   &                                        &                                        &                                        &                                        &                                        &                                        &                                        &  \\
    41:         Se elemento. OrigemInalterado ().índice[i] == elemento.vetorDestino().índice[j] &                                        &                                        &                                        &                                        &                                        &                                        &                                        &  \\
    42:            novaEquivalencia[i] $\leftarrow$ 1; &                                        &                                        &                                        &                                        &                                        &                                        &                                        &  \\
    43:            Senão Se elemento.OrigemInalterado ().índice[i] != elemento.vetorDestino().índice[j] &                                        &                                        &                                        &                                        &                                        &                                        &                                        &  \\
    44:                 novaEquivalencia[i] $\leftarrow$ 0; &                                        &                                        &                                        &                                        &                                        &                                        &                                        &  \\
    45:            Fim se                  &                                        &                                        &                                        &                                        &                                        &                                        &                                        &  \\
    46:       Fim Se                       &                                        &                                        &                                        &                                        &                                        &                                        &                                        &  \\
    47:     Fim para                       &                                        &                                        &                                        &                                        &                                        &                                        &                                        &  \\
    48: Fim para                           &                                        &                                        &                                        &                                        &                                        &                                        &                                        &  \\
    49: resultadoNovaEquivalência $\leftarrow$ 0;     &                                        &                                        &                                        &                                        &                                        &                                        &                                        &  \\
    50: Para i = 0 até novaEquivalencia () faça  &                                        &                                        &                                        &                                        &                                        &                                        &                                        &  \\
    51:      resultadoNovaEquivalência +$\leftarrow$ novaEquivalencia[i] / novaEquivalencia.totalElementos; &                                        &                                        &                                        &                                        &                                        &                                        &                                        &  \\
    52: Fim para                           &                                        &                                        &                                        &                                        &                                        &                                        &                                        &  \\
    53: Imprimir (“FMI – Vetor de Comparação Sintatico e Semântico:”) + novaEquivalencia[i];  &                                        &                                        &                                        &                                        &                                        &                                        &                                        &  \\
    54: Imprimir (“FMI – Grau de Equivalência :”) + resultadoNovaEquivalência;     &                                        &                                        &                                        &                                        &                                        &                                        &                                        &  \\
    55: Imprimir (“FMI – Modelo de Feature Pretendido:”) + vetorDestino;   &                                        &                                        &                                        &                                        &                                        &                                        &                                        &  \\
    \bottomrule
    \end{tabular}%
  \label{tab:alg71}%
  \fonte{Elaborado pelo Autor.}
\end{table}%

O Algoritmo  7 requer a intervenção dos analistas e dos desenvolvedores; para isso da linhas 1 até a linha 4, executam importação dos arquivos de comparação, isto é, o modelo  referência, MF$_{R}$, e o modelo comparado, MF$_{C}$. As linhas subsequentes, 5 e 6 iniciam laço para verificar conflitos nos modelos em análise, mais especificamente na linha 9, exibindo em sua saída as questões inerentes às diferenças encontradas, nas linhas 10, 11 e 12. Em meio às linhas 14 a 19, mais especificamente nas linhas 17 e 19, o algoritmo substitui os elementos dos modelos conflitantes que virão a produzir como saída modelo  pretendido, MF$_{AB}$. O próximo laço, na linha 36, armazena os dados do modelo referência, para futuramente calcular similaridade do novo modelo. Os laços seguintes, designadamente nas linhas subsequentes, 39 a 52 determinam a equivalência entre o modelo referência, MF$_{R}$, e o novo modelo produzido, isto é, modelo pretendido, MF$_{AB}$. Por fim, as linhas finais 53 a 55 exibem os seguintes resultados: (1) vetor de comparação dos dados que foram alterados, representados por 0 e 1, onde zero representa os elementos alterados ou não similares e hum representa os elementos inalterados ou similares; exibe o novo (2)  grau de equivalência  entre a comparação dos modelos MF$_{R}$ e MF$_{AB}$; e finalmente (3) exibe o modelo de \textit{features} pretendido.

\begin{figure}[!ht]
	\caption{Exemplo de integração semiautomática.} 
	\label{exemploIntSemi}
	\centering%
	\begin{minipage}{1\textwidth}
		\includegraphics[width=\textwidth]{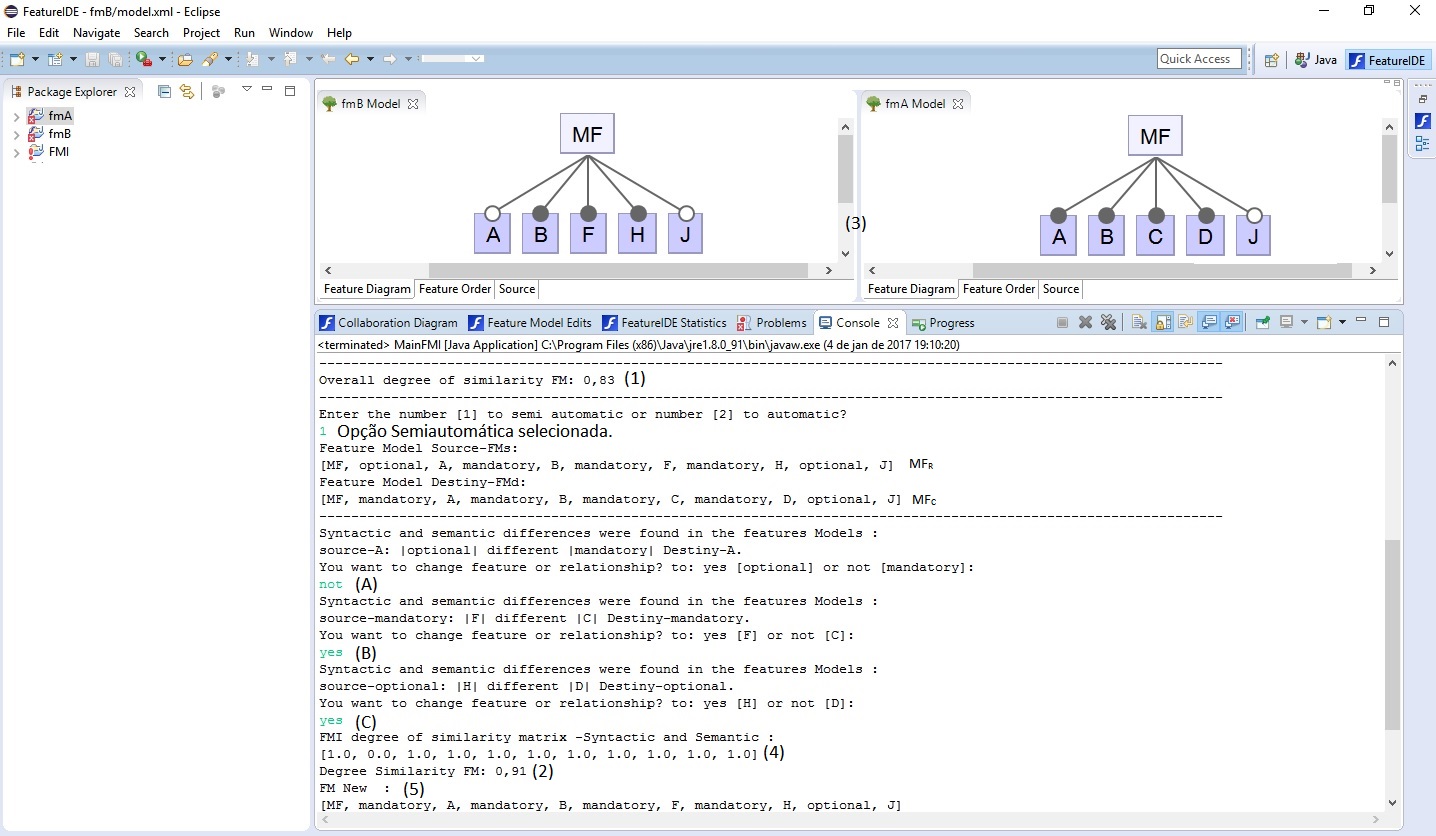}
		\fonte{Elaborado pelo Autor.}
	\end{minipage}
\end{figure}

A Figura \ref{exemploIntSemi} ilustra como ocorre a interação do protótipo com os analistas e desenvolvedores aplicando a estratégia de integração semiautomática. Conforme o algoritmo 5 descrito anteriormente responsável pelo o (1) cálculo de equivalência entre os modelos comprados. A figura abaixo exibe  uma equivalência de 83\%.  Após a alteração do modelos pode-se observar que o novo (2) cálculo de equivalência passou para 91\% de similaridade. Como se pôde perceber na imagem a estratégia de integração selecionada é a semiautomática; logo abaixo é exibido os modelos de \textit{features}, MF$_{R}$ e MF$_{C}$. A interação com o usuário do protótipo é exibida logo abaixo, sendo identificados os seguintes conflitos nos (3) diagramas exibidos também graficamente: (A) conflito de relacionamento, conforme o exemplo manteve-se o elemento obrigatório; (B) o conflito entre as \textit{features} permanecendo o elemento F, o último conflito encontrado entre as \textit{features} permanece o elemento H. Por fim, é exibo um novo vetor (4) reflexo das substituições dos conflitos alterados e o (5) modelo pretendido, MF$_{AB}$.

\section{Tecnologias Utilizadas} \label{AspTecnologias}

O protótipo desenvolvido é um \textit{plug-in} para Eclipse, utilizando a linguagem de programação em Java. O Eclipse possibilita ao desenvolvedor estender suas funcionalidades à IDE na forma de \textit{plug-in}.  O \textit{plug-in} é um componente de software usado para adicionar funções a programas maiores, provendo uma funcionalidade específica ou sob demanda \cite{zibran2011}. A ferramenta adotada como suporte para o desenvolvimento do protótipo foi a FeatureIDE,é \textit{framework} baseado na IDE Eclipse para a modelagem de \textit{features}, o FeatureIDE \cite{kastner2009}. A ferramenta foi desenvolvida para automatizar o processo de modelagem e configurações dos artefatos reusáveis.  A FeatureIDE foi escolhida por dois motivos: (1) flexibilidade na importação e exportação dos modelos produzidos, possibilitando sua edição em tempo de execução; (2) o segundo fato se deve a notação da ferramenta suporta a técnica FODA \cite{kang1990}, bem como atende os critérios exclusão e as dependências entre as \textit{features}.

FeatureIDE fornece um editor para modelagem de \textit{features} e permite a construção textual e gráfica dos modelos. A principal característica da ferramenta é a cobertura de todas  as atividades de desenvolvimento e a incorporação de ferramentas para a implementação de LPS. A proposta apresentada pelo FeatureIDE é a redução de esforço para construção de ferramentas extensíveis a LPS \cite{thum2014, kastner2009, apel2009}, análise e implementação de domínio, análise de requisitos e geração de software. 

\begin{figure}[ht]
	\caption{Particularidades do framework featureIDE.} 
	\label{particularidadesFeatureIDE}
	\centering%
	\begin{minipage}{.7\textwidth}
		\includegraphics[width=\textwidth]{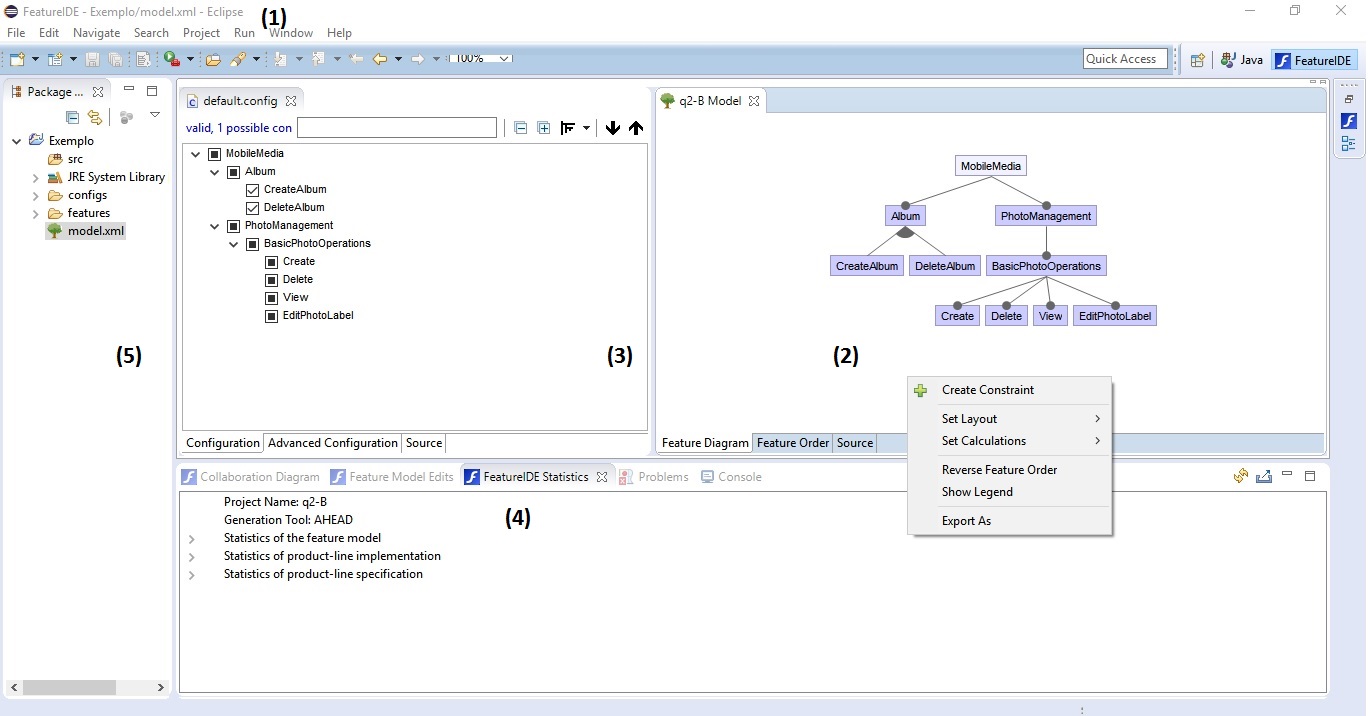}
		\fonte{Elaborado pelo Autor.}
	\end{minipage}
\end{figure}

A ferramenta apresenta também as seguintes particularidades: (1) ampla integração com o Eclipse; (2) editor de modelo de \textit{features}; (3) editor de configurações e restrições; (4) dados estatísticos; e, por fim, (5) possibilita a integração com diversas linguagens de programação (por exemplo, Java, C$++$, XML ), tais como ferramentas que dão suporte à modelagem de \textit{features} (por exemplo, AHEAD, FeatureHOUSE, FeatureC$++$, entre outros). A Figura \ref{particularidadesFeatureIDE} exibe um exemplo através da captura de tela do Eclipse mostrando algumas das principais funcionalidades do \textit{framework }descritas anteriormente.

\chapter{Avaliação da Solução Proposta} \label{AvaliaçãoDaSolução}

Este estudo tem como objetivo avaliar o impacto da técnica proposta no Capítulo \ref{TécnicaDeIntegração}, assim como atender parte dos objetivos específicos descritos na Seção \ref{I_Objetivos}. Através desta avaliação será possível comparar a técnica proposta, suportada pela ferramenta FMIT (semiautomática), como a forma tradicional, suportada pela FeatureIDE (manual). Para isso, realizou-se dois experimentos de integração de modelos de \textit{features} contendo  seis cenários de avaliação intercalados entre cada um dos experimentos. No primeiro cenário fez-se o uso da ferramenta FeatureIDE, porém, sem o suporte  ferramental  que auxilia na identificação de conflitos  entre os modelos compostos. Já o segundo experimento é realizado com o uso do protótipo,  FMIT, auxiliando aos analistas e desenvolvedores na identificação de conflitos, conforme descrito no Capítulo \ref{AspectosDeImplementação} deste trabalho.

Este Capítulo é organizado da seguinte forma. A Seção \ref{AvObjetivo} apresenta as questões de pesquisa e objetivos abordados no experimento. A Seção \ref{AvHipotese} define as hipóteses do estudo, as quais são baseadas nas questões de pesquisa definidos na Seção \ref{IntQuestõesdePesquisa}. A Seção \ref{AvVariáveis} descreve as variáveis e os métodos de quantificação utilizados. A Seção \ref{AvContexto} explica o contexto dos experimentos, bem como a seleção dos participantes. A Seção \ref{AvAnálise} apresenta os dados inerentes aos experimentos executados, juntamente com uma análise comparativa. Por fim. A Seção \ref{AvAmeaças} discute as ameaças à validade do estudo. A execução deste estudo experimental segue as boas práticas para a realização de estudos experimentais encontradas em \citetexto{wohlin2000}. 

\section{Objetivo e Questões de Pesquisa} \label{AvObjetivo}

Este estudo avalia os efeitos das técnicas de integração de modelos de \textit{features} em duas variáveis: o (1) esforço dos analistas e desenvolvedores na detecção de conflitos e a (2)  exatidão na correção dos modelos produzidos. Estes efeitos são investigados a partir de seis cenários de evolução envolvendo composições de modelos. Com isto em mente, o objetivo deste experimento segue o modelo \textit{GQM (Goal, Questions, Metrics)} \cite{wohlin2000}, procurando:

\begin{center}
\textbf{Analisar} as técnicas de integração de modelos de \textit{features} \\
\textbf{com a finalidade} de investigar os efeitos \\
\textbf{no que diz respeito ao} esforço e corretude\\
\textbf{a partir da perspectiva} de analistas e desenvolvedores\\
\textbf{no contexto de} evolução do modelo de \textit{features}.\\
\end{center}

Dessa forma, este experimento procura avaliar o efeito da técnica proposta no esforço empregado para integrar modelos, bem como na corretude das integrações realizadas. Assim, o experimento busca mensurar a utilização da técnica manual em relação a utilização da técnica semiautomáticas, descritas anteriormente na Seção \ref{IntQuestõesdePesquisa}, tendo em vista que estes objetivos definem as hipóteses para o experimento. 

\section{Formulação das Hipóteses} \label{AvHipotese}

Com o embasamento dos experimentos executados neste estudo, que buscam mensurar o esforço e a corretude dos modelos produzidos de forma manual e semiautomática, duas hipóteses foram formuladas para avaliar a técnica proposta no Capítulo \ref{AspectosDeImplementação} .

\textbf{Hipótese 1.} A técnica de integração do modelo de \textit{features} tende em um primeiro momento a ser amigável, devido a  sua simplicidade. Entretanto, à medida em que as possíveis características e combinações se ampliam podem tornar-se onerosas, uma vez que envolvem procedimentos de análise dos modelos de entrada MF$_{R}$  e MF$_{C}$ para identificar as possíveis equivalências entre os elementos de todos os modelos.  

As equipes de desenvolvimento podem investir grande esforço para compor modelos, e esses esforços muitas vezes não são convertidos no modelo pretendido, ou seja, apresenta um modelo inconsistente \cite{farias2015,kleinner2012}.  Portanto, a primeira hipótese avalia se a técnica de integração semiautomática, com base no protótipo proposto, reduz o esforço de integração, auxiliando os analistas e desenvolvedores a comporem os modelos consumindo menor tempo. Com base nesta conjectura, as hipóteses nula e alternativa sobre o Esforço de Integração (EI) são apresentadas a seguir:
\begin{quotation}
  	\textbf{Hipótese Nula 1, (H1-0):}  A técnica semiautomática e o protótipo propostos não reduzem o esforço de integração ao produzir o modelo pretendido, MF$_{AB}$, a partir dos modelos referência, MF$_{R}$, e modelo comparado, MF$_{C}$.
  	
	\textbf{H1-0:} EI semiautomática (MF$_{R}$, MF$_{C}$) $\geq$ EI manual  MF$_{R}$, MF$_{C}$).\\
	
	\textbf{Hipótese Alternativa 1, (H1-1):} A técnica semiautomática e o protótipo propostos reduzem o esforço de integração ao produzir o modelo pretendido, MF$_{AB}$, a partir dos modelos referência, MF$_{R}$, e modelo comparado, MF$_{C}$ .
	
	\textbf{H1-1:} EI semiautomática (MF$_{R}$, MF$_{C}$) $<$ EI manual (MF$_{R}$, MF$_{C}$).
\end{quotation}

Ao analisar a primeira hipótese, pretende-se avaliar se a técnica de integração proposta reduz os esforços dos analistas e desenvolvedores na produção do modelo pretendido, MF$_{AB}$, gerando assim evidências empíricas sobre como as técnicas propostas acomodam as mudanças futuras de MF$_{R}$, MF$_{C}$. 

\emph{Hipótese 2.} As inconsistências incidem na composição de modelos devida às alterações conflitantes, afetando as propriedades sintáticas e semânticas do modelo, sendo que sua integração não coincide com o modelo pretendido \cite{oliveira2009, oliveira2009b, weber2016}.   A correção das integrações é influenciada pela presença, ou não, das inconsistências no modelo integrado de saída, assim é necessário avaliar a corretude das alterações efetuadas. Esta hipótese avalia se a integração realizada de forma semiautomática, com base na técnica e no protótipo proposto, auxilia os analistas e desenvolvedores a reduzir em sua saída um modelo com inconsistência, ou seja, modelos que não possuem erros em comparação ao modelo pretendido. Com base nesta declaração, definem-se as hipóteses nula e alternativa comparando a Integração Correta (IC) apresentada a seguir:
\begin{quotation} 
	\textbf{Hipótese Nula 2, (H2-0):}  A técnica e o protótipo proposto não amparam significativamente na extinção de falhas  ao produzir o modelo pretendido, MF$_{AB}$, a partir dos modelos referência, MF$_{R}$., e modelo comparado, MF$_{C}$.
	
	\textbf{H2-0:} IC semiautomática (MF$_{AB}$) $\leq$ IC manual(MF$_{AB}$).\\

	\textbf{Hipótese Alternativa 2, (H2-1):} A técnica e o protótipo proposto auxiliem significativamente na extinção de falhas  ao produzir o modelo pretendido, MFAB, a partir dos modelos referência, MFR, e modelo comparado, MFC .

	\textbf{H2-1:} IC semiautomática (MF$_{AB}$) $>$ IC manual (MF$_{AB}$).
\end{quotation}

Ao analisar a segunda hipótese, se produz conhecimento empírico sobre a técnica de integração semiautomática, impactando na correção do modelo, visto que a mesma pode inferir no nível de assertividade das alterações reduzindo os conflitos inerentes de falhas produzidas durante a atividade de integração manual e possivelmente um aumento de produtividade e maior precisão da informação, bem como na qualidade final dos modelos de \textit{features }produzidos, e, por fim, acredita-se em uma amortização ou eliminação dos esforços de retrabalho na resolução de más integrações.

A Tabela \ref{tab:resumoHipoteses} apresenta um resumo das hipóteses nulas e alternativas deste experimento, que visam elencar indícios que sejam suficientes para responder à questão de pesquisa deste estudo. A partir dos testes de hipóteses, pretende-se confirmar se a técnica e o protótipo proposto auxiliam na eficiência das integrações assim como propõem se em melhorar a eficácia do modelo de \textit{features} final produzido.

\begin{table}[htbp]
  \centering
  \caption{Resumo das hipóteses.} \label{tab:resumoHipoteses}
    \begin{tabular}{lll}
    \toprule
    \multicolumn{1}{l}{\textbf{Hipótese}}            & \textbf{Tipo}                                     & \multicolumn{1}{l}{\textbf{Representação}} \\
    \midrule
    \multirow{4}[1]{*}{H1}                            & \multirow{2}[1]{*}{Nula}                          & \multirow{2}[1]{*}{H1-0: EI semiautomática (MF$_{R}$, MF$_{C}$)  $\geq$ EI manual (MF$_{R}$, MF$_{C}$)} \\
                                                      &                                                   &  \\
                                                      & \multirow{2}[0]{*}{Alternativa}                   & \multirow{2}[0]{*}{H1-1: EI semiautomática (MF$_{R}$, MF$_{C}$) $<$ EI manual (MF$_{R}$, MF$_{C}$)} \\
                                                      &                                                   &  \\
    \multirow{4}[1]{*}{H2}                            & \multirow{2}[0]{*}{Nula}                          & \multirow{2}[0]{*}{H2-0: IC semiautomática (MF$_{AB}$) $\leq$ manual (MF$_{AB}$)} \\
                                                      &                                                   &  \\
                                                      & \multirow{2}[1]{*}{Alternativa}                   & \multirow{2}[1]{*}{H2-1: IC semiautomática (MF$_{AB}$) $>$ IC manual (MF$_{AB}$)} \\
                                                      &                                                   &  \\
    \bottomrule
    \end{tabular}%
  \label{tab:addlabel}%
   \fonte{Elaborada pelo autor.}
\end{table}%

\section{Variáveis do Estudo} \label{AvVariáveis}

Como variável independente esta pesquisa identifica o modelo de diagrama baseado em feature, já as variáveis nominais são representadas através da experiência, habilidade, e do grau de escolarização dos participantes, demostrando as questões de avaliação que incidem sobre as vaiáveis dependentes: Integração Correta (IC) e Esforço de Integração (EI) que norteiam esta pesquisa \cite{wohlin2012, farias2015, oliveira2008b}. Esforço de integração, “EI” representa o tempo médio (minutos) em que os participantes levaram para solucionar as questões de integração entre os modelos propostos.  A integração correta, “IC” demonstra  a assertividade, ou não da questão em analise e identifica o modelo determinado como ideal na pesquisa. Em síntese, verifica se o modelo retornado como saída, isto é, o modelo pretendido, MF$_{AB}$, é correto ou incorreto. A Tabela \ref{VariáveisEstudo} mostra o resumo das variáveis investigadas neste estudo.

A variável dependente na primeira hipótese é o Esforço de Integração exibido na Tabela \ref{VariáveisEstudo}, que inclui o esforço global para integrar dois modelos de entrada, MF$_{R}$ e MF$_{C}$, considerado o tempo para aplicar as técnicas de integração e detecção de diferenças sintáticas, semânticas e estruturais entre os modelos comparados, assim como o tempo para resolver os conflitos que surgem durante o processo de integração, no intuito de produzir o modelo pretendido, MF$_{AB}$, sendo esta variável medida em minutos. O motivo principal pelo qual investiga-se esta variável é determinado por ser uma das tarefas mais importantes realizadas pelos analistas e desenvolvedores para integrar dois modelos de entrada em configurações realistas \cite{wohlin2012, farias2015}. A análise desta variável permite medir o impacto sobre a variável, independente em cada uma das questões elencadas, comparando os valores (em minutos) assumidos por essas variáveis. Pode-se também entender como as técnicas de integração analisadas tendem a ser mais eficientes na obtenção de seus resultados desejados.

\begin{table}[htbp]
  \centering
  \caption{Variáveis do estudo.}\label{VariáveisEstudo}
    \begin{tabular}{lll}
    \toprule
    \multicolumn{2}{l}{\textbf{Variável}}                                                       & \textbf{Escala} \\
    \midrule
    \multirow{4}[1]{*}{Variáveis independentes}  & Principal                                    & Nominal: {analistas, desenvolvedores } \\
                                                 & Experiência                                  & Nominal: {Iniciante, Experiente} \\
                                                 & Habilidade                                   & Nominal: {Baixa, Alta} \\
                                                 & Educação                                     & Nominal: {Graduação, Pós-graduação} \\
    \multirow{2}[1]{*}{Variáveis dependentes}    & Corretude                                    & Ordinal: {0 ou 1 } \\
                                                 & Esforço                                      & Intervalo [0..60] \\
    \bottomrule
    \end{tabular}%
  \label{tab:addlabel}%
  \fonte{Elaborada pelo autor.}
\end{table}%

A variável independente na segunda hipótese é a Integração Correta do modelo. Considerando o primeiro, modelo integrado produzido é correto se este estiver em conformidade com as solicitações de mudanças elencadas no experimento, isto é, modelo de \textit{features} integrado é igual ao modelo de \textit{features} pretendido, MF$_{I}$ $=$ MF$_{AB}$, onde a correção total de uma integração é parcialmente assegurada. O modelo integrado poder ser classificado como Correto, considerando que todas as mudanças elencadas foram realizadas, neste caso a variável assume o valor igual 1 e Incorreto quando alguma mudança não fora realizada, a variável assume o valor igual 0. Pretende-se assim avaliar qual o impacto resultante da técnica semiautomática, visto que na ocorrência de conflitos os mesmo são indicados aos analistas e desenvolvedores, os quais tem a competência de tomada de decisão, apoiados pelo protótipo, FMIT, que exibe informações dos conflitos sejam estes semânticos sintáticos ou estruturais,  determinando assim a futura integração do modelo pretendido. Com isso em mente, procura-se analisar a qualidade dos modelos produzidos, a efetividade e a pertinência do conhecimento agregado, buscando assim elevar o nível de eficácia das integrações para ser capaz de reduzir os custos, aumentar a qualidade, e produtividade dos modelos de \textit{features}.

\section{Contexto do Experimento e Seleção dos Participantes} \label{AvContexto}

Com o intuito de mensurar o esforço e a corretude das integrações de modelos de \textit{features}, aplicou-se um experimento para avaliar como os analistas e desenvolvedores realizam esta tarefa, ou seja, a apuração das principais dificuldades durante o processo de integração avaliando-se o desempenho, assim como efetuar um comparativo entre os modelos produzidos, visto assegurar as mudanças das funcionalidades elencadas.  Para isso, foram criados seis cenários de avaliação, sendo que cada cenário compreende dois modelos, identificados pelo conjunto M = {(MF1R; MF1C)}; {( MF2R; MF2C)}; ... {(MF6R; MF6C)}, onde MF$_{R}$  indica o primeiro modelo de entrada e MF$_{C}$ , o segundo modelo e cada cenário é identificado pelos números de 1 a 6. Além disso, cada cenário apresenta uma descrição das funcionalidades estabelecidas para conduzir os analistas e desenvolvedores na produção do modelo de \textit{features} desejado.  

A Tabela \ref{CenárioExperimento} apresenta as atividades, bem como os cenários de evolução aplicados a este experimento. Para o desenvolvimento do experimento utilizou-se 06 atividades elaboradas para verificar o esforço e habilidade dos analistas e desenvolvedores na resolução das atividades conflitantes, sejam estas sintáticas, semânticas ou estruturais. Os cenários elaborados na Tabela \ref{CenárioExperimento} apresentam uma pequena descrição destes e os elementos contidos em cada modelo de \textit{features}, isto é, o número de \textit{features}, o número de relacionamentos e o número de conflito quando existir entre os modelos que virão a ser integrados, bem como o número total dos elementos que compõem o modelo. Os modelos de entrada, 06 modelos  MF$_{R}$ e 06 modelos  MF$_{C}$, totalizando em 12 modelos produziram um total de 205 elementos entre \textit{features} e relacionamentos, isto é notações.

\begin{table}[htbp]
  \centering
  \caption{Cenário do experimento.} \label{CenárioExperimento}
    \begin{tabular}{llllllllll}
    \toprule
    \multicolumn{1}{c}{\multirow{2}[2]{*}{Cenário de Evolução}} & \multicolumn{4}{c}{ MFR}                                                                                                                                                  & \multicolumn{5}{c}{MFC} \\
                                             & \textbf{Nome}                            & \multicolumn{1}{c}{\textbf{NF}}          & \multicolumn{1}{c}{\textbf{NR}}          & \multicolumn{1}{c}{\textbf{NE}}          & \multicolumn{1}{r}{\textbf{Nome}}        & \multicolumn{1}{c}{\textbf{NF}}          & \multicolumn{1}{c}{\textbf{NR}}          & \multicolumn{1}{c}{\textbf{NE}}          & \multicolumn{1}{c}{\textbf{NC}} \\
    \midrule
    01- Não sofre mudanças.                  & MF1R                                     & \multicolumn{1}{c}{7}                    & \multicolumn{1}{c}{4}                    & \multicolumn{1}{c}{11}                   & \multicolumn{1}{r}{MF1C}                 & \multicolumn{1}{c}{7}                    & \multicolumn{1}{c}{4}                    & \multicolumn{1}{c}{11}                   & \multicolumn{1}{c}{0} \\
    02- Mudanças sintáticas.                 & MF2R                                     & \multicolumn{1}{c}{10}                   & \multicolumn{1}{c}{8}                    & \multicolumn{1}{c}{18}                   & \multicolumn{1}{r}{MF2C}                 & \multicolumn{1}{c}{10}                   & \multicolumn{1}{c}{8}                    & \multicolumn{1}{c}{18}                   & \multicolumn{1}{c}{4} \\
    03- Mudanças estruturais.                & MF3R                                     & \multicolumn{1}{c}{14}                   & \multicolumn{1}{c}{12}                   & \multicolumn{1}{c}{26}                   & \multicolumn{1}{r}{MF3C}                 & \multicolumn{1}{c}{14}                   & \multicolumn{1}{c}{12}                   & \multicolumn{1}{c}{26}                   & \multicolumn{1}{c}{1} \\
    04- Mudanças de semânticas.              & MF4R                                     & \multicolumn{1}{c}{13}                   & \multicolumn{1}{c}{11}                   & \multicolumn{1}{c}{24}                   & \multicolumn{1}{r}{MF4C}                 & \multicolumn{1}{c}{13}                   & \multicolumn{1}{c}{11}                   & \multicolumn{1}{c}{24}                   & \multicolumn{1}{c}{6} \\
    05- Inclusão de novas \textit{features}.          & MF5R                                     & \multicolumn{1}{c}{7}                    & \multicolumn{1}{c}{5}                    & \multicolumn{1}{c}{12}                   & \multicolumn{1}{r}{MF5C}                 & \multicolumn{1}{c}{3}                    & \multicolumn{1}{c}{2}                    & \multicolumn{1}{c}{5}                    & \multicolumn{1}{c}{0} \\
    06- Mudanças sintáticas e semânticas.    & MF6R                                     & \multicolumn{1}{c}{9}                    & \multicolumn{1}{c}{6}                    & \multicolumn{1}{c}{15}                   & \multicolumn{1}{r}{MF6C}                 & \multicolumn{1}{c}{9}                    & \multicolumn{1}{c}{6}                    & \multicolumn{1}{c}{15}                   & \multicolumn{1}{c}{7} \\
    Total de Elementos                       & \multicolumn{4}{c}{106}                                                                                                                                                   & \multicolumn{5}{c}{99} \\
    \midrule
    Legenda                                  &                                          &                                          &                                          &                                          &                                          &                                          &                                          &                                          &  \\
    \multicolumn{10}{l}{NF: nº de \textit{features}, NR: nº de relacionamentos, } \\
    \multicolumn{10}{l}{NE: nº de elementos, NC: nº de conflitos.} \\
    \end{tabular}%
  \label{tab:addlabel}%
  \fonte{Elaborada pelo autor.}
\end{table}%

Além dos modelos para integração, os participantes receberam um questionário com perguntas para obter dados inerentes ao perfil, tempo de experiência, formação acadêmica, sexo, idade entre outros. Desta forma, será possível realizar uma análise dos dados dos participantes.

Após a definição das funcionalidades a serem tratadas: (1) sintática; (2) semântica e (3) estrutural, elaborou-se as atividades a serem realizadas pelos participantes, conforme a Tabela \ref{CenárioExperimento}, sendo que cada atividade foi definida para um fim específico. As atividades foram subdivididas em duas tarefas, sendo uma executada manualmente, isto é, sem o auxílio do protótipo, FMIT, e a outra com o apoio do protótipo, FMIT.  As questões foram separadas em dois grupos cada uma contendo três modelos de \textit{features} para análise dos participantes. Formularam-se os seguintes conjuntos de questões: AQ1 = ({Questão 1, Questão 2, e Questão 4})  e AQ2 = ({Questão 3, Questão 5, e Questão 6}) , sendo estes conjuntos distribuídos entre os participantes conforme demostrada na Tabela \ref{questõesPorParticipante}, intercalando entre os participantes o agrupamento das questões, executando ora de forma manual e ora de forma semiautomática assim buscando equilibrar as questões entre os participantes. O experimento foi conduzido com dez participantes na maioria alunos do mestrado de Computação Aplicada da Unisnos.

\begin{table}[htbp]
  \centering
  \caption{Distribuição das questões por participante.} \label{questõesPorParticipante}
    \begin{tabular}{lll}
    \toprule
    \multicolumn{1}{r}{\textbf{Participantes}} & \multicolumn{1}{c}{\textbf{Agrupamento Questões   AQ1}} & \multicolumn{1}{c}{\textbf{Agrupamento Questões   AQ2}} \\
    \midrule
    \multicolumn{1}{c}{P01} & \multicolumn{1}{c}{MS-FMIT } & \multicolumn{1}{c}{SC-FMIt} \\
    \multicolumn{1}{c}{P02} & \multicolumn{1}{c}{SC-FMIT}  & \multicolumn{1}{c}{MS-FMIt} \\
    \multicolumn{1}{c}{P03} & \multicolumn{1}{c}{MS-FMIT } & \multicolumn{1}{c}{SC-FMIt} \\
    \multicolumn{1}{c}{P04} & \multicolumn{1}{c}{SC-FMIT}  & \multicolumn{1}{c}{MS-FMIt} \\
    \multicolumn{1}{c}{P05} & \multicolumn{1}{c}{MS-FMIT } & \multicolumn{1}{c}{SC-FMIt} \\
    \multicolumn{1}{c}{P06} & \multicolumn{1}{c}{SC-FMIT}  & \multicolumn{1}{c}{MS-FMIt} \\
    \multicolumn{1}{c}{P07} & \multicolumn{1}{c}{MS-FMIT } & \multicolumn{1}{c}{SC-FMIt} \\
    \multicolumn{1}{c}{P08} & \multicolumn{1}{c}{SC-FMIT}  & \multicolumn{1}{c}{MS-FMIt} \\
    \multicolumn{1}{c}{P09} & \multicolumn{1}{c}{MS-FMIT } & \multicolumn{1}{c}{SC-FMIt} \\
    \multicolumn{1}{c}{P10} & \multicolumn{1}{c}{SC-FMIT}  & \multicolumn{1}{c}{MS-FMIt} \\
    \midrule
    Legenda                &                        &  \\
    \multicolumn{3}{l}{MS-FMIT : Manual- sem auxílio protótipo.} \\
    \multicolumn{3}{l}{SC-FMIT: Semiautomático- com auxílio protótipo.} \\
    \end{tabular}%
  \label{tab:addlabel}%
  \fonte{Elaborada pelo autor.}
\end{table}%

\begin{figure}[ht]
	\caption{Processo experimental.} \label{processoExperimento}
	\label{fig:bolha}
	\centering%
	\begin{minipage}{.8\textwidth}
		\includegraphics[width=\textwidth]{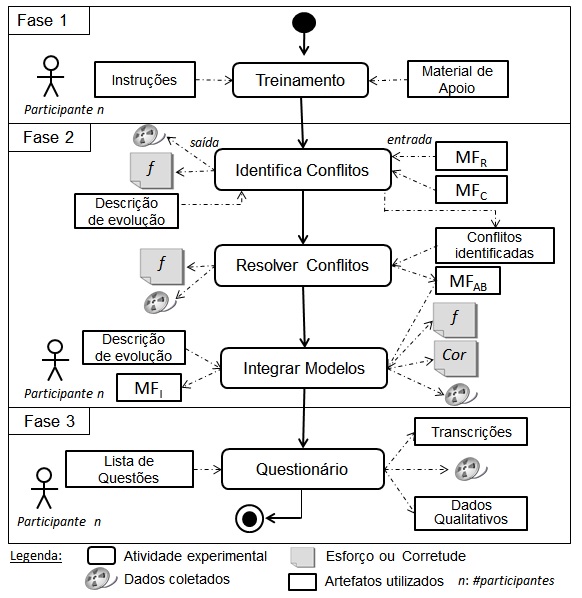}
		\fonte{Adaptado de Farias (2015).}
	\end{minipage}
\end{figure}

Ao elaborar o experimento para integração de modelos de \textit{features}, houve a preocupação em alinhar o nível de conhecimento dos participantes sobre as técnicas de modelagem de \textit{features}, visto que os participantes não tinham nenhum conhecimento sobre a aplicação desta. A condução do processo para realização do experimento ocorreu em três fases, conforme é exibido na Figura \ref{processoExperimento}. As atividades são descritas como segue:

\textbf{Treinamento.} Todos os participantes receberam treinamento para assegurar que adquiriram familiaridade necessária com as técnicas de integração de modelos. Nesta etapa é demostrada uma apresentação contendo os passos para realizar a integração de forma manual sem o apoio do protótipo, FMIT, com auxilio de uma documentação, explicitando os casos desejados, em conjunto com um pequeno tutorial ilustrando os modelos de \textit{features}, suas notações e configurações de relacionamento. Por fim é demostrado o uso do protótipo, FMIT, com um exemplo  de uma integração de modelos \textit{features}.

A próxima etapa consiste na aplicação do experimento entre os participantes, na perspectiva de analistas e desenvolvedores com a finalidade de colocar em prática a técnica proposta no Capítulo \ref{AspectosDeImplementação}, facilitando na resolução dos conflitos assim como na redução dos esforços aplicados nos casos propostos, bem como na condução em realizar a integração pretendida. Os participantes realizam individualmente todas as atividades para evitar qualquer ameaça ao processo experimental. 

\textbf{Detectar Conflitos.} A segunda fase constitui em analisar os modelos de entrada MF$_{R}$ e MF$_{C}$ de cada cenário com base nas descrições de mudanças (Tabela \ref{CenárioExperimento}) licitadas em cada questão os quais definem como os elementos do modelo MF$_{R}$ foram alterados. Os participantes detectam os conflitos. A mediada do esforço de detecção (tempo em minutos) foi coletada durante esta atividade, assim como uma lista dos conflitos identificados, sendo estes registros gravados em vídeo e áudio. Os registros serão utilizados para realizar as análises qualitativas. 

\textbf{Resolver Conflitos.} Os participantes devem resolver os conflitos conforme as solicitações elencadas em cada questão para produzir o modelo pretendido, MF$_{AB}$. O esforço da resolução também é medido (tempo em minutos) bem como o vídeo e áudios são gravados.

\textbf{Integrar Modelos.} Esta atividade constitui em realizar a integração dos modelos MF$_{R}$ e MF$_{C}$ de cada cenário com base na identificação e resolução dos conflitos. Os modelos são compostos produzindo assim o modelo integrado, MF$_{I}$. A medida do esforço de aplicação (tempo em minutos) é coletada durante esta atividade, armazenados em áudio e vídeo. Nesta fase também é realizada uma análise entre o modelo produzido, isto é, o modelo integrado MF$_{I}$, e modelo pretendido, MF$_{AB}$, verificando assim se o mesmo encontra-se correto ou não. 

\textbf{Questionário.} Os participantes preenchem um questionário, o que permite coletar informações sobre a experiência profissional, formação acadêmica, experiência em modelagem e desenvolvimento, sexo, idade, etc.. 

\textbf{Material.} Os modelos utilizados neste experimento foram diagramas de \textit{features} com cerca de 10 \textit{features}, 7 relacionamentos, 3 níveis de profundidade, e 3 conflitos em média por modelo de \textit{features} apresentado.  A aplicação de pequenos modelos ocorre por conta da limitação de tempo do experimento controlado, bem como mantém sob controle as variáveis em análise.

Para realizar este experimento, foram convidados 10 participantes todos com formação superior completa, dos cursos de Ciência da Computação, Sistemas de Informação, Licenciatura em Informática, Análise de Sistemas e Jogos Digitais. Todos os participantes estão cursando pós-graduação em nível de mestrado, conforme detalhado na Tabela \ref{escolaridade}.

\begin{table}[htbp]
  \centering
  \caption{Grau de escolaridade dos participantes.} \label{escolaridade}
    \begin{tabular}{lcc}
    \toprule
    \textbf{Curso}         & \textbf{Número Participantes} & \multicolumn{1}{l}{\textbf{Mestrando}} \\
    \midrule
    Ciência da Computação  & 3                      & 3 \\
    Sistemas de Informação & 2                      & 2 \\
    Licenciatura em Informática & 1                      & 1 \\
    Análise de Sistemas    & 2                      & 2 \\
    Jogos Digitais         & 2                      & 2 \\
    \bottomrule
    \end{tabular}%
  \label{tab:addlabel}%
  \fonte{Elaborada pelo autor.}
\end{table}%

Desses 10 voluntários, 3 atuam como estudante/pesquisador em nível de mestrado,  5 atuam como desenvolvedor de sistemas, 01 atua como analista de negócio e por último,  1 atua como analista de qualidade, conforme descrito na Tabela \ref{ocupacao}. A Tabela \ref{idade} ilustra a faixa etária dos participantes, que compreende 3 entre 20 e 25 anos,  6 entre 26 a 30 anos e 1 com 31 anos de idade.

\begin{table}[!ht]
  \centering
  \caption{Ocupação dos participantes.} \label{ocupacao} 
    \begin{tabular}{lc}
    \toprule
    \textbf{Cargo}         & \textbf{Número Participantes} \\
    \midrule
    Desenvolvedor de sistemas & 5 \\
    Analista de negócios   & 1 \\
    Analista de qualidade  & 1 \\
    Estudante/pesquisador  & 3 \\
    \bottomrule
    \end{tabular}%
  \label{tab:addlabel}%
  \fonte{Elaborada pelo autor.}
\end{table}%

\begin{table}[!ht]
  \centering
  \caption{Faixa etária dos participantes.} \label{idade}
    \begin{tabular}{lc}
    \toprule
    \textbf{Idade}         & \textbf{Número Participantes} \\
    \midrule
    Entre 20 e 25 anos   & 3 \\
    Entre 26 e 30 anos   & 6 \\
    31 anos              & 1 \\
    \bottomrule
    \end{tabular}%
  \label{tab:addlabel}%
  \fonte{Elaborada pelo autor.}
\end{table}%

Sobre a experiência em desenvolvimento de software, 2 participantes possuem mais de 8 anos, 3 participantes possuem entre 7 e 8 anos,  2 participantes possuem entre 5 e 6 anos, 2 participantes possuem entre 3 a 4 anos e por fim 1 participante tem menos de 2 anos de experiência, conforme ilustrado na Figura 31.

\begin{figure}[ht]
	\caption{Experiência em desenvolvimento de software.} \label{figExperiencia}
	\label{fig:bolha}
	\centering%
	\begin{minipage}{.5\textwidth}
		\includegraphics[width=\textwidth]{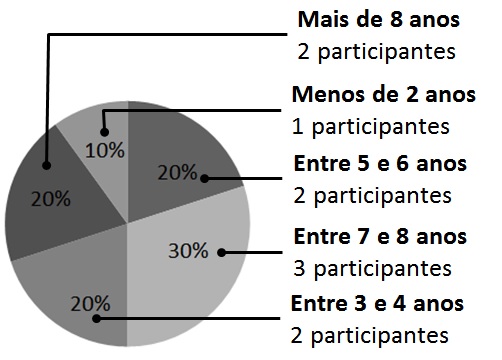}
		\fonte{Elaborada pelo autor.}
	\end{minipage}
\end{figure}

Em relação à experiência de modelagem de software, 1 participante tem mais de 8 anos, 4 participantes possuem entre 7 e 8 anos,  1 participante tem entre 5 e 6 anos, 2 participantes possuem entre 3 a 4 anos e por fim 2 participantes possuem menos de 2 anos de experiência, conforme ilustrado na Figura \ref{figExperiencia}.

\begin{figure}[ht]
	\caption{Experiência em modelagem de software.} \label{figModelagem}
	\label{fig:bolha}
	\centering%
	\begin{minipage}{.5\textwidth}
		\includegraphics[width=\textwidth]{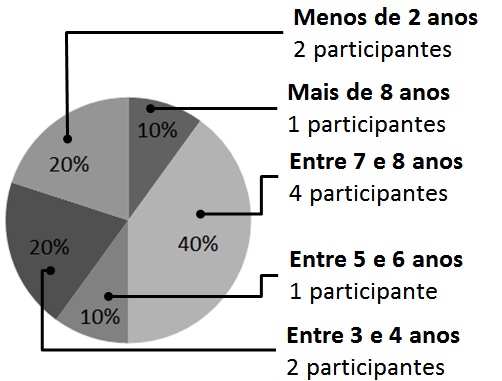}
		\fonte{Elaborada pelo autor.}
	\end{minipage}
\end{figure}

Ao analisar o perfil dos participantes deste experimento, identificou-se que todos os selecionados possuem relação direta com a área de tecnologia, mais especificamente em desenvolvimento, análise e modelagem de software, tanto em sua formação acadêmica como em suas áreas de atuação profissional. Além disso, verificou-se que a maioria dos participantes possui entre 26 e 30 anos de idade, tem amplo conhecimento em desenvolvimento de software, ou seja, 50\% dos participantes possui mais de 5 anos de experiência, outro fator decisivo para este experimento se deve a que 80\% dos participantes tem mais de 3 anos de experiência em modelagem, porém nenhum destes tinha conhecimento sobre modelagem de \textit{featuares}, caracterizando, assim, que a amostra selecionada possui um perfil adequado para a execução do experimento.

\section{Análise, Resultados e Discussão dos Dados Obtidos} \label{AvAnálise}

Após a submissão dos resultados inerentes ao questionário através dos participantes, iniciou a verificação dos dados, buscando investigar os impactos relativos ao esforço necessário para interpretar os modelos de \textit{features} e realizar sua integração, bem como quantificar a corretude das respostas. Os métodos estatísticos balizaram esta pesquisa, onde buscou-se examinar todas as ocorrências dentro de uma população, isto é, a amostra a ser investigada para evidenciar o seu comportamento na aplicação da técnica de modelagem de \textit{features}. Os resultados coletados foram submetidos à ferramenta RStudio \footnote{https://www.rstudio.com/}, tornando assim mais simples a visualização das variações dos resultados e sua parametrização.

Conforme descrito na Seção \ref{AvContexto}, este experimento consta de 12 modelos de \textit{features} (MF$_{R}$ e MF$_{C}$) e em seguida os modelos são compostos, produzindo 6 modelos de \textit{features}, MF$_{I}$. Esses modelos foram analisados e corrigidos pelos participantes, assim produzindo o modelo pretendido, bem como se propõe ao mensurar o esforço aplicado na produção dos modelos. A Tabela \ref{resultadoManual} exibe os resultados manuais, isto é, sem o uso do protótipo individualizado por participantes para cada cenário avaliado, onde é possível visualizar o tempo que cada participante utilizou na construção dos modelos, assim como verificar se o modelo produzido é o correto ou o incorreto. 

\begin{table}[htbp]
  \centering
  \caption{Resultados individuais por participante- manual.} \label{resultadoManual}
    \begin{tabular}{ccccccccccccc}
    \toprule
    \multicolumn{13}{l}{\textbf{Manual – sem uso do protótipo}} \\
    \midrule
    \multirow{2}[2]{*}{\begin{sideways}Participante\end{sideways}} & \multicolumn{2}{c}{\textbf{Cenário 1}} & \multicolumn{2}{c}{\textbf{Cenário 2}} & \multicolumn{2}{c}{\textbf{Cenário 3}} & \multicolumn{2}{c}{\textbf{Cenário 4}} & \multicolumn{2}{c}{\textbf{Cenário 5}} & \multicolumn{2}{c}{\textbf{Cenário 6}} \\
               & \begin{sideways}Esforço (min.)\end{sideways} & \begin{sideways}Corretude\end{sideways} & \begin{sideways}Esforço (min.)\end{sideways} & \begin{sideways}Corretude\end{sideways} & \begin{sideways}Esforço (min.)\end{sideways} & \begin{sideways}Corretude\end{sideways} & \begin{sideways}Esforço (min.)\end{sideways} & \begin{sideways}Corretude\end{sideways} & \begin{sideways}Esforço (min.)\end{sideways} & \begin{sideways}Corretude\end{sideways} & \begin{sideways}Esforço (min.)\end{sideways} & \begin{sideways}Corretude\end{sideways} \\
    \midrule
    P01        & 5          & 1          & 6          & 1          & -          & -          & 8          & 0          & -          & -          & -          & - \\
    P02        & -          & -          & -          & -          & 5          & 0          & -          & -          & 4          & 1          & 7          & 0 \\
    P03        & 5          & 1          & 7          & 0          & -          & -          & 8          & 0          & -          & -          & -          & - \\
    P04        & -          & -          & -          & -          & 5          & 1          & -          & -          & 4          & 1          & 8          & 0 \\
    P05        & 6          & 1          & 7          & 0          & -          & -          & 9          & 1          & -          & -          & -          & - \\
    P06        & -          & -          & -          & -          & 6          & 0          & -          & -          & 5          & 1          & 12         & 1 \\
    P07        & 5          & 1          & 5          & 0          & -          & -          & 12         & 0          & -          & -          & -          & - \\
    P08        & -          & -          & -          & -          & 5          & 0          & -          & -          & 5          & 1          & 8          & 0 \\
    P09        & 5          & 1          & 5          & 1          & -          & -          & 7          & 1          & -          & -          & -          & - \\
    P10        & -          & -          & -          & -          & 5          & 0          & -          & -          & 6          & 1          & 6          & 0 \\
    \bottomrule
    \end{tabular}%
  \label{tab:addlabel}%
  \fonte{Elaborada pelo autor.}
\end{table}%

A Tabela \ref{resultadoSemiautomático} exibe os resultados semiautomáticos, ou seja, com o uso do protótipo individualizado por participantes para cada cenário avaliado, onde é possível visualizar o tempo que cada participante utilizou na construção dos modelos, assim como verificar se o modelo produzido é o correto ou incorreto. 

\begin{table}[htbp]
  \centering
  \caption{Resultados individuais por participante- semiautomático.} \label{resultadoSemiautomático}
    \begin{tabular}{ccccccccccccc}
    \toprule
    \multicolumn{13}{l}{\textbf{Semiautoático – sem uso do protótipo}} \\
    \midrule
    \multirow{2}[2]{*}{\begin{sideways}Participante\end{sideways}} & \multicolumn{2}{c}{\textbf{Cenário 1}} & \multicolumn{2}{c}{\textbf{Cenário 2}} & \multicolumn{2}{c}{\textbf{Cenário 3}} & \multicolumn{2}{c}{\textbf{Cenário 4}} & \multicolumn{2}{c}{\textbf{Cenário 5}} & \multicolumn{2}{c}{\textbf{Cenário 6}} \\
               & \begin{sideways}Esforço (min.)\end{sideways} & \begin{sideways}Corretude\end{sideways} & \begin{sideways}Esforço (min.)\end{sideways} & \begin{sideways}Corretude\end{sideways} & \begin{sideways}Esforço (min.)\end{sideways} & \begin{sideways}Corretude\end{sideways} & \begin{sideways}Esforço (min.)\end{sideways} & \begin{sideways}Corretude\end{sideways} & \begin{sideways}Esforço (min.)\end{sideways} & \begin{sideways}Corretude\end{sideways} & \begin{sideways}Esforço (min.)\end{sideways} & \begin{sideways}Corretude\end{sideways} \\
    \midrule
    P01        & -          & -          & -          & -          & 2          & 1          & -          & -          & 0          & 1          & 2          & 1 \\
    P02        & 0          & 1          & 1          & 1          & -          & -          & 2          & 1          & -          & -          & -          & - \\
    P03        & -          & -          & -          & -          & 3          & 1          & -          & -          & 0          & 1          & 2          & 1 \\
    P04        & 0          & 1          & 3          & 1          & -          & -          & 4          & 1          & -          & -          & -          & - \\
    P05        & -          & -          & -          & -          & 2          & 1          & -          & -          & 0          & 1          & 4          & 1 \\
    P06        & 0          & 1          & 2          & 1          & -          & -          & 3          & 1          & -          & -          & -          & - \\
    P07        & -          & -          & -          & -          & 4          & 1          & -          & -          & 0          & 1          & 5          & 0 \\
    P08        & 0          & 1          & 3          & 1          & -          & -          & 4          & 0          & -          & -          & -          & - \\
    P09        & -          & -          & -          & -          & 3          & 1          & -          & -          & 0          & 1          & 5          & 1 \\
    P10        & 0          & 1          & 1          & 1          & -          & -          & 2          & 1          & -          & -          & -          & - \\
    \bottomrule
    \end{tabular}%
  \fonte{Elaborada pelo autor.}
\end{table}%

Os dados para análise qualitativa foram coletadas das seguintes fontes: registros do questionário, áudio, vídeo, transcrições e comentários, possibilitando obter evidências para complementar os resultados quantitativos e posteriormente fundamentar as conclusões a partir destas informações. Para a análise quantitativa foi utilizada a estatística descritiva para analisar sua distribuição e para os testes de hipótese, aplicou-se a inferência estatística. O total de amostra deste experimento é igual a 60, ou seja, 30 amostras manuais, sem interferência do protótipo e 30 amostras semiautomáticas, com suporte do protótipo, FMIT. Para analisar a normalidade das amostras, optou-se em executar dois testes: \textit{Kolmogorov-Smirnov}, utilizado para grandes amostras (n $\geq$ 30) e \textit{Shapiro-Wilk} que apresenta um melhor desempenho em amostras reduzidas (n $<$30) (LEVINE et. al., 2005). 

A Tabela \ref{testeNormalidade} apresenta o resultado do nível descritivo - a estatística de\textit{ Kolmogorov-Smirnov}, com um nível de significância de \textit{Lilliefors} para teste de normalidade, e \textit{Shapiro-Wilk}, os resultados retornados da significância, também conhecida como valor p < 0,05 para ambos os testes nas amostras analisadas. Assim, pode-se afirmar que nível de significância de 5\% não provém de uma população normal, tanto pelos indicadores de esforço como os de corretude.

A Tabela \ref{esforcoAgrupado}, assim como a Tabela \ref{corretudeAgrupado}, apresenta um os resultados agrupado por cenário de avaliação, com base nos itens avaliados: média, mediana, desvio padrão, máximo, mínimo primeiro quartil, e terceiro quartil, separadas pelo esforço e corretude. Nos cenários Q4 e Q6 verifica-se que foi necessário aplicar o maior esforço tanto manual como semiautomático, esta questão neste experimento apresenta o maior grau de dificuldade, pois em sua elaboração foram inseridos conflitos semânticos e sintáticos ou ambos.  Quanto a variável corretude, no cenário Q1 e Q5, verificou-se que todos os participantes acertaram, tanto na forma manualmente como na semiautomática, cabendo uma análise mais profunda nesta questão mo que refere-se ao esforço aplicado, que manualmente é de 5 minutos e 2 segundos para Q1 e 4 minutos e 8 segundos para Q5. Já com o uso do protótipo, FMIT, o tempo consumido é zero e isto ocorre em virtude da técnica detectar que no primeiro cenário não há qualquer alteração entre os modelos de \textit{features} comparados , ou seja, são idênticos bem como  no quinto cenário ocorre somente a inclusão de novas \textit{features} e, nestes casos, a integração ocorre de forma automática sem há intervenção dos participantes.

\begin{table}[htbp]
  \centering
  \caption{Teste de normalidade.} \label{testeNormalidade}
    \begin{tabular}{lllllll}
    \toprule
    \multirow{2}[2]{*}{\textbf{Esforço}}   & \multicolumn{3}{c}{\textbf{Kolmogorov-Smirnov a}}                                                                        & \multicolumn{3}{c}{\textbf{Shapiro-Wilk}} \\
                                           & \multicolumn{1}{c}{\textbf{Estatística}} & \multicolumn{1}{c}{\textbf{df}}        & \multicolumn{1}{c}{\textbf{Sig.}}      & \multicolumn{1}{c}{\textbf{Estatística}} & \multicolumn{1}{c}{\textbf{df}}        & \multicolumn{1}{c}{\textbf{Sig.}} \\
    \midrule
    Semiautomático                         & \multicolumn{1}{c}{0,206}              & \multicolumn{1}{c}{30}                 & \multicolumn{1}{c}{0,002}              & \multicolumn{1}{c}{0,881}              & \multicolumn{1}{c}{30}                 & \multicolumn{1}{c}{0,003} \\
    Manual                                 & \multicolumn{1}{c}{0,219}              & \multicolumn{1}{c}{30}                 & \multicolumn{1}{c}{0,001}              & \multicolumn{1}{c}{0,814}              & \multicolumn{1}{c}{30}                 & \multicolumn{1}{c}{0,000} \\
    \textbf{Corretude}                     & \multicolumn{1}{c}{\textbf{Estatística}} & \multicolumn{1}{c}{\textbf{df}}        & \multicolumn{1}{c}{\textbf{Sig.}}      & \multicolumn{1}{c}{\textbf{Estatística}} & \multicolumn{1}{c}{\textbf{df}}        & \multicolumn{1}{c}{\textbf{Sig.}} \\
    \midrule
    Semiautomático                         & \multicolumn{1}{c}{0,354}              & \multicolumn{1}{c}{30}                 & \multicolumn{1}{c}{0,000}              & \multicolumn{1}{c}{0,637}              & \multicolumn{1}{c}{30}                 & \multicolumn{1}{c}{0,000} \\
    Manual                                 & \multicolumn{1}{c}{0,537}              & \multicolumn{1}{c}{30}                 & \multicolumn{1}{c}{0,000}              & \multicolumn{1}{c}{0,275}              & \multicolumn{1}{c}{30}                 & \multicolumn{1}{c}{0,000} \\
    \midrule
    Legenda                                & \multicolumn{6}{l}{a. Correlação de Significância de Lillieforrs} \\
    \end{tabular}%
    \fonte{Elaborada pelo autor.}
\end{table}%

\begin{table}[htbp]
  \centering
  \caption{Resultados agrupados por cenário - Esforço.} \label{esforcoAgrupado}
    \begin{tabular}{cclccrrrr}
\cmidrule{4-9}                                           &                                        &                                        & \multicolumn{1}{l}{\begin{sideways}Cenário 1\end{sideways}} & \multicolumn{1}{l}{\begin{sideways}Cenário 2\end{sideways}} & \multicolumn{1}{l}{\begin{sideways}Cenário 3\end{sideways}} & \multicolumn{1}{l}{\begin{sideways}Cenário 4\end{sideways}} & \multicolumn{1}{l}{\begin{sideways}Cenário 5\end{sideways}} & \multicolumn{1}{l}{\begin{sideways}Cenário 6\end{sideways}} \\
    \midrule
    \multicolumn{3}{l}{Participantes}                                                                                        & 10                                     & 10                                     & \multicolumn{1}{c}{10}                 & \multicolumn{1}{c}{10}                 & \multicolumn{1}{c}{10}                 & \multicolumn{1}{c}{10} \\
    \midrule
    \multirow{7}[2]{*}{\begin{sideways}Esforço (temp\textbackslash{}min.)\end{sideways}} & \multirow{7}[2]{*}{\begin{sideways}Manual\end{sideways}} & Média                                  & 5,2                                    & 6                                      & \multicolumn{1}{c}{5,2}                & \multicolumn{1}{c}{8,8}                & \multicolumn{1}{c}{4,8}                & \multicolumn{1}{c}{8,8} \\
                                           &                                        & Mediana                                & 5                                      & 6                                      & \multicolumn{1}{c}{5}                  & \multicolumn{1}{c}{8}                  & \multicolumn{1}{c}{5}                  & \multicolumn{1}{c}{8} \\
                                           &                                        & Desvio Padrão                          & 0,4                                    & 0,9                                    & \multicolumn{1}{c}{0,4}                & \multicolumn{1}{c}{1,7}                & \multicolumn{1}{c}{0,7}                & \multicolumn{1}{c}{2} \\
                                           &                                        & Mínimo                                 & 5                                      & 5                                      & \multicolumn{1}{c}{5}                  & \multicolumn{1}{c}{7}                  & \multicolumn{1}{c}{4}                  & \multicolumn{1}{c}{6} \\
                                           &                                        & Máximo                                 & 6                                      & 7                                      & \multicolumn{1}{c}{6}                  & \multicolumn{1}{c}{12}                 & \multicolumn{1}{c}{6}                  & \multicolumn{1}{c}{12} \\
                                           &                                        & 1º Quartil                             & 5                                      & 5                                      & \multicolumn{1}{c}{5}                  & \multicolumn{1}{c}{8}                  & \multicolumn{1}{c}{4}                  & \multicolumn{1}{c}{7} \\
                                           &                                        & 3º Quartil                             & 5                                      & 7                                      & \multicolumn{1}{c}{5}                  & \multicolumn{1}{c}{9}                  & \multicolumn{1}{c}{5}                  & \multicolumn{1}{c}{8} \\
    \midrule
    \multirow{7}[2]{*}{\begin{sideways}Esforço (temp\textbackslash{}min.)\end{sideways}} & \multirow{7}[2]{*}{\begin{sideways}Semiautomático\end{sideways}} & Média                                  & 0                                      & 2                                      & \multicolumn{1}{c}{2,8}                & \multicolumn{1}{c}{3}                  & \multicolumn{1}{c}{0}                  & \multicolumn{1}{c}{3,6} \\
                                           &                                        & Mediana                                & 0                                      & 2                                      & \multicolumn{1}{c}{3}                  & \multicolumn{1}{c}{3}                  & \multicolumn{1}{c}{0}                  & \multicolumn{1}{c}{4} \\
                                           &                                        & Desvio Padrão                          & 0                                      & 0,9                                    & \multicolumn{1}{c}{0,7}                & \multicolumn{1}{c}{0,9}                & \multicolumn{1}{c}{0}                  & \multicolumn{1}{c}{1,4} \\
                                           &                                        & Mínimo                                 & 0                                      & 1                                      & 2                                      & 2                                      & \multicolumn{1}{c}{0}                  & \multicolumn{1}{c}{2} \\
                                           &                                        & Máximo                                 & 0                                      & 3                                      & 4                                      & 4                                      & \multicolumn{1}{c}{0}                  & \multicolumn{1}{c}{5} \\
                                           &                                        & 1º Quartil                             & 0                                      & 1                                      & 2                                      & 2                                      & \multicolumn{1}{c}{0}                  & \multicolumn{1}{c}{2} \\
                                           &                                        & 3º Quartil                             & 0                                      & 3                                      & 3                                      & 4                                      & 0                                      & 5 \\
    \bottomrule
    \end{tabular}%
  \fonte{Elaborada pelo autor.}
\end{table}%

\begin{table}[htbp]
  \centering
  \caption{Resultados agrupados por cenário – Corretude.} \label{corretudeAgrupado}
    \begin{tabular}{cclcccccc}
\cmidrule{4-9}                                           &                                        &                                        & \multicolumn{1}{l}{\begin{sideways}Cenário 1\end{sideways}} & \multicolumn{1}{l}{\begin{sideways}Cenário 2\end{sideways}} & \multicolumn{1}{l}{\begin{sideways}Cenário 3\end{sideways}} & \multicolumn{1}{l}{\begin{sideways}Cenário 4\end{sideways}} & \multicolumn{1}{l}{\begin{sideways}Cenário 5\end{sideways}} & \multicolumn{1}{l}{\begin{sideways}Cenário 6\end{sideways}} \\
    \midrule
    \multicolumn{3}{l}{Participantes}                                                                                        & 10                                     & 10                                     & 10                                     & 10                                     & 10                                     & 10 \\
    \midrule
    \multirow{7}[2]{*}{\begin{sideways}Corretude\end{sideways}} & \multirow{7}[2]{*}{\begin{sideways}Manual\end{sideways}} & Média                                  & 1                                      & 0,4                                    & 0,2                                    & 0,4                                    & 1                                      & 0,2 \\
                                           &                                        & Mediana                                & 1                                      & 0                                      & 0                                      & 0                                      & 1                                      & 0 \\
                                           &                                        & Desvio Padrão                          & 0                                      & 0,5                                    & 0,4                                    & 0,5                                    & 0                                      & 0,4 \\
                                           &                                        & Mínimo                                 & 1                                      & 0                                      & 0                                      & 0                                      & 1                                      & 0 \\
                                           &                                        & Máximo                                 & 1                                      & 1                                      & 1                                      & 1                                      & 1                                      & 1 \\
                                           &                                        & 1º Quartil                             & 1                                      & 0                                      & 0                                      & 0                                      & 1                                      & 0 \\
                                           &                                        & 3º Quartil                             & 1                                      & 1                                      & 0                                      & 1                                      & 1                                      & 0 \\
    \midrule
    \multirow{7}[2]{*}{\begin{sideways}Corretude\end{sideways}} & \multirow{7}[2]{*}{\begin{sideways}Semiautomático\end{sideways}} & Média                                  & 1                                      & 1                                      & 1                                      & 0,8                                    & 1                                      & 0,8 \\
                                           &                                        & Mediana                                & 1                                      & 1                                      & 1                                      & 1                                      & 1                                      & 1 \\
                                           &                                        & Desvio Padrão                          & 0                                      & 0                                      & 0                                      & 0,4                                    & 0                                      & 0,4 \\
                                           &                                        & Mínimo                                 & 1                                      & 1                                      & 1                                      & 0                                      & 1                                      & 0 \\
                                           &                                        & Máximo                                 & 1                                      & 1                                      & 1                                      & 1                                      & 1                                      & 1 \\
                                           &                                        & 1º Quartil                             & 1                                      & 1                                      & 1                                      & 1                                      & 1                                      & 1 \\
                                           &                                        & 3º Quartil                             & 1                                      & 1                                      & 1                                      & 1                                      & 1                                      & 1 \\
    \bottomrule
    \end{tabular}%
  \fonte{Elaborada pelo autor.}
\end{table}%

Além disso, a Tabela \ref{quantificacao} abstrai as informações por cenário aplicado, esforço de integração e integração correta, demostrando a media dos resultados individual das variáveis e o percentual de diferença entre cada tratamento.

\begin{table}[htbp]
  \centering
  \caption{Relação das variáveis de quantificação.} \label{quantificacao}
    \begin{tabular}{cllcc}
    \toprule
    \textbf{Questões}                      & \multicolumn{1}{c}{\textbf{Variáveis}} & \multicolumn{1}{c}{\textbf{Tratamento}} & \textbf{Média}                         & \textbf{\% Dif} \\
    \midrule
    \multirow{4}[1]{*}{Todos}              & \multicolumn{1}{c}{\multirow{2}[1]{*}{Corretude}} & Manual                                 & 0,53                                   & \multirow{2}[1]{*}{43,01} \\
                                           &                                        & Semiautomático                         & 0,93                                   &  \\
                                           & \multirow{2}[0]{*}{Esforço}            & Manual                                 & 6,36                                   & \multirow{2}[0]{*}{70,13} \\
                                           &                                        & Semiautomático                         & 1,90                                   &  \\
    \multirow{4}[0]{*}{C1}                 & \multicolumn{1}{c}{\multirow{2}[0]{*}{Corretude}} & Manual                                 & 1,0                                    & \multirow{2}[0]{*}{0,00} \\
                                           &                                        & Semiautomático                         & 1,0                                    &  \\
                                           & \multirow{2}[0]{*}{Esforço}            & Manual                                 & 5,2                                    & \multirow{2}[0]{*}{100} \\
                                           &                                        & Semiautomático                         & 0,0                                    &  \\
    \multirow{4}[0]{*}{C2}                 & \multicolumn{1}{c}{\multirow{2}[0]{*}{Corretude}} & Manual                                 & 0,40                                   & \multirow{2}[0]{*}{60,00} \\
                                           &                                        & Semiautomático                         & 1,00                                   &  \\
                                           & \multirow{2}[0]{*}{Esforço}            & Manual                                 & 6,00                                   & \multirow{2}[0]{*}{66,66} \\
                                           &                                        & Semiautomático                         & 2,00                                   &  \\
    \multirow{4}[0]{*}{C3}                 & \multicolumn{1}{c}{\multirow{2}[0]{*}{Corretude}} & Manual                                 & 0,20                                   & \multirow{2}[0]{*}{80,00} \\
                                           &                                        & Semiautomático                         & 1,00                                   &  \\
                                           & \multirow{2}[0]{*}{Esforço}            & Manual                                 & 5,20                                   & \multirow{2}[0]{*}{46,15} \\
                                           &                                        & Semiautomático                         & 2,80                                   &  \\
    \multirow{4}[0]{*}{C4}                 & \multicolumn{1}{c}{\multirow{2}[0]{*}{Corretude}} & Manual                                 & 0,40                                   & \multirow{2}[0]{*}{50,00} \\
                                           &                                        & Semiautomático                         & 0,80                                   &  \\
                                           & \multirow{2}[0]{*}{Esforço}            & Manual                                 & 8,80                                   & \multirow{2}[0]{*}{65,9} \\
                                           &                                        & Semiautomático                         & 3,00                                   &  \\
    \multirow{4}[0]{*}{C5}                 & \multicolumn{1}{c}{\multirow{2}[0]{*}{Corretude}} & Manual                                 & 1,00                                   & \multirow{2}[0]{*}{0,00} \\
                                           &                                        & Semiautomático                         & 1,00                                   &  \\
                                           & \multirow{2}[0]{*}{Esforço}            & Manual                                 & 4,80                                   & \multirow{2}[0]{*}{100} \\
                                           &                                        & Semiautomático                         & 0,00                                   &  \\
    \multirow{4}[1]{*}{C6}                 & \multicolumn{1}{c}{\multirow{2}[0]{*}{Corretude}} & Manual                                 & 0,20                                   & \multirow{2}[0]{*}{75,00} \\
                                           &                                        & Semiautomático                         & 0,80                                   &  \\
                                           & \multirow{2}[1]{*}{Esforço}            & Manual                                 & 8,82                                   & \multirow{2}[1]{*}{59,18} \\
                                           &                                        & Semiautomático                         & 3,60                                   &  \\
    \bottomrule
    \end{tabular}%
  \fonte{Elaborada pelo autor.}
\end{table}%

Igualmente, foi possível averiguar a compreensão dos modelos de \textit{features} entre participantes, devido a estes não conhecerem a técnica de modelagem de \textit{features} proposta, evidenciado ser um facilitador para análise e implementação no desenvolvimento de software, no contexto de formalizar um entendimento único entre todos. 

Realizando uma análise geral, em relação ao esforço, bem como a corretude, aplicada entre os participantes em compor os modelos de \textit{features}, apesar do fácil entendimento destes, dos quais são exemplos considerados como simples, a aplicação da técnica manual exige um esforço considerável, sendo que o tempo médio para solucionar todos os seis cenários proposto ficou em aproximadamente trinta e seis minutos, tendo uma assertividade de cinquenta três por cento.

Entretanto, a aplicação com os mesmos cenários utilizando o protótipo, FMIT, demostrou-se mais eficiente e eficaz. O esforço médio aplicado na condução de todo o experimento foi de aproximadamente doze minutos, tendo uma assertividade de noventa três por cento.

\begin{figure}[ht]
	\caption{Tempo de execução por cenário.} 
	\label{fig:tmpCenario}
	\centering%
	\begin{minipage}{.7\textwidth}
		\includegraphics[width=\textwidth]{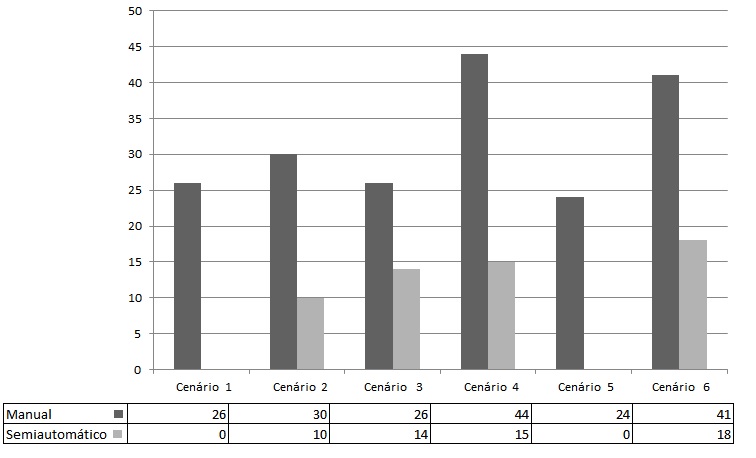}
		\fonte{Elaborada pelo autor.}
	\end{minipage}
	\fonte{Elaborada pelo autor.}
\end{figure}

A Figura \ref{fig:tmpCenario} exibe o comparativo em minutos de cada cenário deste experimento. Na figura é possível visualizar o tempo utilizado pelos participantes, com o auxilio do protótipo inferior ao tempo quando executado manualmente. Ressalta-se que apesar de comparar a aplicação de técnica manual versus a técnica semiautomática, a mesma requer a intervenção dos participantes, exceto quando é detectado pela técnica proposta que ambos os modelos são totalmente equivalentes ou quando não há qualquer similaridade entre os modelos comparados, assim aplicando a inclusão de forma automática.  Desse modo, pode afirmar que os participantes empenharam menos esforço para realizar a técnica semiautomática em relação à técnica manual.

\begin{figure}[ht]
	\caption{Corretude por experimento.} 
	\label{fig:graficoCorretude}
	\centering%
	\begin{minipage}{.7\textwidth}
		\includegraphics[width=\textwidth]{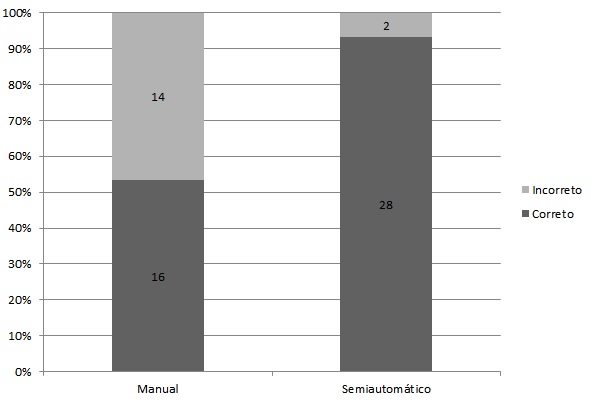}
		\fonte{Elaborada pelo autor.}
	\end{minipage}
	\fonte{Elaborada pelo autor.}
\end{figure}

A Figura \ref{fig:graficoCorretude} exibe o comparativo da corretude aplicada de forma manual em relação à corretude aplicada de forma semiautomática, sendo produzido um total de 60 modelos de \textit{features} pelos participantes, dos quais 30 modelos de \textit{features} foram produzidos de forma manual e 30 modelos de forma semiautomática. É possível visualizar que manualmente produziu-se 16 modelos de \textit{features} corretamente, conforme os requisitos estabelecidos nos cenários, contra 28 Modelos de \textit{Features} produzidos com o auxílio do protótipo, FMIT. Já dos modelos produzidos incorretamente, 14 modelos de \textit{features}  são manuais contra 2 modelos produzidos de forma semiautomática. Assim é possível concluir que a técnica proposta auxilia os analistas e desenvolvedores na detecção de conflitos, bem como melhora a precisão dos modelos integrados.

Aplicando os dados coletados com a ferramenta RStudio obteve-se os cálculos de cada variável, sendo estas subdivididas conforme o tratamento, em categorias, tendo em vista diferentes resultados incorporados a Tabela \ref{testeEstatistico}. Nesta tabela incluem-se os resultados para variáveis de quantificação, Integração Correta (IC), que apresenta a corretude para a integração entre modelos de \textit{features}, ou seja, quanto maior o índice, melhor sua capacidade de retornar a resposta correta. Igualmente a variável Esforço de Integração (EI), apresenta o tempo médio que os participantes levaram para responder os cenários de composição dos modelos de \textit{features} propostos.  O principal fato em aplicar os cálculos pela estatística descritiva com o uso da ferramenta  RStudio advém deste retornar  dos resultados de testes estáticos e testes de hipóteses: desvio padrão, mínimo e máximo, média, mediana, e por fim a aplicação dos testes de \textit{Wilcoxon}  e o \textit{McNemar}.

Os testes de normalidades de \textit{Kolmogorov-Smirnov – (Lilliefors)} e \textit{Shapiro-Wilk }\cite{razali2011, kleinner2012} executados anteriormente indicam que os dados não se encontram normalmente distribuídos, assim se aplica os testes não paramétricos. Logo, o teste de \textit{Wilcoxon} é aplicado para analisar o esforço, bem como foi utilizado o teste de \textit{McNemar} para certificar a validade da técnica proposta. Esses testes serão utilizados para analisar os testes das hipóteses abordados na Seção \ref{AvHipotese}. Assim, de posse dos dados coletados e distribuídos nas tabelas para potencializar sua visualização, será aplicada a análise de cada questão e suas hipóteses investigadas.

O teste de \textit{Wilcoxon} é um teste de hipótese estático não paramétrico, aplicado para confrontar a média de duas amostras relacionadas, ou seja, é um teste de diferença emparelhado, pode ser utilizado como alternativa ao \textit{t-test}.  O \textit{t-test} é considerado um teste de hipótese que utiliza conceitos estatísticos para rejeitar ou não uma hipótese nula, empregado para confrontar médias de amostras diferentes, normalmente o teste-t é aplicado quando segue uma distribuição normal \cite{barros2005}.

 O teste de McNemar é usado para analisar a eficiência de determinada técnica, isto é, tem como objetivo avaliar a eficiência de situações, “o antes e o depois”, em que cada amostra é utilizada e a mensuração se faz ao nível de uma escala nominal ou ordinal. Este teste é aplicado a variáveis dicotômicas, ou seja, à amostras que apenas tomam dois valores, por exemplo, 0 e 1 \cite{farias2015, kleinner2012, firmino2015}.

\begin{table}[htbp]
\scalefont{0.9}
  \centering
  \caption{Estatística descritiva e testes estatísticos.}\label{testeEstatistico}
    \begin{tabular}{lllllllllccc}
    \toprule
    \multicolumn{1}{c}{\textbf{Variáveis}} & \multicolumn{1}{c}{\textbf{Tratanento}} & \multicolumn{1}{c}{\textbf{DP}}        & \multicolumn{1}{c}{\textbf{Mín}}       & \multicolumn{1}{c}{\textbf{25th}}      & \multicolumn{1}{c}{\textbf{MD}}        & \multicolumn{1}{c}{\textbf{75th}}      & \multicolumn{1}{c}{\textbf{Máx}}       & \multicolumn{1}{c}{\textbf{Méd}}       & \textbf{\%dif}                         & \textbf{W}                             & \textbf{N} \\
                                           &                                        &                                        &                                        &                                        &                                        &                                        &                                        &                                        &                                        & p-v                                    & p-v  \\
    \midrule
    Esforço                                & Manual                                 & 2,00                                   & 4,00                                   & 5,00                                   & 6,00                                   & 7,00                                   & 12,0                                   & 6,36                                   & \multirow{2}[2]{*}{70,13}              & \multirow{2}[2]{*}{<0,001}             & \multirow{2}[2]{*}{-} \\
                                           & Semiaut.                               & 1,66                                   & 0,00                                   & 0,00                                   & 2,00                                   & 3,00                                   & 5,00                                   & 1,90                                   &                                        &                                        &  \\
    \midrule
    Corretude                              & Manual                                 & 0,50                                   & 0,00                                   & 0,00                                   & 1,00                                   & 1,00                                   & 1,00                                   & 0,53                                   & \multirow{2}[2]{*}{43,01}              & \multirow{2}[2]{*}{-}                  & \multirow{2}[2]{*}{0,002} \\
                                           & Semiaut.                               & 0,25                                   & 0,00                                   & 1,00                                   & 1,00                                   & 1,00                                   & 1,00                                   & 0,93                                   &                                        &                                        &  \\
    \bottomrule
    \end{tabular}%
	\fonte{Elaborada pelo autor.}
\end{table}%
\textbf{RQ1:QP1: Esforço de Integração.} A primeira demanda investigada procura saber o impacto do protótipo, FMIT, analisando o esforço dos analista e desenvolvedores na detecção e resolução de conflitos para uma composição correta.  Ao verificar os dados contidos na Tabela \ref{esforcoAgrupado} e Tabela \ref{quantificacao}, diagnosticou-se que o esforço aplicado pelos participantes para integrar modelos de \textit{features} foi inferior na aplicação da técnica manual em relação à técnica semiautomática.  A tabela \ref{quantificacao} exibe o esforço médio por cenário, bem como o registro acumulado em “Todas” na linha de esforço, a média de esforço empregada para a identificação e resolução de conflitos é de 6,36 (minutos) na aplicação da técnica manual e 1,90 (minutos) na aplicação da técnica semiautomática. A redução do esforço aplicado no uso da técnica semiautomática, bem como no protótipo, FMIT, também é observado na Tabela \ref{testeEstatistico} onde exibe a diferença de 70,13\% no esforço, ou seja, o esforço empregado para atingir o resultado  pretendido com o protótipo, FMIT, é 70,13\% menor do que o registrado quando usando a técnica manual. Este esforço também pode ser observado comparando as medianas de 6,00 (minutos) para a técnica automática e apenas 2,00 (minutos) para a técnica semiautomática, que se repetem nas médias.

\begin{table}[htbp]
  \centering
  \caption{Resultado do esforço – teste \textit{Wilcoxon}.}\label{testeWilcoxon}
    \begin{tabular}{llcc}
    \toprule
    \multicolumn{4}{c}{\textbf{Teste de Wilcoxon}} \\
    \midrule
    \multicolumn{2}{l}{Pares(n)} & \multicolumn{2}{c}{30} \\
    \multicolumn{2}{l}{Z}   & \multicolumn{2}{c}{-4,7821} \\
    \multicolumn{2}{l}{p-Valor} & \multicolumn{2}{c}{< 0,001} \\
    \bottomrule
    \end{tabular}%
  \label{tab:addlabel}%
  \fonte{Elaborada pelo autor.}
\end{table}%

Analisando a H1-0 tem-se para a hipótese nula o teste de \textit{wilcoxon}, não paramétrico que pode ser visto na Tabela \ref{testeEstatistico} e Tabela \ref{testeWilcoxon}. É possível verificar que a estatística coletada da significância é < 0,001, com um intervalo de confiança de 95\% em uma amostra de 30 pares. 

Para analisar a H1-1 são considerados os dados da Tabela \ref{esforcoAgrupado}, Tabela \ref{quantificacao} e Tabela \ref{testeEstatistico}. Pode-se concluir que o tempo médio para chegar a uma resposta é menor na aplicação da técnica semiautomática, ficando em 1,90 (minutos), enquanto que na técnica manual o tempo médio para a identificação e resolução de conflitos ficou em 6,36 minutos. O percentual de esforço é exibida na Tabela \ref{quantificacao}, onde se pode verificar que o esforço na aplicação da técnica manual por cenário é superior a técnica semiautomática. Assim, é possível identificar que os participantes ao utilizar o protótipo, FMIT,  conseguiram chegar a uma conclusão com maior eficácia. Verifica-se nessa hipótese que o esforço na identificação de conflitos, com a técnica semiautomática é menor 61,03\% comparada a técnica manual. Assim pode-se concluir que essa hipótese é válida e os analistas e desenvolvedores aplicam menos esforços com a aplicação da técnica semiautomática.

Colocando em escala os resultados desta análise do esforço empregado por cenário, conforme as Tabelas \ref{resultadoManual} e Tabela \ref{resultadoSemiautomático} obtém-se o gráfico da Figura \ref{fig:tmpCenario}. A figura evidencia mais nitidamente a diferença de esforço para detecção e análise de conflitos na comparação de modelos de \textit{features}, entre a aplicação da técnica manual e semiautomática. Observa-se que no caso da técnica manual o esforço decorrido é bem superior, confirmando as discussões anteriores.

Respondendo então a primeira questão de pesquisa o protótipo, FMIT, com base na técnica semiautomática, influencia nos esforços investidos pelos analistas e desenvolvedores na identificação e resolução de conflitos durante o processo de integração, reduzindo o esforço empregado.

\textbf{RQ2:QP2: Respostas Corretas.} A segunda demanda investigada procura saber o impacto na aplicação técnica semiautomática, se os modelos de \textit{features} produzidos foram corretamente integrados, determinando assim a eficácia e efetividade na qualidade do processo de integração, uma vez que os conflitos existentes prejudicam a compreensibilidade dos modelos aumentando o risco de atrasos em projetos de software devido a retrabalho, bem como poderá elevar os custos de produção. Visualizando os dados na da Tabela \ref{testeEstatistico} pode-se observar que as médias da corretude ficou superior com o uso do protótipo, FMIT, comparado à técnica manual, isto é, uma média de 0,53 (técnica manual), em comparação com a média de 0,93 (técnica semiautomática), entretanto as medianas são equivalente, isto é, são iguais a 1. Este fato ocorre devido à condução do experimento, onde o valor igual a 1 considera o modelo correto e valor igual a 0 considera o modelo incorreto. Na Tabela \ref{quantificacao} é possível ver os percentuais dos modelos \textit{features} corretamente integrados com a técnica semiautomática é superior se comparado com a técnica manual, ou seja, os participantes do experimento melhoram a assertividade dos modelos em 43\%.

\begin{table}[htbp]
  \centering
  \caption{Resultado do esforço – teste \textit{McNemar}.}\label{testeMcNemar}
    \begin{tabular}{llcc}
    \toprule
    \multicolumn{4}{c}{\textbf{Teste de Wilcoxon}} \\
    \midrule
    \multicolumn{2}{l}{Pares(n)} & \multicolumn{2}{c}{30} \\
    \multicolumn{2}{l}{$x^{2}$}   & \multicolumn{2}{c}{10,08} \\
    \multicolumn{2}{l}{p-Valor} & \multicolumn{2}{c}{0,002} \\
    \bottomrule
    \end{tabular}%
  \label{tab:addlabel}%
  \fonte{Elaborada pelo autor.}
\end{table}%

Analisando a H2-0, tem-se para investigar a hipótese nula, o teste de \textit{McNemar} não paramétrico que pode ser observado na Tabela \ref{testeEstatistico} e Tabela \ref{testeMcNemar}. É possível verificar que a estatística coletada da significância é 0,002, com um intervalo de confiança de 95\% em uma amostra de 30 pares. Este valor indica que a segunda hipótese nula (H2-0) pode ser rejeitada. Como o valor-p é menor que 0,05, pode-se concluir que há evidências de que a técnica semiautomática é significativamente mais eficaz do que a técnica manual. 

Analisando a H2-1 são considerados os dados da Tabela \ref{esforcoAgrupado}, Tabela \ref{corretudeAgrupado} e Tabela \ref{testeEstatistico}. Pode-se concluir que a técnica semiautomática, a corretude em média é 0,93 (acertos), superior a técnica manual em média é 0,53 (acertos). Os percentuais de corretude são exibidos na Tabela \ref{quantificacao}, onde pode-se confirmar o melhor desempenho da aplicação semiautomática em relação a técnica manual. Igualmente é confirmado um aproveitamento de 43,01\% realizando as integrações de maneira semiautomática, com base na técnica e no protótipo proposto, aumentando a corretude dos modelos de \textit{features}. Portanto, vê-se que a grande maioria dos participantes concluiu a integração dos modelos corretamente. A busca pela eficácia juntamente com a efetividade de compor os modelos é facilitada com o uso do protótipo, FMIT, ou seja a técnica semiautomática conduz melhor o nível de assertividade de integração comparada a técnica manual. Assim pode-se concluir que essa hipótese é válida e os analistas e desenvolvedores obtiveram maior êxito na assertividade dos modelos compostos, aplicando uso da técnica semiautomática.

Colocando em escala os resultados desta análise de corretude, conforme as Tabelas \ref{resultadoManual} e a Tabela \ref{resultadoSemiautomático} obtém-se o gráfico da Figura \ref{fig:graficoCorretude}. A figura evidencia mais nitidamente a diferença na obtenção dos modelos de \textit{features} pretendido, isto é, o modelo produzido corretamente entre a aplicação da técnica semiautomática e manual, confirmando os resultados acima.

Então respondendo a segunda questão de pesquisa pode-se concluir que o protótipo proposto, FMIT, com base na técnica semiautomática, afeta a eficácia e efetividade dos modelos produzidos entre os analistas e desenvolvedores  de forma a aperfeiçoar  a integração de modelos, positivamente aumentando o número de modelos corretamente produzidos. 

\textbf{Discussão.} Nessa avaliação procura-se em cada questão explorar diferentes situações para integrar modelos de \textit{features}, com o intuito de verificar as necessidades de aperfeiçoar a técnica para integração entre dois modelos de entrada, assim como calcular o esforço empregado para solucionar conflitos e validar a corretude das integrações. 

Com base nos testes estatísticos executados durante este experimento, foi possível avaliar as hipóteses testadas nesta Seção, que se referem ao esforço e a corretude das integrações, relacionados às questões de pesquisa descritas na Seção \ref{IntQuestõesdePesquisa}. Os resultados evidenciam claramente a rejeição das duas hipóteses nulas formuladas. Assim como, foram demostrados indícios para justificar a validade das hipóteses alternativas, os resultados contribuíram para o atendimento dos objetivos específicos, citados na Seção \ref{I_Objetivos} deste estudo. 

No entanto, fazendo uma visão geral é possível concluir que de acordo com os resultados de todos os cenários o protótipo proposto, FMIT, baseado na técnica semiautomática aumentou o índice de respostas corretas em 43,01\% e reduziu o esforço empregado  em 70,13\%. Os resultados apontam para o beneficio na execução das integrações automatizadas, porém cabe aprofundar os estudos, visto ser necessário investigar a aplicação da técnica proposta com outras ferramentas de automatização, assim como aplicar os testes no setor industrial em grandes escalas de produção, para de fato tomar conhecimento de sua aplicabilidade.

Todos os participantes desta pesquisa submeteram-se anteriormente a participar de um pequeno \textit{workshop} que durou aproximadamente 20 minutos, onde se realizou uma explanação sobre os modelos de \textit{features}, como se comporta, quais relacionamentos existentes e por fim exemplos de integração. Assim buscou-se balizar o conhecimento entre todos os indivíduos, sendo estes em sua maioria de sexo masculino 90\%, 10\% tem experiência de menos de dois anos como desenvolvedor e analista, e 70\% por cento atuam com profissionais e estudantes em tempo integral.  O experimento contou com a participação de 10 indivíduos, todos os estudantes do curso de pós-graduação em Computação Aplicada da universidade Unisinos. Para desempenhar uma análise mais criteriosa será de suma importância expandir o número de participantes na condução deste experimento.

Quanto à validade da conclusão estatística, pode-se afirmar que diretrizes experimentais foram seguidas para eliminar ameaças pressupostas aos testes estatísticos (teste de \textit{Wilcoxon }e o teste de \textit{McNemar}). A homogeneidade dos indivíduos foi assegurada. O método de quantificação empregado neste experimento se fez com o uso de ferramentas estatísticas. Os testes de \textit{Shapiro-Wilk} e \textit{Kolmogorov-Smirnov} foram aplicados e os mesmos não aderem a uma distribuição normal, assim utilizaram-se testes não paramétricos.

Finalizando pode-se concluir que a técnica e o protótipo proposto, FMIT, nesta  pesquisa demostraram que, com a execução da avaliação e seus resultados melhoraram significativamente  a eficácia no esforço  empregado durante o processo de integração na mitigação de conflitos, assim como aumentaram  a eficiência e efetividade  na corretude dos modelos produzidos, respondendo de forma satisfatória às questões de pesquisa elaboradas neste estudo. 

\section{Ameaças à Validade do Estudo} \label{AvAmeaças}

Esta seção discute algumas possíveis limitações e ameaças a validade do estudo. A aplicação de experimento demanda da necessidade de analisar o quão valido são os métodos e os participantes selecionados, a forma como foram aplicados e avaliados, e os resultados deparados. Conforme \citetexto{wohlin2012}, essa constatação ocorre através da avaliação de quatro tipos de validade: validade da conclusão, validade da construção e por fim a validade as ameaças internas e externas. 

\textbf{Validade de conclusão.} A validade considera a capacidade de obter a conclusão correta apoiada nos resultados oriundos do experimento, através das escolhas dos métodos estatísticos pressupondo o tamanho da amostra e a confiabilidade das medidas \cite{wohlin2012}. Nesse critério o experimento produziu uma amostra de 60 modelos de \textit{features}, cuja análise estatística descritiva identificou que os dados não aderem a uma distribuição normal, assim aplicando os testes não paramétricos, ou seja, o teste de \textit{Wilcoxon}, para mensurar o esforço empregado na detecção de conflitos e o teste de \textit{McNemar}, para mensurar a corretude dos modelos produzidos, com um intervalo de confiança de 95\% conforme detalhado na Seção \ref{AvAnálise}.

\textbf{Validade da construção.} Leva em consideração a relação entre a teoria e a observação, ou seja, os resultados obtidos através de questionários ou experimentos e possuem relação com as expectativas da teoria estudada \cite{kleinner2012, farias2015, wohlin2012, kitchenham2010}. Assim, os experimentos realizado nesta pesquisa foram planejados para mensurar o esforço de integração, bem como quantificar a corretude dos modelos produzidos pelos participantes, sendo está estratégia já adotada na literatura como, por exemplo, o trabalho de \citetexto{farias2015, kleinner2012}. A fim de atender os procedimentos de execução e correção do experimento, foram cuidadosamente planejados, seguindo as boas práticas de quantificação conforme \citetexto{wohlin2012, kitchenham2010, kitchenham2011}.

\emph{Validade externa.} Leva em consideração as condições que permitem generalizar os resultados do experimento para a prática industrial ou para o mais realista possível \cite{wohlin2012, farias2013}. Nesse contexto, selecionaram-se os participantes que possuem formação adequada para a prática da atividade relacionada ao experimento, isto é, todos os participantes são mestrandos em Computação Aplicada, além de possuir boa experiência na área de desenvolvimento ou modelagem de software, conforme detalhado na Seção \ref{AvContexto}. Além disso, as ferramentas e equipamentos utilizados no experimento são equivalentes ao praticado na indústria, sendo que a aplicação de todo o experimento ocorreu nas dependências da universidade.   

Por fim, apesar dos cuidados tomados em relação às ameaças de validade ao estudo, citados nesta Seção, não se pode afirmar que os resultados encontrados podem ser generalizados para modelos maiores, uma vez que os selecionados para este experimento são considerados pequenos, conforme detalhado na Tabela \ref{CenárioExperimento}.

 \chapter{Conclusões} \label{Conclusões}
 
 No decorrer do desenvolvimento deste trabalho é essencial o entendimento sobre conceito de Modelos de \textit{Features}. Conforme ressaltado no Capítulo \ref{I_introdução}, a integração de modelos apresenta desafios para a Engenharia de Software devido à dinamicidade e adaptabilidade dos requisitos necessários para o desenvolvimento de novos produtos. Nesse cenário, o reuso revelar-se como um recurso eficiente para o desenvolvimento de aplicações. Entretendo, ao executar a composição de modelos de \textit{features} demostra-se que este não cobre as eventuais necessidades devidas as alterações comportamentais oriunda de novos requisitos.
 
 A condução desta proposta procurou identificar as técnicas aplicadas para realizar a integração de modelos de \textit{features}, através de um mapeamento sistemático da literatura, assim como realizar um experimento controlado para verificar o comportamento de estudantes e profissionais, sejam estes analistas ou desenvolvedores, quando um conjunto de modelos de \textit{features} sofre adaptações e como estas mudanças poderão intervir na integração de modelo de \textit{features}, ressaltando que a criação de um novo modelo poderá causar impacto na derivação de novos produtos.  

Após a revisão dos trabalhos relacionados identificou-se trabalhos comprovando que o esforço necessário para integrar modelos de \textit{features} ocorre de forma empírica sendo que a maior parte dos estudos executa esta transição de forma generalizada e automática, as regras de integração já se encontram estabelecidas. As técnicas aplicadas para compor o modelo de \textit{features} acabam originando como saída um modelo final em alguns casos indesejado, sem levar em consideração necessidades específicas dos analistas e desenvolvedores.

Através do experimento controlado foi possível determinar a qualidade final dos modelos produzidos, ou seja, sua corretude e o esforço empregado entre analistas e desenvolvedores na detecção de conflitos, apesar de considerar o tempo diagnosticado manualmente como sendo baixo, isto é, 6,36 minutos por questão composta, pois com a aplicação da técnica semiautomática os participantes conseguiram uma redução significativa, passando para 1,90 minutos por questão, o que representa um percentual de 70,13\%, mais eficiente com a aplicação da técnica semiautomática, elevando assim a capacidade produtiva dos participantes. O percentual de corretude é considerado eficaz, a técnica semiautomática produziu 93\% de modelos corretos, contra 53\% aplicando a técnica manual, assegurando um aumento de 43,01\% na qualidade dos modelos produzidos. Assim, evidenciando a efetividade na resolução dos conflitos surgidos no decorrer de sua execução tem-se como melhorar significativamente a precisão dos modelos corretamente integrados. 

Os resultados do estudo revelam que as intervenções da técnica proposta  e as iterações empregadas melhoram significativamente o tempo de conclusão da compreensão e a integração dos modelos de \textit{features}. Além disso, de acordo com os resultados as intervenções propostas são fáceis de usar e fáceis de aprender para os participantes.

O modelo de \textit{features} é um dos principais artefatos da Engenharia \cite{ghanam2010, acher2010}, consequentemente a análise deste modelo ocorre nos primeiros estágios do processo de desenvolvimento, sendo essencial para o sucesso no desenvolvimento das linhas de produto. Assim, conclui-se que o protótipo, FMIT, reduz o esforço empregado, bem como aumenta qualidade final das integrações de modelos de \textit{features}, minimizando os conflitos propagados durante sua composição e reduzindo os riscos iniciais do projeto.

\section{Mapeamento dos Trabahos Relacionados} \label{sec:CON_mapeamentoTrabahosRelacionados}

Neste trabalho conduziu-se um estudo de mapeamento sistemático da literatura, sobre a integração de modelos de \textit{features},onde a pesquisa inicial retornou 775 publicações e 34 publicações foram selecionadas como estudos primários após um cuidadoso processo de filtragem, nos quais se investigaram seis questões de pesquisa. Resumidamente este estudo consiste em apresentar um conjunto de características para definição das técnicas de integração entre modelos de \textit{features} aplicadas ao protótipo. Os estudos primários indicaram as oportunidades de pesquisa, assim como demonstraram como estas características se manifestam ao longo dos trabalhos investigados.

\begin{table}[htbp]
  \centering
  \caption{Comparação das técnicas investigadas.} \label{tab:compTecInv}
    \begin{tabular}{llrl}
    \toprule
    \multicolumn{3}{c}{\textbf{Técnicas Investigada}}                                                                                                         & \multicolumn{1}{c}{\textbf{Protótipo - FMIT}} \\
    \midrule
    \multirow{2}[1]{*}{1 – Técnica de Automatização } & Automático                                        & Sim                                               & Sim \\
                                                      & Semiautomático                                    & Sim                                               & Sim \\
    \multirow{4}[0]{*}{2 - Técnicas de Comparação}    & Sintática                                         & Sim                                               & Sim \\
                                                      & Semântica                                         & Sim                                               & Sim \\
                                                      & Estrutural                                        & Não                                               & Sim \\
                                                      & Equivalência                                      & Não                                               & Sim \\
    \multirow{4}[0]{*}{3 - Técnicas de Integração}    & União                                             & Sim                                               & Sim \\
                                                      & Intersecção                                       & Sim                                               & Sim \\
                                                      & Diferença                                         & Sim                                               & Sim \\
                                                      & Complemento                                       & Não                                               & Sim \\
    \multirow{3}[0]{*}{4 - Notações}                  & FODA                                              & Sim                                               & Sim \\
                                                      & FORML                                             & Sim                                               & Não \\
                                                      & Outros                                            & Sim                                               & Não \\
    \multirow{3}[0]{*}{5 - Técnicas de Configuração}  & CSP                                               & Sim                                               & Não \\
                                                      & SAT                                               & Sim                                               & Não \\
                                                      & Multicritério                                     & Sim                                               & Não \\
    \multirow{6}[1]{*}{6 – Métodos de Pesquisa}       & Proposta de Solução                               & Sim                                               & Não \\
                                                      & Pesquisa de Avaliação                             & Sim                                               & Sim \\
                                                      & Pesquisa de Validação                             & Sim                                               & Não \\
                                                      & Artigos de Opinião                                & Sim                                               & Não \\
                                                      & Artigos Filosóficos                               & Sim                                               & Não \\
                                                      & Artigos de Experiência                            & Sim                                               & Não \\
    \bottomrule
    \end{tabular}%
  \label{tab:addlabel}%
  \fonte{Elaborada pelo autor.}
\end{table}%

A Tabela \ref{tab:compTecInv} apresenta as variáveis das 06 questões de pesquisa descritas na Seção \ref{TR_SubQuestões}, bem como os resultados obtidos na investigação, comparados as técnicas implementadas ao protótipo, FMIT, proposto.

\textbf{Técnica de Automatização.} A primeira questão procura investigar o grau de automatização, a maioria dos trabalhos aplica-se a técnica automática, e o protótipo proposto implementa ambas as técnicas.  

\textbf{Técnicas de Comparação.}  A segunda questão investiga as técnicas de comparação, boa parte dos trabalhos executa comparações sintáticas e semânticas durante a composição de modelos, sendo que há duas lacunas em aberto nos trabalhos, em \citetexto{kastner2009, thum2014} viabilizam a ordem em que \textit{features} ou produtos encontram-se, devido as possíveis configurações que possam vir a existir, assim a estrutura da \textit{features} neste estudo é considerada relevante para aprimorar a qualidade das integrações.  Finalmente o grau de equivalência calcula a diferença entre modelos comparados, ou seja, sintático semântico, e estrutural \cite{oliveira2008}. Ao aplicar a técnicas de similaridade, esta viabiliza facilmente verificar nível ou taxa de conflitos ao comparar os modelos, assim tornando possível executar uma ação, sejam ela implementada pelo protótipo ou pelos analistas e desenvolvedores. 

\textbf{Técnicas de Integração.} As técnicas de integração aplicadas normalmente são derivadas do grau de equivalência e executadas automaticamente. A técnica de união ocorre quando a necessidade de incluir novos elementos ao modelo e a interseção, quando é necessário excluir elementos incomuns no modelo analisado, estas são as técnicas mais utilizadas na literatura, por fim a técnica de diferença e complemento que é pouco aplicada na literatura ou mesmo não utilizada. No protótipo ambas as técnicas são consideradas e aplicadas na escolha automática, assim dando a possibilidade dos analistas e desenvolvedores verificarem um  gama maior de composições, ou até mesmo para analisar as diferenças entre os modelos comparados , no caso da aplicação das técnicas de diferença e complemento.

\textbf{Notações.}  A notação FODA \cite{kang1990} é a mais aplicada na literatura, sendo desta técnica as demais variações existentes, há um vasto número de notações descritas na literatura. O protótipo seguiu o modelo FODA, por ser amplamente conhecido, bem como a sua facilidade de interpretação e a ferramenta incorporada ao protótipo, FeatureIDE, também faz uso desta notação. 
 
\textbf{Técnicas de Configuração.}  As técnicas de configurações e validação são amplamente investigadas na literatura, pois o nível de maturação é considerado elevado \cite{benavides2010}. Há um número significativo de ferramentas que forcem suporte para a criação, edição, análise e configuração dos modelos de \textit{features}, como por exemplo, SPLOT \cite{mendonca2009}, FeatureIDE \cite{thum2014}, fmp \cite{czarnecki2005}, pure::variants \cite{beuche2012}. Assim optou-se em utilizar a ferramenta FeatureIDE é framework baseado em Eclipse a qual abrange várias atividade no processo de desenvolvimentos das LPS.  

\textbf{Métodos de Pesquisa. } O trabalho em questão é classificado de acordo com métodos de pesquisa descritos na Seção \ref{TR_SubExtração}. Este estudo é classificado como uma pesquisa de avaliação, onde as técnicas e soluções são implementadas e avaliadas na prática, e as consequências investigadas. 

\section{Contribuições} \label{CONcontribuições}  

Esta pesquisa propôs a aplicação de uma técnica de integração de modelos de \textit{features}, que compreende um conjunto de técnicas propostos na literatura, bem como o desenvolvimento de um protótipo, \textit{Feature Model Integration Tool – FMIT}, assim, procura-se aperfeiçoar  as técnicas de integração e flexibilizar o processo de composição. Com o auxílio do protótipo, os analistas e desenvolvedores realizaram as tarefas com maior eficiência, impactando diretamente em sua capacidade produtiva, isto é, otimizando melhor o tempo de produção, assim como possibilitou um aumento na precisão dos modelos produzidos, oriundos da diminuição de conflitos. Com estes resultados acredita-se em uma redução nos custos de produção e uma importante melhoria decorrida do processo de integração que determinam a qualidade final do produto gerado, estabelecendo assim a primeira contribuição científica deste estudo.

A produção de conhecimento empírico sobre integração de modelos de \textit{features}, apresentada pelos estudos experimentais de integração realizados manualmente ou de forma semiautomática, proporcionaram medir os esforços gerados e o taxa de modelos de \textit{features }corretamente produzidos, é a segunda contribuição científica deste estudo.

Por último, através de um mapeamento sistemático da literatura, construiu-se uma base de conhecimento referente às técnicas de integração de modelos de \textit{features} aplicadas na literatura, abrangendo 34 estudos primários. Este estudo reúne em um único trabalho uma descrição dos principais avanços e desafios para futuras pesquisas relacionadas neste campo de atuação.

%

\section{Limitações do Trabalho} \label{LimitaçõesTrabalho}  

Este estudo investigou no seu decorrer questões essencialmente sobre integração de modelos de \textit{features}. Embora as contribuições deste estudo sejam percebíveis, sua abordagem necessita de melhorias, uma vez que o tempo de desenvolvimento deste estudo não é o suficiente para cobrir todas as lacunas existentes. Assim, são apresentadas as principais limitações identificadas neste estudo, bem como sugestões de trabalhos futuros em virtude de cobrir estes pontos.

\textbf{Limitações Experimentais.}  Há necessidade de expandir o número de participantes bem como ampliar a participação de profissionais ligados ao setor industrial nas pesquisas.  Os modelos \textit{features} utilizados nos cenários apresentados possuem uma quantidade reduzida de elementos, ou seja, são pequenos modelos que em sua maioria não condiz com a atualidade no âmbito industrial. Portando, como trabalho futuro verifica-se a necessidade implantar modelos mais ricos, que apresentem o número maior de elementos, com isso, possibilitando realizar testes de escalabilidade e desempenho de integração. 

\textbf{Limitações do Protótipo.}  Apesar, do protótipo realizar a integração de forma  ampla, o mesmo encontra-se limitado a uma única notação, ou seja, a FODA \cite{kang1990}, sendo que neste estudo identificou-se um grande número de notações. Destacam-se entre as mais conhecidas 14 notações derivadas, conforme ilustrado na Figura \ref{fig:hierarquiaca}, no Capítulo \ref{TrabalhosRelacionados} deste estudo. Assim, sugere-se como trabalho futuro ampliar o suporte a um número maior de notações, por exemplo, FORM, CBFM, e PLUSS. Após, a composição dos modelos, o protótipo retorna como saída um arquivo textual com informações sobre as equivalências entre os modelos comparados, percentual de diferença sintática, semântica e estrutural individualizado, bem como similaridade geral dos modelos. Por fim, o modelo pretendido é apresentado no formato textual, ou seja, um vetor. É condizente melhorar a forma de armazenagem para a coleta de buscas futuras, assim como exibir graficamente o modelo produzido. 

\textbf{Limitações da Técnica.} Apesar, da técnica semiautomática para a integração de modelo de \textit{features} apresentar-se eficaz em relação a técnica manual, encontra-se a necessidade de ampliar e aprofundar a fase dos estudos, isso ocorre devido ao fato das decisões sofrerem interferência humana e estarem sujeitas ao erro, conforme demostrado no experimento, ou seja, aproximadamente 7\% dos modelos foram produzidos incorretamente. Assim, como trabalho futuro observa-se a necessidade de novos experimentos face à execução da técnica automática, tendo ênfase em analisar sua precisão e acurácia em relação a técnica proposta, e por fim acoplar outras técnicas ao protótipo, como por exemplo  o uso da inteligência artificial, visto o desafio de tomar melhores e mais precisas as decisões dos  modelos de \textit{features} integrados.

\appendix
\chapter{Lista de estudos primários}
Os 34 artigos selecionados como estudos primários na revisão do mapeamento sistemático estão listados abaixo.
\begin{description}
\item[S01.]Moisan, S., Rigault, J.-P., and Acher, M. (2012) A feature-based approach to system deployment and adaptation. \textit{Proceedings of the 4th International Workshop on Modeling in Software Engineering}, pp. 84-90.
\item[S02.]Andersen, N., Czarnecki, K., She, S., and Wasowski, A. (2012) Efficient synthesis of feature models.
\textit{Proceedings of the 16th International Software Product Line Conference-Volume 1}, pp. 106-115. 
\item[S03.]Acher, M., Collet, P., Lahire, P., and France, R. (2009) Composing feature models. \textit{International Conference on Software Language Engineering}, pp. 62-81. 
\item[S04.]Urli, S., Blay-Fornarino, M., Collet, P., and Mosser, S.(2012) Using composite feature models to support agile software product line evolution.\textit{ Proceedings of the 6th International Workshop on Models and Evolution}, pp. 21-26.
\item[S05.]Sayyad, A. S., Menzies, T., and Ammar, H. (2013) On the value of user preferences in search-based software engineering: a case study in software product lines. \textit{2013 35th International Conference on Software Engineering (ICSE)}, pp. 492-501. 
\item[S06.]Khalfaoui, K., Kerkouche, E., Chaoui, A., and Foudil, C. (2015) Automatic generation of spl structurally valid products: An approach based on progressive composition of partial configurations. \textit{Information and Communication Systems (ICICS), 2015 6th International Conference on}, pp. 25-31. 
\item[S07.]Beidu, S., Atlee, J. M., and Shaker, P. (2015) Incremental and commutative composition of state-machine models of features. \textit{2015 IEEE/ACM 7th International Workshop on Modeling in Software Engineering}, pp. 13-18. 
\item[S08.]Schw\"agerl, F., Uhrig, S., and Westfechtel, B. (2015) A graph-based algorithm for three-way merging of ordered collections in emf models. \textit{Science of Computer Programming}, 113, pp. 51-81.
\item[S09.]Acher, M., Heymans, P., Collet, P., Quinton, C., Lahire, P., and Merle, P. (2012) Feature model differences. \textit{International Conference on Advanced Information Systems Engineering}, pp. 629-645. 
\item[S10.]Dao, T. M. and Kang, K. C. (2010) Mapping features to reusable components: A problem frames based approach. \textit{International Conference on Software Product Lines}, pp. 377-392. 
\item[S11.]Guo, J. and Wang, Y. (2010) Towards consistent evolution of feature models. \textit{International Conference on Software Product Lines}, pp. 451-455. 
\item[S12.]Westfechtel, B. (2014) Merging of emf models. \textit{Software and Systems Modeling}, 13, pp. 757-788.
\item[S13.]Scholz, W., Th\"um, T., Apel, S., and Lengauer, C. (2011) Automatic detection of feature interactions using the java modeling language: an experience report. \textit{Proceedings of the 15th International Software Product Line Conference, Volume 2} p. 7. 
\item[S14.]Quinton, C., Romero, D., and Duchien, L. (2013) Cardinality-based feature models with constraints: a pragmatic approach. \textit{Proceedings of the 17th International Software Product Line Conference}, pp. 162-166. 
\item[S15.]Batory, D., Hofner, P., and Kim, J. (2011) Feature interactions, products, and composition. \textit{ACM SIGPLAN Notices}, pp. 13-22.
\item[S16.]Asadi, M., Bagheri, E., Mohabbati, B., and Gasevic, D. (2012) Requirements engineering in feature oriented software product lines: an initial analytical study. \textit{Proceedings of the 16th International Software Product Line Conference-Volume 2}, pp. 36-44. 
\item[S17.]Sarinho, V. T. and Apolin\'ario, A. L. (2010) Combining feature modeling and object oriented concepts to manage the software variability. \textit{Information Reuse and Integration (IRI), 2010 IEEE International Conference on}, pp. 344-349. 
\item[S18.]Kolesnikov, S., von Rhein, A., Hunsen, C., and Apel, S. (2013) A comparison of product-based, feature-based, and family-based type checking. \textit{ACM SIGPLAN Notices}, pp. 115-124. 
\item[S19.]Soltani, S., Asadi, M., Gasevic, D., Hatala, M., and Bagheri, E. (2012) Automated planning for feature model configuration based on functional and non-functional requirements. \textit{Proceedings of the 16th International Software Product Line Conference Volume 1}, pp. 56-65. 
\item[S20.]Benavides, D., Segura, S., and Ruiz-Cort\'es, A. (2010) Automated analysis of feature models 20 years later: A literature review. \textit{Information Systems}, 35, pp. 615-636.
\item[S21.]Th\"um, T., K\"astner, C., Benduhn, F., Meinicke, J., Saake, G., and Leich, T. (2014) Featureide: An extensible framework for feature-oriented software development. \textit{Science of Computer Programming}, 79, pp. 70-85.
\item[S22.]Ensan, F., Bagheri, E., and Gasevic, D. (2012) Evolutionary search-based test generation for software product line feature models. \textit{International Conference on Advanced Information Systems Engineering}, pp. 613-628. 
\item[S23.]Budiardjo, E. K., Zamzami, E. M., et al. (2014) Feature modeling and variability modeling syntactic notation comparison and mapping. \textit{Journal of Computer and Communications}, \textbf{2}, p. 101.
\item[S24.]Acher, M., Collet, P., Lahire, P., and France, R. B. (2011) Decomposing feature models: language, environment, and applications. \textit{Proceedings of the 2011 26th IEEE/ACM International Conference on Automated Software Engineering}, pp. 600-603. 
\item[S25.]Benavides, D., Felfernig, A., Galindo, J. A., and Reinfrank, F. (2013) Automated analysis in feature modelling and product configuration. \textit{International Conference on Software Reuse}, pp. 160-175. 
\item[S26.]Czarnecki, K., Helsen, S., and Eisenecker, U. (2004) Staged configuration using feature models. \textit{International Conference on Software Product Lines}, pp. 266-283. 
\item[S27.]B\'ecan, G., Behjati, R., Gotlieb, A., and Acher, M. (2015) Synthesis of attributed feature models from product descriptions. \textit{Proceedings of the 19th International Conference on Software Product Line}, pp. 1-10. 
\item[S28.]Passos, L., Czarnecki, K., Apel, S., Wasowski, A., K\"astner, C., and Guo, J. (2013) Feature-oriented software evolution. \textit{Proceedings of the Seventh International Workshop on Variability Modelling of Software-intensive Systems} p. 17.  
\item[S29.]Teixeira, L., Borba, P., and Gheyi, R. (2013) Safe composition of configuration knowledge-based software product lines. \textit{Journal of Systems and Software}, \textbf{86}, pp. 1038-1053.
\item[S30.]Eichelberger, H., Kr\"oher, C., and Schmid, K. (2013) An analysis of variability modeling concepts: Expressiveness vs. analyzability. \textit{International Conference on Software Reuse}, pp. 32-48.
\item[S31.]K\"astner, C., Apel, S., and Ostermann, K. (2011) The road to feature modularity? \textit{Proceedings of the 15th International Software Product Line Conference, Volume 2}, p. 5. 
\item[S32.]Classen, A., Boucher, Q., and Heymans, P. (2011) A text-based approach to feature modelling: Syntax and semantics of tvl. \textit{Science of Computer Programming}, \textbf{76}, pp. 1130-1143.
\item[S33.]Dehmouch, I. (2014) Towards an agile feature composition for a large scale software product
lines. \textit{2014 IEEE Eighth International Conference on Research Challenges in Information Science (RCIS)}, pp. 1-6. 
\item[S34.]Lesta, U., Schaefer, I., and Winkelmann, T. (2015) Detecting and explaining conflicts in attributed feature models. \textit{arXiv preprint arXiv:1504.03483},  pp. 31-43.
\end{description}
\chapter{Modelos de \textit{Features} Experimento}
Modelos propostos para integração de modelos de \textit{features} aplicados na realização do experimento, manual e semiautomático. Questionário aplicado para coleta dos dados dos participantes.
\includepdf[pages=1-4]{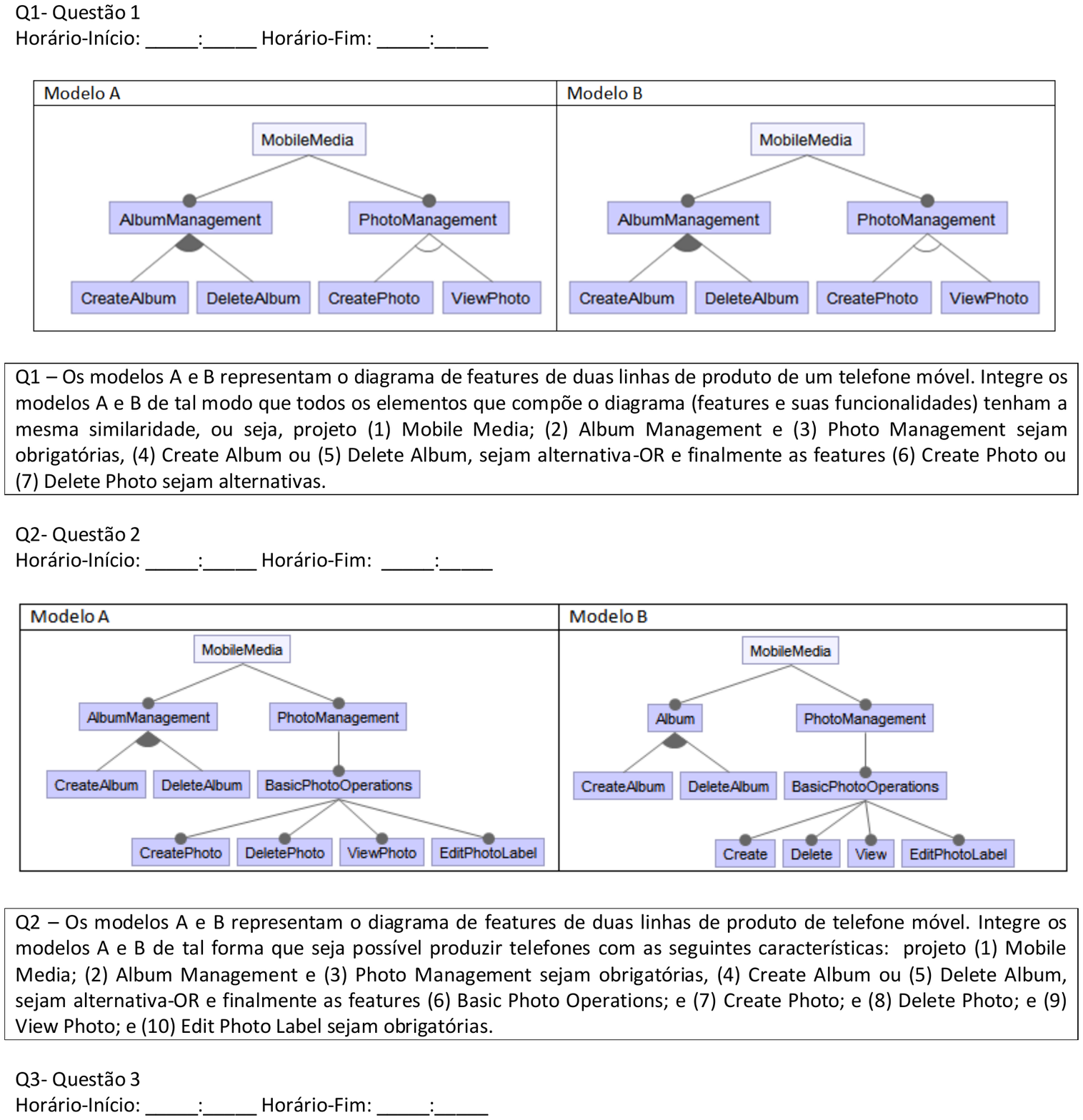}

\bibliography{reference}


\end{document}